%% file: OverfittingSurvivalAnalysis2017_revised.tex
\documentclass[10pt]{iopart}
\usepackage{epsfig,graphicx,color}
\usepackage[mathscr]{euscript}

\input commands.tex

\newcommand{\Prob}{\mathscr{P}}
\newcommand{\Data}{\mathscr{D}}

\newcommand{\rmD}{{\rm D}}
\newcommand{\fs}{\footnotesize}
\newcommand{\red}{}

\begin{document}

\title[Replica analysis of overfitting in time-to-event regression]{Replica analysis of overfitting in regression models for time-to-event data}
\author{ACC Coolen$^{\dag\ddag}$, JE Barrett$^\S$, P Paga$^{\dag\ddag}$, CJ Perez-Vicente$^\P$}

\address{
$\dag$ Department of Mathematics, King's College London, \\
\hspace*{1.5mm} The Strand, London WC2R 2LS, UK
\\[1mm]
$\ddag$  Institute for Mathematical and Molecular Biomedicine, King's College London,
\\\hspace*{1.5mm} Hodgkin Building, Guy's Campus,
London SE1 1UL, UK
\\[1mm]
$\S$ Department of Primary Care and Public Health Sciences, King's College\\
\hspace*{1.5mm} London, Addison House, Guy's Campus, London SE1 1UL, UK
\\[1mm]
$\P$ Departament de F\'{i}sica Fonamental, Universitat de Barcelona,\\
\hspace*{1.5mm} 08028 Barcelona, Spain
}


\ead{ton.coolen@kcl.ac.uk, james.barrett@kcl.ac.uk,\\
\hspace*{9.5mm}  pierre.paga@kcl.ac.uk, conrad@ffn.ub.es}

\begin{abstract}{
Overfitting, which happens when the number of parameters in a model is too large compared to the number of data points available for determining these parameters,  is a serious and growing problem in survival analysis. While modern medicine presents us with data of unprecedented dimensionality, these data cannot yet be used effectively for clinical outcome prediction. Standard error measures in maximum likelihood regression, such as p-values and z-scores, are blind to overfitting, and even for Cox's proportional hazards model (the main tool of medical statisticians), one finds in literature only rules of thumb on the number of samples  required to avoid overfitting. 
In this paper we present a mathematical theory of overfitting in regression models for time-to-event data, which aims to increase our quantitative understanding of the problem and provide practical tools with which to correct regression outcomes for the impact of overfitting.  
It is based on  the replica method, a statistical mechanical technique for the analysis of heterogeneous many-variable systems  that has been used successfully for several decades in physics, biology, and computer science, but not yet in medical statistics. We develop the theory initially for arbitrary regression models for time-to-event data, and verify its predictions in detail for the popular Cox model. }
\end{abstract}

\clearpage
\vsp

\section*{Contents}

\begin{tabbing}
zzzz\=zzzz\=zzzzzzzzzzzzzzzzzzzzzzzzzzzzzzizzzzzzzzzzzzzzzzzzzzzzzzzzzzzzzzzzzzzzzzzzz\=\kill
{\bf 1} \> {\bf Introduction} \>\> ~{\bf 3}
\\[3mm]
{\bf 2} \> {\bf Overfitting in Maximum Likelihood models for survival analysis} \>\> ~{\bf 7}
\\
\> 2.1 \> Definitions \> ~7 \\
\> 2.2 \> An information-theoretic measure of under- and over-fitting \> ~8 \\
\> 2.3 \> Analytical evaluation of the average over data sets \> ~9 \\
\> 2.4 \> Application to Cox regression \> 10
\\[3mm]
{\bf 3} \> {\bf Asymptotic analysis of overfitting in the Cox model} \>\> {\bf 12}
\\
\> 3.1 \> Conversion to a saddle-point problem \> 12 \\
\> 3.2 \> Replica symmetric extrema \> 13 \\
\> 3.3 \> Physical interpretation of order parameters \> 14 \\
\> 3.4 \> Derivation of RS saddle point equations \> 16
\\[3mm]
{\bf 4} \> {\bf Analysis of the RS equations for the Cox model} \>\> {\bf 16}
\\
\> 4.1 \> RS equations in the limit $\gamma\to\infty$  \> 16 \\
\> 4.2 \> Numerical and asymptotic solution of RS equations \> 19 \\
\> 4.3 \> Variational approximation \> 21
\\[3mm]
{\bf 5} \> {\bf Tests and applications} \>\> {\bf 26}
\\[3mm]
{\bf 6} \> {\bf Discussion} \>\> {\bf 27}
\\[3mm]
\> {\bf References} \>\> {\bf 31}
\\[3mm]
\> {\bf Appendix A: Covariate correlations in Cox regression} \>\> {\bf 31}\\
\> {\bf Appendix B: Derivation of the replica symmetric equations} \>\> {\bf 32}\\
\> {\bf Appendix C: The limits $\zeta\to 0$ and $\zeta\to 1$} \>\> {\bf 35}\\
\> {\bf Appendix D: Asymptotic form of the event time distribution} \>\> {\bf 36}
\end{tabbing}

\clearpage

\section{Introduction}

In the simplest possible scenario, survival analysis is concerned with data of the following form.   We consider a cohort of  $N$ individuals, each of whom are at risk of a specified irreversible event, such as the onset  of a given disease or death.  For each individual $i$ in this cohort we are given $p$ specific measurements $\bz_i=(z_{i1},\ldots,z_{ip})$  (the covariates) which were  taken at a baseline time $t=0$, as well as the time $t_i>0$ at which for individual $i$ we either observed the irreversible event, or we ceased  our observation without having observed the event yet (the latter case is called `censoring'). More complex scenarios could involve e.g. having multiple distinct risk types, such as distinct causes of death, or interval censoring, where rather than $t_i$ itself, one is given an interval that contains $t_i$. The theory developed in this paper can be generalised without serious difficulty  to include such extensions, but in the interest of transparency we will  focus for now strictly on the simplest case. 
\begin{eqnarray*}
&&\nonumber \\
\hspace*{-14mm}
\bz_i\in\R^p\!: &\hspace*{3mm} &p~{\it covariates~of ~individual}~i,~{\it measured~at}~t=0
\\
\hspace*{-14mm}
t_i>0\!: &\hspace*{3mm} &{\it event~time~of ~individual}~i~{\it (death, ~onset~of~disease, ...)}
\end{eqnarray*}

\hspace*{3mm}
\unitlength=0.05mm
\begin{picture}(2000,700)
\thinlines
\put(-5,83){$\bullet$}\put(0,100){\line(1,0){2000}}\put(995,84){x}\put(0,320){$\vdots$}
\put(2040,82){$i\!=\!1$}\put(2040,182){$i\!=\!2$}\put(2090,320){$\vdots$}
\put(-5,183){$\bullet$}\put(0,200){\line(1,0){2000}}\put(295,184){x}
\put(-5,483){$\bullet$}\put(0,500){\line(1,0){2000}}\put(1695,484){x}
\put(-5,583){$\bullet$}\put(0,600){\line(1,0){2000}}\put(695,584){x}
\put(2040,582){$i\!=\!N$}\put(2040,482){$i\!=\!N\!-\!1$} 
\put(-55,0){$t\!=\!0$}
\thicklines
\put(0,100){\vector(1,0){1000}}\put(0,200){\vector(1,0){400}}\put(0,500){\vector(1,0){1700}}\put(0,600){\vector(1,0){700}}
\end{picture}
\vspace*{7mm}

\noindent
The aim of  survival analysis is regression, i.e. to use our data for detecting and quantifying probabilistic patterns (if any) that relate an individual's  failure time $t$ to their covariates $\bz$. Such patterns may allow us to predict individual patients' clinical outcomes, distinguish between high-risk and low-risk patients, reveal general disease mechanisms, or design new data-driven therapeutic interventions (by changing the values of modifiable covariates). For general  reviews of the considerable survival analysis literature we refer to textbooks such as \cite{Hougaard,KleinBook,Ibrahim,Crowder}\footnote{Non-medical applications of survival analysis include e.g. the study of the time to component failure in manufacturing, or of the duration of unemployment in economics.}. Being able to use the extracted patterns to predict clinical outcomes for {\em unseen} patients is the only reliable test of whether our regression results represent true knowledge. Accurate prediction requires that we use as much of the available covariate information as possible, so our focus must  be on multivariate regression methods.

Most multivariate survival analysis methods are  based on postulating a suitable and plausible parametrisation of the covariate-conditioned event time distribution, whose parameters are estimated from the data via either the maximum likelihood protocol (ML), or (following  Bayesian reasoning) via maximum a posteriori probability (MAP). The most popular parametrisation is undoubtedly the proportional hazards model of Cox \cite{Cox}, which uses ML inference, and assumes the event time distribution to be of the so-called proportional hazards form $p(t|\bz)=-\frac{\rmd}{\rmd t}\exp[-\exp(\bbeta\!\cdot\!\bz)\Lambda(t)]$. MAP versions of \cite{Cox} are the so-called `penalised Cox' or `ridge' regression models (with Gaussian parameter priors), see e.g.  
\cite{Witten1,Witten2}. More complex parametrisation proposals, such as frailty or random effects models \cite{KeidingAndersenKlein,Vaida,Duchateau,Wienke} or latent class models \cite{MR}, 
still tend to have proportional hazards type formulae as their building blocks. 
In all such models   the number of parameters is always larger than or equal to the number $p$ of covariates. Hence, to avoid overfitting they can be used safely only when $N\gg p$. This limitation was harmless in the 1970s and 1980s, when many of the currently used models were devised, and where one would typically have datasets with  $p\sim 10^2$ at most.  For the data of  post-genome medicine, however, where we regularly have $p\sim 10^{4-6}$, it poses a serious problem which  has for instance prevented us from using genomic covariates in rigorous multivariate regression protocols, forcing us instead to work with `gene signatures'.   

\begin{figure}[t]
\unitlength=0.3mm

\hspace*{40mm}\begin{picture}(180,183)
\put(4,3){\includegraphics[height=182\unitlength]{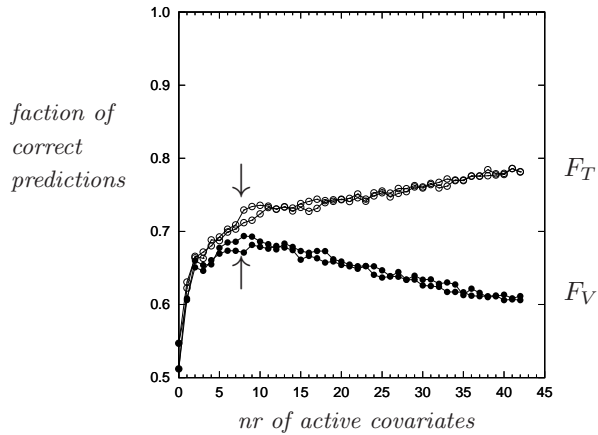}}
\put(128,-4){\small\em\here{nr of active covariates}}
\put(-25,130){\small\em faction of}
\put(-25,115){\small\em correct }
\put(-25,100){\small\em predictions}
\put(73,58){\Large{$\uparrow$}}
\put(73,99){\Large{$\downarrow$}}
\put(220,105){$F_T$}
\put(220,50){$F_V$}
\end{picture}
\vspace*{3mm}

\caption{\small Illustration of overfitting in Cox-type regression. A breast cancer data set \cite{Anita} containing $N=309$ samples (129 with recorded events, 180 censored), with clinical and immunological covariates, and disease relapse chosen as event time,  was randomly divided into training and validation sets (of roughly equal sizes). 
L2-regularised Cox regression was used to infer regression coefficients and base hazard rates  from the training set (via Breslow's formula \cite{Breslow}), upon which the model was used to predict survival at time $t=8$ years, for the samples in the training set and for those in the validation set. The fractions of correct predictions are $F_T$ and $F_V$, respectively. This was repeated multiple times, initially with all covariates, and following repeated iterative removal of the last relevant covariate after each regression. The resulting curves exhibit the standard fingerprints of overfitting: initially the validation performance improves as the number $p$ of retained covariates increases, up to a critical point (here around $p=6$, see arrows), followed by deterioration as $p$ increases further. 
}
\label{fig:overfitting}
\end{figure}

Overfitting in survival analysis models \cite{Concato,Babyak} can be visualized effectively by combining regression with cross-validation. For the Cox model, for instance, one can use the inferred association parameters $\bbeta$ of \cite{Cox} in combination with Breslow's \cite{Breslow} estimator for the base hazard rate  (which is the canonical estimator for \cite{Cox}),  to predict whether an event will have happened by a given cutoff time, and compare the fraction of correct predictions in the training set (the data used for regression) to those  in a validation set (the unseen data). When drawn as functions of the number of covariates used, the resulting curves typically exhibit the standard fingerprints of overfitting \cite{MacKay,NNbook}; see Figure \ref{fig:overfitting}. {\red Simulations with synthetic data \cite{Peduzzi} showed that the optimal number of covariates in Cox regression (see arrows in Figure  \ref{fig:overfitting}) tends to be roughly proportional to the number of samples $N$. }
Given this observed phenomenology, it seems vital before doing multivariate regression to have a tool for estimating the minimum number of samples or events needed to avoid the overfitting regime.  To our knowledge, there is no theory in the literature yet for predicting this number, not even for the Cox model \cite{Cox}. One finds only rules of thumb -- e.g. the number of failure events must exceed 10 times the number of independent covariates -- and empirical bootstrapping protocols, often based on relatively small scale simulation data \cite{Peduzzi,Kawada,Dobbin}. 
This situation is not satisfactory. 

\begin{figure}[t]
\unitlength=0.34mm
\hspace*{-10mm}
\begin{picture}(400,199)

\put(0,100){\includegraphics[width=143\unitlength]{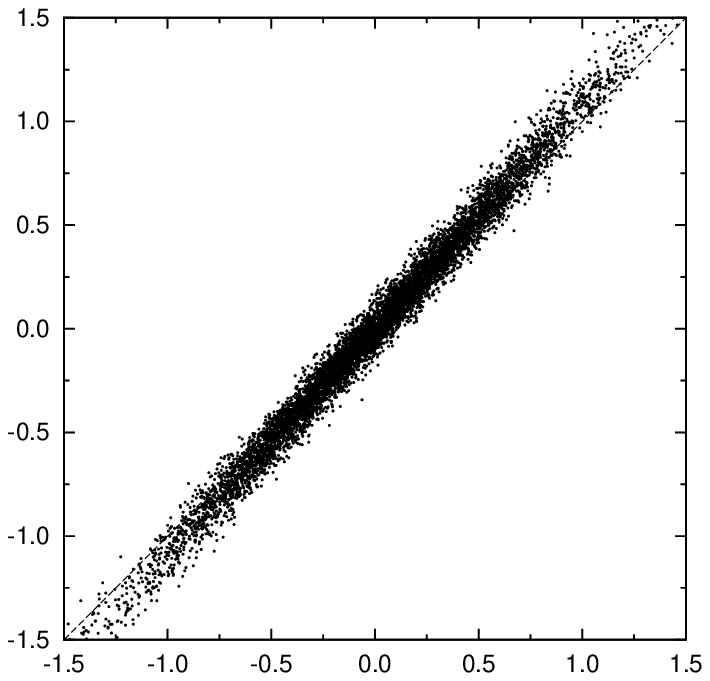}}
\put(100,100){\includegraphics[width=143\unitlength]{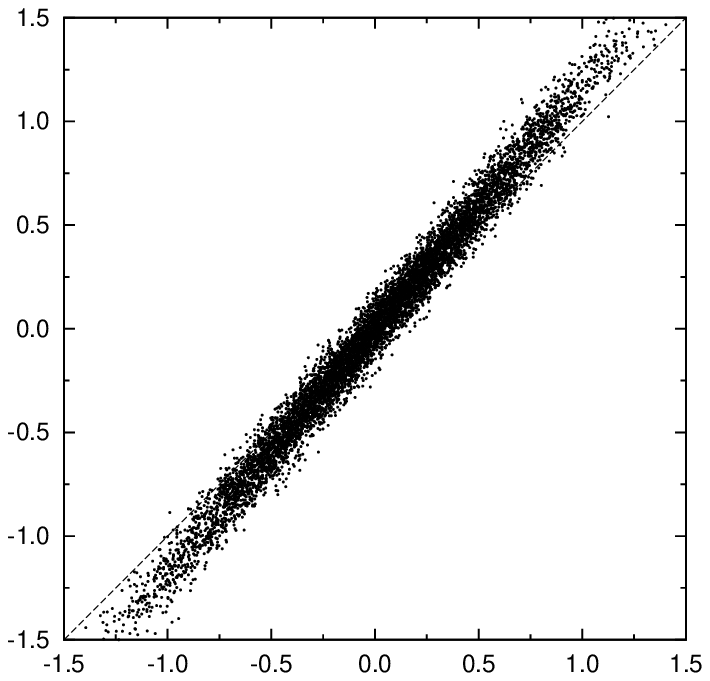}}
\put(200,100){\includegraphics[width=143\unitlength]{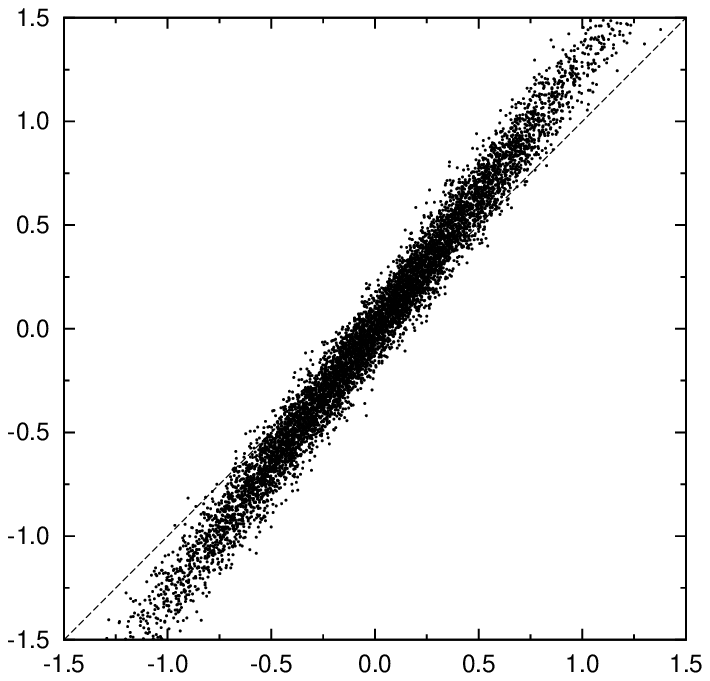}}
\put(300,100){\includegraphics[width=143\unitlength]{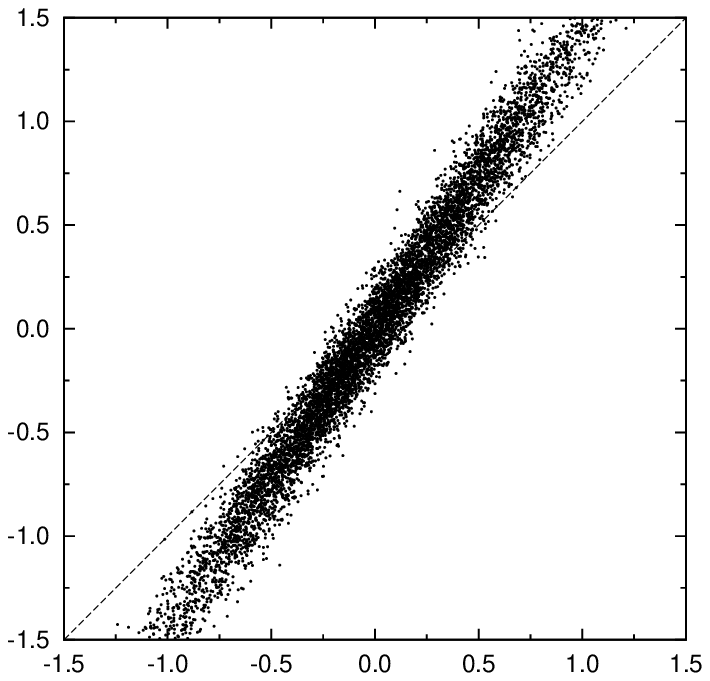}}

\put(0,0){\includegraphics[width=143\unitlength]{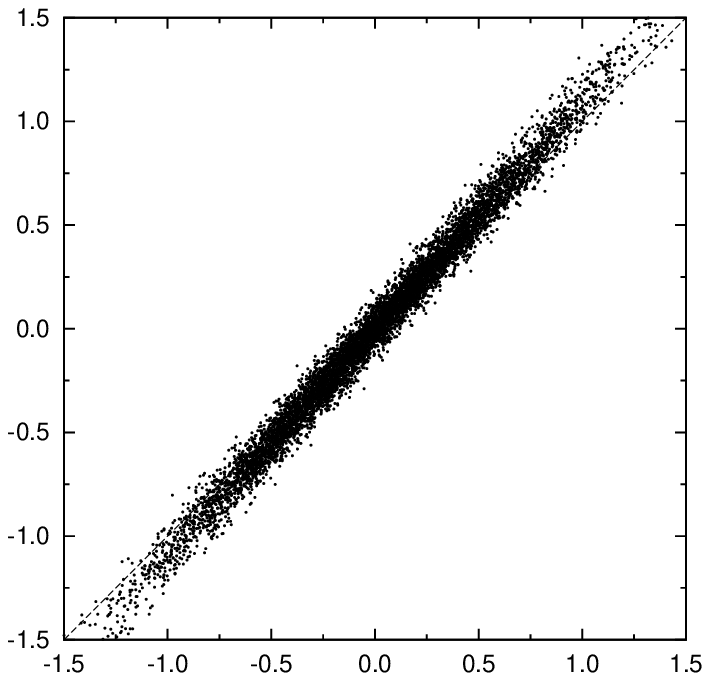}}
\put(100,0){\includegraphics[width=143\unitlength]{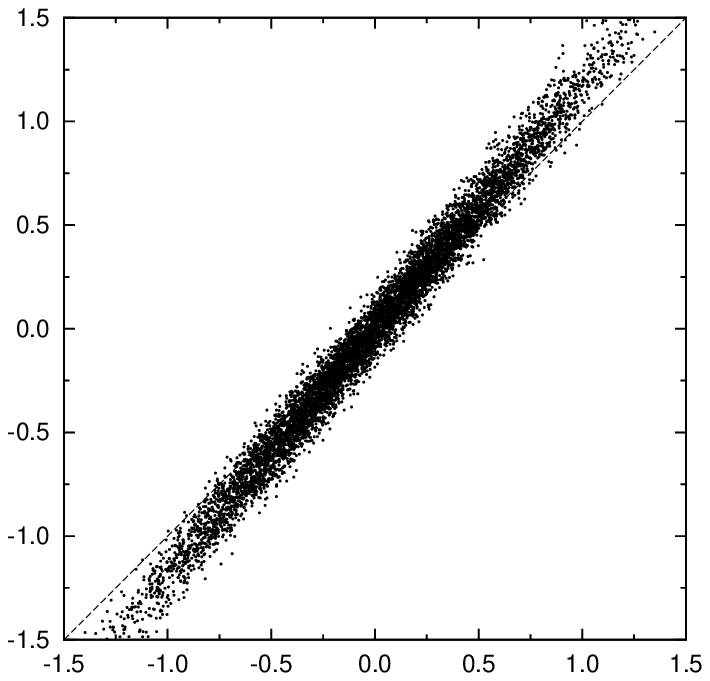}}
\put(200,0){\includegraphics[width=143\unitlength]{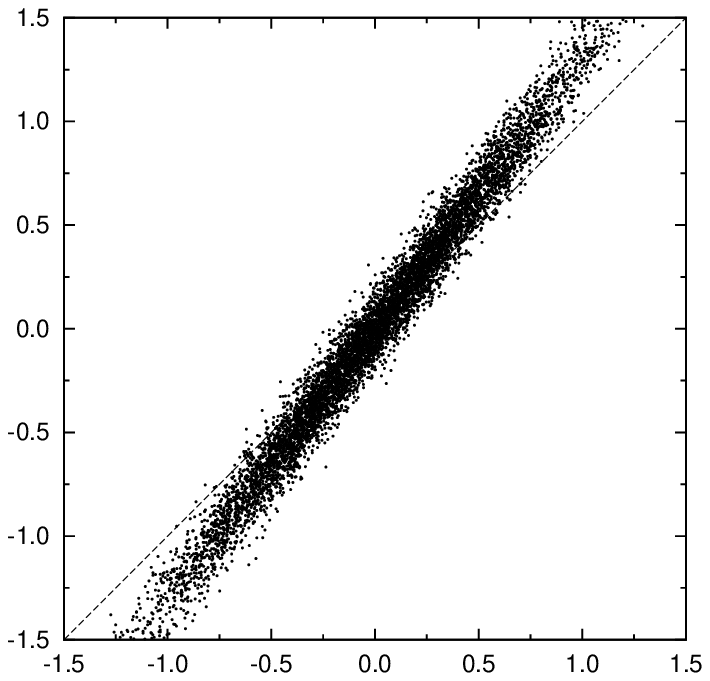}}
\put(300,0){\includegraphics[width=143\unitlength]{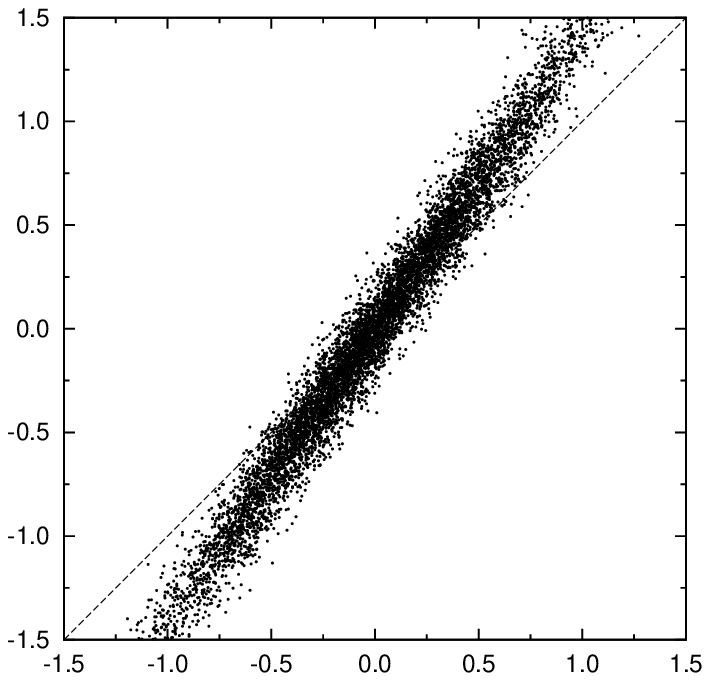}}

\put(60,-8){\small\em $\beta_\mu$}
\put(160,-8){\small\em $\beta_\mu$}
\put(260,-8){\small\em $\beta_\mu$}
\put(360,-8){\small\em $\beta_\mu$}
\put(30,86){\fs $p/N\!=\!0.1$}\put(30,75){\fs $\lambda(t)\!= \! a/\sqrt{t}$}
\put(130,86){\fs $p/N\!=\!0.2$}\put(130,75){\fs $\lambda(t)\!= \! a/\sqrt{t}$}
\put(230,86){\fs $p/N\!=\!0.3$}\put(230,75){\fs $\lambda(t)\!= \! a/\sqrt{t}$}
\put(330,86){\fs $p/N\!=\!0.4$}\put(330,75){\fs $\lambda(t)\!= \! a/\sqrt{t}$}
\put(30,186){\fs $p/N\!=\!0.1$}\put(30,175){\fs $\lambda(t)\!=\!1$}
\put(130,186){\fs $p/N\!=\!0.2$}\put(130,175){\fs $\lambda(t)\!=\!1$}
\put(230,186){\fs $p/N\!=\!0.3$}\put(230,175){\fs $\lambda(t)\!=\!1$}
\put(330,186){\fs $p/N\!=\!0.4$}\put(330,175){\fs $\lambda(t)\!=\!1$}
\put(5,55){\small\em $\hat{\beta}_\mu$}
\put(5,155){\small\em $\hat{\beta}_\mu$}

\end{picture}
\vsp

\caption{\small 
Inferred association parameters (vertical axis) versus true association parameters (horizontal axis) for synthetic survival data generated according to the Cox model, 
and subsequently analysed with the Cox model. Covariates and true association parameters were drawn randomly from zero-average Gaussian distributions. In all cases $N=400$, $\bra \beta_\mu^2\ket=0.25$ for all $\mu$,  and experiments were repeated such that the total number of points in each panel is identical. Top row: time-independent base hazard rate $\lambda(t)=1$. Bottom row: time-dependent base hazard rate $\lambda(t)=a/\sqrt{t}$ (dashed), with $a>0$ chosen such that the average event time is  $\bra t\ket=1$. 
The errors in the association parameters  induced by overfitting are more dangerous than finite sample size errors, since they mainly take the form of a consistent bias and therefore cannot be `averaged out'. Moreover, they appear to be independent of the true base hazard rate. }
\label{fig:overfitting2}
\end{figure}

\begin{figure}[t]
\unitlength=0.38mm
\hspace*{-8mm}
\begin{picture}(400,140)
\put(20,0){\includegraphics[width=203\unitlength]{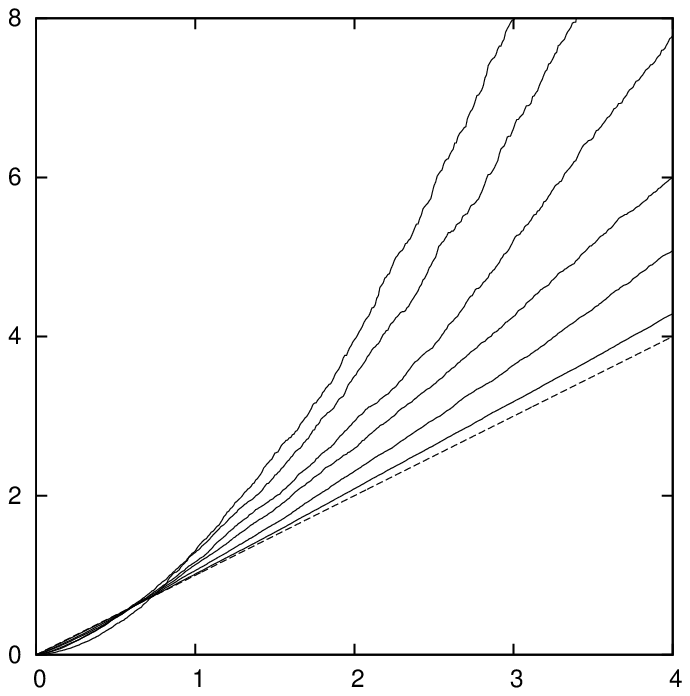}}
\put(180,0){\includegraphics[width=203\unitlength]{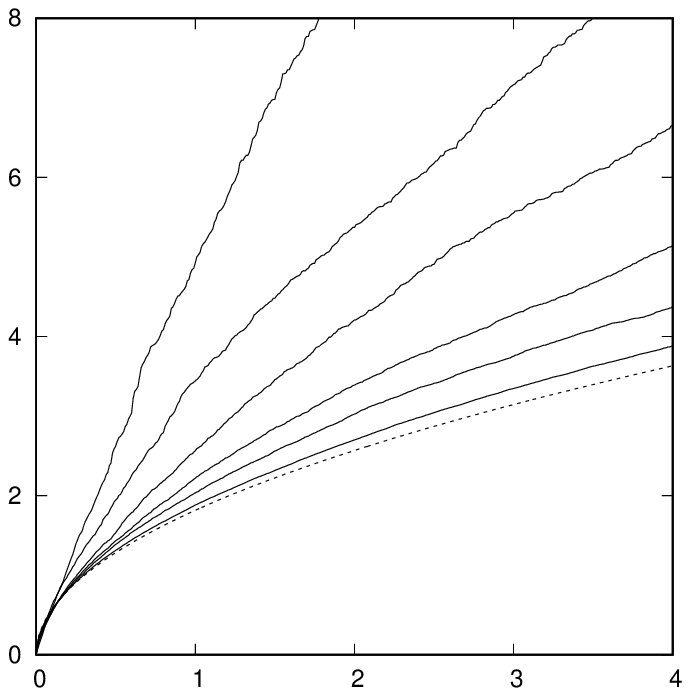}}

\put(120,3){\small\em $t$}
\put(280,3){\small\em $t$}
\put(32,85){\small $\hat{\Lambda}(t)$}
\put(192,85){\small $\hat{\Lambda}(t)$}

\end{picture}
\vspace*{-5mm}

\caption{\small 
Inferred integrated base hazard rates $\hat{\Lambda}(t)=\int_0^t\!\rmd t^\prime~\hat{\lambda}(t^\prime)$ (solid curves, averaged over multiple experiments) for synthetic survival data, generated and subsequently analysed with the Cox model. Covariates and true association parameters were drawn randomly from zero-average Gaussian distributions. In all cases $N=400$, $\bra \beta_\mu^2\ket=0.25$ for all $\mu$, and $p/N\in\{0.05, 0.15, 0.25, 0.35, 0.45, 0.55\}$ (lower to upper solid curves).  Left: data generated with $\lambda(t)=1$ (dashed). Right: data generated with $\lambda(t)=a/\sqrt{t}$ (dashed), with $a>0$ chosen such that the average event time is  $\bra t\ket=1$. The errors induced by overfitting again take the form of a consistent bias: for very short time the base hazard rate is always under-estimated, whereas for large times it is always over-estimated.
}
\label{fig:overfitting3}
\end{figure}

To increase our intuition for the problem, we first explore via simple simulation studies the relation between inferred and true parameters in  Cox's model \cite{Cox}.  The parameters of \cite{Cox} are the vector $\bbeta=(\beta_1,\ldots,\beta_p)$ of regression coefficients (where $p$ is the number of covariates), and the  base hazard rate $\lambda(t)=\rmd\Lambda(t)/\rmd t$. We generated  association parameters and covariates randomly from zero-average Gaussian distributions, and corresponding synthetic survival data using Cox's model without censoring (so all $N$ samples correspond to failure events), for different base hazard rates.  To understand the nature of the overfitting-induced regression errors we plotted   the $p$ pairs $(\beta_\mu,\hat{\beta}_\mu)$ as points in the plane, where $\beta_\mu$ 
and $\hat{\beta}_\mu$ are the true and inferred association parameters of covariate $\mu$, respectively, calculated via the recipes of \cite{Cox}. This resulted in  scatterplots as shown in Figure \ref{fig:overfitting2}. Simulations were done for different values of the ratio $p/N$, with multiple independent runs such that the number of points in each panel is identical. The true association parameters were drawn independently from a zero-average Gaussian distribution with $\bra \beta_\mu^2\ket=0.25$ for all $\mu$. Perfect regression would imply finding all points to lie on the diagonal. Rather than a widening of the variance (as with finite sample size regression errors)  overfitting-induced errors are somewhat surprisingly seen to manifest themselves mainly as a reproducible tilt of the data cloud, which increases with $p/N$, and implies a consistent over-estimation of associations: both positive and negative $\beta_\mu$ will always be reported as more extreme than their true values. These observed errors in association parameters appear to be independent of the form of the true base hazard rate. 
Similarly, we show in Figure \ref{fig:overfitting3}  the inferred integrated base hazard rates $\hat{\Lambda}(t)$ versus time (solid lines), together with the true values (dashed), which again shows consistent and reproducible overfitting errors. 
A quantitative theory of overfitting that can predict both the observed tilt and width of the data clouds of Figure \ref{fig:overfitting2} 
and the deformed inferred  hazard rates of Figure \ref{fig:overfitting3} 
 would enable us to {\em correct} the inferred  parameters of the Cox model  for overfitting, and thereby enable reliable regression   up to hitherto forbidden ratios of $p/N$. 

There are mathematical obstacles to the development of a theory of overfitting in survival analysis, which probably explain why it has so far remained an open problem. First, unlike discriminant analysis, it is not immediately clear which error measure to study when outcomes to be predicted are event times. Second, in most survival analysis models (including Cox regression) the estimated parameters are to be solved from coupled transcendental equations, and cannot therefore be written  in explicit form. Third, in the overfitting regime one will by definition find even for large $N$ that  the inferred parameters depend on the realisation of the data set, while at the more macroscopic level of prediction accuracy there is no such dependence. It is thus not a priori clear which quantities to focus on in analytical studies of the regression process, and at which stage in the calculation (if any) averages over possible realisations of the data set may be performed safely. 

 Our present approach to the problem consists of distinct stages, each removing a specific obstacle, and this is reflected in the structure of our paper. We adapt to time-to-event regression the strategy proposed and executed several decades ago for binary classifiers in the  groundbreaking paper by Gardner \cite{Gardner}.  { \red We first translate the problem of modelling overfitting into the calculation of a specific information-theoretic generating function, from which we can extract the information we need. Next we use Laplace's argument to eliminate the  maximisation over model parameters that comes with all ML methods, which is equivalent to writing the ground state energy of a statistical mechanical  system as the zero temperature limit of the free energy.  The third stage is devoted to making the resulting calculation of the generating function feasible, using the so-called replica method. This method has an impressive  track record of several decades in the analysis of complex heterogeneous many-variable systems in physics \cite{SK,Parisi,ParisiMezardVirasoro,FinConReplicas,VanMourikCoolen}, computer science \cite{Gardner,Nishimoribook}, biology \cite{AGS,Rabello,immune}, and economics \cite{MGreplicas1,MGreplicas2}, and enables us to carry out analytically the average of the generating function over all possible realisations of the data set. Finally we exploit steepest descent integration for $N\to\infty$, leading to the identification of  the `natural' macroscopic order parameters of the problem, for which we derive closed equations within the replica symmetric (RS) ansatz. Some technical arguments are placed in appendices, to improve  the flow of the paper. We develop our methods initially for generic time-to-event regression models, and then specialise to  the Cox model. The final RS  equations obtained for the Cox model involve  a small number of scalar order parameters, from which we can compute the link between true and inferred regression parameters, and the inferred base hazard rate.  The functional saddle point equation for the base hazard rate is rather nontrivial; while we can calculate the asymptotic form of its solution analytically, we limit ourselves mostly to a variational approximation, which already turns out to be quite accurate.}
 We close  with a discussion of our results, their implications and applications, and avenues for future work.

\section{Overfitting in Maximum Likelihood models for survival analysis}

\subsection{Definitions}

We assume we have simple time-to-event data $\Data$ of the standard type,   consisting of $N$ independently drawn samples $i=1\ldots N$, with just one active risk and no censoring. Each sample consists of a covariate vector $\bz_i\in\R^p$, drawn independently from a distribution $P(\bz)$, and an associated time to event $t_i\in[0,\infty)$, drawn from $P(t|\bz,\btheta^{\star})$:
\begin{eqnarray}
\Data=\{(\bz_1,t_1),\ldots,(\bz_N,t_N)\}
\end{eqnarray}
 Here  $P(t|\bz,\btheta^{\star})$ describes a parametrised time-generating model, with $q$ unknown real-valued parameters collected in a vector  $\btheta^{\star}\in\R^q$ that we seek to estimate from the data $\Data$. We are not interested in estimating $P(\bz)$, so we take the covariate vectors $\{\bz_1,\ldots,\bz_N\}$ as given.  
The data probability for each parameter choice $\btheta$ is
\begin{eqnarray}
P(\Data|\btheta)= \prod_{i=1}^N P(t_i|\bz_i,\btheta)
\end{eqnarray}
We next define  the empirical distribution of covariates and event times,  given the observed data:
\begin{eqnarray}
\hat{P}(t,\bz|\Data)&=&\frac{1}{N}\sum_{i=1}^N \delta(t-t_i)\delta(\bz-\bz_i)
\end{eqnarray}
This allows us to write
\begin{eqnarray}
\frac{1}{N}\log P(\Data|\btheta)&=&\int\!\rmd t\rmd\bz~\hat{P}(t,\bz|\Data)\log P(t|\bz,\btheta)
\nonumber
\\
&=& \int\!\rmd t\rmd\bz~\hat{P}(t,\bz|\Data)\log \Big(\frac{P(t|\bz,\btheta)}{\hat{P}(t|\bz,\Data)}\Big)
\nonumber
\\
&&\hspace*{20mm}
+\int\!\rmd t\rmd\bz~\hat{P}(t,\bz|\Data)\log \hat{P}(t|\bz,\Data)
\nonumber
\\
&=& -~ H(t|\bz,\Data)-D(\hat{P}_{\Data}||P_{\btheta})
\label{eq:KL}
\end{eqnarray}
with the conditional differential Shannon entropy of the event time distribution, and the Kullback-Leibler distance \cite{infotheory} between the empirical distribution $\hat{P}(t|\bz,\Data)$ and the parametrised form $P(t|\bz,\btheta)$:
\begin{eqnarray}
H(t|\bz,\Data)&=& 
-\int\!\rmd\bz~\hat{P}(\bz|\Data)\int\!\rmd t~\hat{P}(t|\bz,\Data)\log\hat{P}(t|\bz,\Data)
\\
D(\hat{P}_{\Data}||P_{\btheta})&=& 
\int\!\rmd\bz~\hat{P}(\bz|\Data)
\int\!\rmd t~\hat{P}(t|\bz,\Data)\log\Big(\frac{\hat{P}(t|\bz,\Data)}{ P(t|\bz,\btheta)}\Big)
\end{eqnarray}
The parameters $\btheta$ estimated via the ML recipe are those that maximise $P(\Data|\btheta)$. According to (\ref{eq:KL}) they  minimise the Kullback-Leibler distance $D(\hat{P}_{\Data}||P_{\btheta})$ between the empirical covariate-conditioned event time distribution and the parametrised event time distribution  with parameter values $\btheta$:
\begin{eqnarray}
\btheta_{\rm ML}&=& {\rm argmin}_{\btheta}~D(\hat{P}_{\Data}||P_{\btheta})
\label{eq:ML}
\end{eqnarray}
If $N\to\infty$ for fixed $p$ and $q$,  the law of large numbers guarantees that $\lim_{N\to\infty}\hat{P}(t|\bz,\Data)
=P(t|\bz,\btheta^{\star})$ (in a distributional sense), and hence ML regression will indeed estimate the parameters $\btheta$ asymptotically correctly, provided the chosen paramerisation is unambiguous:
\begin{eqnarray}
\lim_{N\to\infty}\btheta_{\rm ML}&=&  {\rm argmin}_{\btheta} ~D(P_{\btheta^{\star}}||P_{\btheta})=\btheta^{\star}
\label{eq:simple_case}
\end{eqnarray}
In this paper, however, we focus on the regime of large datasets with high-dimensional covariate and parameter vectors where overfitting 
occurs, namely $p,q=\order(N)$ and $N\to\infty$. Here $\hat{P}(t|\bz,\Data)$ no longer converges to 
$P(t|\bz,\btheta^{\star})$ for $N\to\infty$ in any mathematical sense,  the identity (\ref{eq:simple_case}) is therefore violated, and minimising $D(\hat{P}_{\Data}||P_{\btheta})$ as per the  ML prescription is no longer appropriate.  This is the information-theoretic description of the overfitting phenomenon in survival analysis.

\subsection{An information-theoretic measure of under- and overfitting} 

Maximum likelihood regression algorithms report those parameters $\btheta$ for which $P(t,\bz|\btheta)$ is as similar as possible to the {\em empirical} distribution $\hat{P}(t|\bz,\Data)$, as opposed  to the true distribution $P(t|\bz,\btheta^{\star})$ from which the data $\Data$ were generated. The optimal outcome of regression  is for  the inferred parameters to be identical to the true ones, i.e. to find 
${\rm argmin}_{\btheta}~D(\hat{P}_{\Data}||P_{\btheta})=\btheta^{\star}$. We therefore define 
\begin{eqnarray}
E(\btheta^{\star}\!,\Data)&=& \min_{\btheta}D(\hat{P}_{\Data}||P_{\btheta})-
D(\hat{P}_{\Data}||P_{\btheta^{\star}})
\nonumber
\\
&=&
\min_{\btheta}\Big\{\frac{1}{N}\sum_{i=1}^N\log \Big[
\frac{P(t_i|\bz_i,\btheta^{\star})}{P(t_i|\bz_i,\btheta)}\Big]\Big\}
\label{eq:Etheta}
\end{eqnarray}
This allows us to interpret the value of $E(\btheta^{\star}\!,\Data)$ as a measure of ML regression performance:
\begin{eqnarray}
E(\btheta^{\star}\!,\Data)>0:&& {\rm underfitting}\\
E(\btheta^{\star}\!,\Data)=0:&& {\rm optimal~parameter~ estimation}\\
E(\btheta^{\star}\!,\Data)<0:&& {\rm overfitting}
\end{eqnarray}
Optimal regression algorithms would reduce $D(\hat{P}_{\Data}||P_{\btheta})$ until 
$D(\hat{P}_{\Data}||P_{\btheta})=D(\hat{P}_{\Data}||P_{\btheta^{\star}})$ and then stop. Maximum likelihood regression will not do this;  if it can reduce the Kullback-Leibler distance further it will do so, and thereby cause overfitting. 
For $N\to\infty$ we expect $E(\btheta^{\star}\!,\Data)$ to depend on the data $\Data$ only via $P(\bz)$ and $\btheta^\star$, this is the fundamental assumption behind any regression. It allows us to focus on the average of  $E(\btheta^{\star}\!,\Data)$  over all realisations of the data, given $P(\bz)$ and $\btheta^\star$:
\begin{eqnarray}
E(\btheta^\star)&=&  \Big\bra \min_{\btheta}\Big\{\frac{1}{N}\sum_{i=1}^N\log \Big[
\frac{P(t_i|\bz_i,\btheta^{\star})}{P(t_i|\bz_i,\btheta)}\Big]\Big\}\Big\ket_{\Data}
\label{eq:Edef_explicit}
\\[-6mm]&&\nonumber
\end{eqnarray}
in which \\[-6mm]
\begin{eqnarray}
\bra F(t_1,\ldots,t_N;\bz_1,\ldots,\bz_N)\ket_{\Data}&=&
\int\!\prod_{i=1}^N\Big[\rmd t_i\rmd\bz_i ~P(\bz_i)P(t_i|\bz_i,\btheta^{\star})\Big]
 \nonumber
\\&&\times
F(t_1,\ldots,t_N;\bz_1,\ldots,\bz_N)
\end{eqnarray}
Evaluating  $E(\btheta^\star)$ analytically for $N\to \infty$ is the focus of this paper. 
Clearly, if the relevant minimum over $\btheta$ corresponds to the true value $\btheta^{\star}$ for all $\Data$, then $E(\btheta^\star)=0$. 

\subsection{Analytical evaluation of the average over data sets}

Working out (\ref{eq:Edef_explicit}) analytically for large $N$ requires first that we deal with the minimisation over $\btheta$. 
This can be done by converting the problem into the calculation of the ground state energy for  a statistical mechanical system with degrees of freedom $\btheta\in\R^q$ and Hamiltonian\footnote{The rescaling with $N$ of the Hamiltonian is done in anticipation of subsequent limits.} $H(\btheta)=N E(\btheta)$:
\begin{eqnarray}
E(\btheta^\star)&=& \lim_{\gamma\to\infty}E_\gamma(\btheta^\star)\\
E_\gamma(\btheta^\star)&=&   - \frac{1}{N}\frac{\partial}{\partial \gamma} \Big\bra
\log\int\!\rmd\btheta~\rme^{-\gamma \sum_{i=1}^N \log 
 \Big[
\frac{P(t_i|\bz_i,\btheta^\star)}{P(t_i|\bz_i,\btheta)}
\Big]}
\Big\ket_{\Data}
\nonumber
\\
&=&
- \frac{1}{N}\frac{\partial}{\partial \gamma}\Big\bra
\log\int\!\rmd\btheta~\prod_{i=1}^N 
 \Big[
\frac{P(t_i|\bz_i,\btheta)}{P(t_i|\bz_i,\btheta^\star)}
\Big]^\gamma
\Big\ket_{\Data}
\label{eq:Edef_nominimum}
\end{eqnarray}
For finite $\gamma$, the quantity $E_\gamma(\btheta^\star)$  can be interpreted as the average result of a {\em stochastic} minimisation, based on carrying out gradient descent on the function  $-\log P(\Data|\btheta)$, supplemented by a Gaussian white noise with variance proportional to $\gamma^{-1}$. 

The remaining obstacle is the logarithm in (\ref{eq:Edef_nominimum}), which prevents the average over all data sets $\Data$ from factorising over the samples. This we handle using the so-called replica method, which is based on the identity $\bra \log Z\ket=\lim_{n\to 0}n^{-1}\log \bra Z^n\ket$, and to our knowledge has not yet been applied in survival analysis. In the replica method the average $\bra Z^n\ket$ is carried out for {\em integer} $n$, and the limit $n\to 0 $ is taken at the end of the calculation via analytical continuation. Application to (\ref{eq:Edef_nominimum}) leads us after some simple manipulations to a new expression in which the average over data sets {\em does} factorise over samples:
\begin{eqnarray}
\hspace*{-5mm}
E_\gamma(\btheta^\star)&=&  -\frac{\partial}{\partial\gamma}\lim_{n\to 0} \frac{1}{Nn}\log\Big\bra
\Big\{\int\!\rmd\btheta~\prod_{i=1}^N 
 \Big[
\frac{P(t_i|\bz_i,\btheta)}{P(t_i|\bz_i,\btheta^\star)}
\Big]^\gamma\Big\}^n
\Big\ket_{\Data}
\nonumber
\\
\hspace*{-5mm}
&=&  -\frac{\partial}{\partial\gamma}\lim_{n\to 0} \frac{1}{Nn}\log
\int\!\rmd\btheta^1\ldots \rmd\btheta^n~\Big\bra\prod_{i=1}^N 
 \prod_{\alpha=1}^n\Big[
\frac{P(t_i|\bz_i,\btheta^\alpha)}{P(t_i|\bz_i,\btheta^\star)}
\Big]^\gamma
\Big\ket_{\Data}
\nonumber
\\
\hspace*{-5mm}
&=&  -\frac{\partial}{\partial\gamma}\lim_{n\to 0} \frac{1}{Nn}\log
\int\!\rmd\btheta^1\ldots \rmd\btheta^n~\Big\{\int\!\rmd\bz\rmd t~P(\bz)P(t|\bz,\btheta^\star)
\nonumber
\\
\hspace*{-5mm}
&&
\hspace*{20mm}\times
 \prod_{\alpha=1}^n\Big[
\frac{P(t|\bz,\btheta^\alpha)}{P(t|\bz,\btheta^\star)}
\Big]^\gamma
\Big\}^N~~
\label{eq:Edef_replicas}
\end{eqnarray}
The average over data sets has now been done, and we are left with a completely general explicit  expression for $E(\btheta^\star)$ in terms of the covariate statistics $P(\bz)$ and the assumed parametrised data generating model $P(t|\bz,\btheta)$. We will now work out and study this expression for Cox's proportional hazards model \cite{Cox} with statistically independent zero-average Gaussian covariates.

\subsection{Application to Cox regression}

In Cox's method \cite{Cox} the model parameters are a  base hazard rate $\lambda(t)\geq 0$ (with $t\geq 0$) and a vector $\bbeta\in\R^p$ of regression coefficients. The assumed event time statistics are then of the following form:
\begin{eqnarray}
\hspace*{-5mm}
P(t|\bz,\bbeta,\lambda)&=& \lambda(t)\rme^{\bbeta\cdot\bz/\sqrt{p}- \exp(\bbeta\cdot\bz/\sqrt{p})\Lambda(t)},
~~~~\Lambda(t)=\int_0^t\!\ds~\lambda(s)~~~~
\end{eqnarray}
The factors $\sqrt{p}$ only induce an irrelevant scaling factor that will make it easier to take the limit $p\to\infty$. 
In fact, for large $p$ it is inevitable that the typical association parameter in the Cox model will scale as $\order(p^{-\frac{1}{2}})$, since otherwise one would not find finite nonzero event times. 

For simplicity we assume that the covariates are distributed according to 
$P(\bz)= (2\pi)^{-p/2}\exp(-\frac{1}{2}\bz^2)$. {\red This restriction of our analysis to uncorrelated covariates is no limitation, since for the Cox model one can always obtain, via a simple mapping,  the regression results for data with correlated covariates from those obtained for uncorrelated covariates. This is demonstrated in \ref{app:correlations}.}

For the Cox model our general result (\ref{eq:Edef_replicas}) takes the following form, involving ordinary integration over $n$-fold replicated vectors $\bbeta^\alpha$ and functional integration over $n$-fold replicated base hazard rates $\lambda^\alpha$:
\begin{eqnarray}
E_\gamma(\bbeta^\star\!,\lambda^\star)&=&  -\frac{\partial}{\partial\gamma}\lim_{n\to 0} \frac{1}{Nn}\log
\int\{\rmd\lambda_1\ldots\rmd\lambda_n\}
\int\!\rmd\bbeta^1\ldots \rmd\bbeta^n~
\label{eq:Edef_Cox}
\\
&& \times \Big\{\int\!\rmd\bz\rmd t~P(\bz)P(t|\bz,\bbeta^\star\!,\lambda^\star)
 \prod_{\alpha=1}^n\Big[
\frac{P(t|\bz,\bbeta^\alpha\!,\lambda_\alpha)}{P(t|\bz,\bbeta^\star\!,\lambda^\star)}
\Big]^\gamma
\Big\}^{\!N}
\nonumber
\end{eqnarray}
To enable efficient further analysis we define the short-hands 
\begin{eqnarray}
p(t|\xi,\lambda)&=&\lambda(t)\rme^{\xi-\exp(\xi)\int_0^t\rmd s~\lambda(s)}
\\
p(\by|\bbeta^0,\ldots,\bbeta^n)&=&
\int\!\rmd\bz~P(\bz)
\prod_{\alpha=0}^n \delta\Big[y_\alpha\!-\!\frac{\bbeta^\alpha\cdot\bz}{\sqrt{p}}\Big]
\end{eqnarray}
and the $n\!+\!1$-dimensional vector $\by=(y_0,\ldots,y_p)$. In addition we rename $(\bbeta^\star,\lambda^\star)=(\bbeta^0,\lambda^0)$, so that 
\begin{eqnarray}
\hspace*{-10mm}
E_\gamma(\bbeta^0\!,\lambda_0)&=&  -\frac{\partial}{\partial\gamma}\lim_{n\to 0} \frac{1}{Nn}\log
\int\{\rmd\lambda_1\ldots\rmd\lambda_n\}
\int\!\rmd\bbeta^1\ldots \rmd\bbeta^n
\\
\hspace*{-10mm}
&&
\hspace*{-5mm}\times
\Big\{\int\!\rmd\by~p(\by|\bbeta^0,\ldots,\bbeta^n)\int\!\rmd t~p(t|y_0,\lambda_0)
 \prod_{\alpha=1}^n\Big[
\frac{p(t|y_\alpha,\lambda_\alpha)}{p(t|y_0,\lambda_0)}
\Big]^\gamma
\Big\}^N
\nonumber
\end{eqnarray}
All $\{y_\alpha\}$ are linear combinations of Gaussian random variables, so also $p(\by|\bbeta^0,\ldots,\bbeta^n)$ will be Gaussian (even for most non-Gaussian covariates this would still hold for large $p$ due to the central limit theorem), giving
\begin{eqnarray}
p(\by|\bbeta^0,\ldots,\bbeta^n)&=& 
\frac{\rme^{-\frac{1}{2}\by\cdot\bC^{-1}[\{\bbeta\}]\by}}{\sqrt{(2\pi)^{n+1}{\rm Det}\bC[\{\bbeta\}]}}
\end{eqnarray}
in which the entries of the $(n\!+\!1)\times (n\!+\!1)$ covariance matrix $\bC[\{\bbeta\}]$ are 
\begin{eqnarray}
C_{\alpha\rho}[\{\bbeta\}]&=& \frac{1}{p}\int\!\rmd\bz~P(\bz)(\bbeta^\alpha\cdot\bz)(\bbeta^\rho\cdot\bz)
=\frac{1}{p}\bbeta^\alpha\cdot\bbeta^\rho
\end{eqnarray}
We introduce integrals over $\delta$-distributions to transport variables to more convenient places, by substituting for each pair $(\alpha,\rho)$:
\begin{eqnarray}
1&=& \int\!\rmd C_{\alpha\rho}~\delta\Big[C_{\alpha\rho}-C_{\alpha\rho}[\{\bbeta\}]\Big]
= \int\! \frac{\rmd C_{\alpha\rho}\rmd \hat{C}_{\alpha\rho}}{2\pi/p}~\rme^{\rmi p\hat{C}_{\alpha\rho}\big[
C_{\alpha\rho}-C_{\alpha\rho}[\{\bbeta\}]\big]}
\nonumber
\\[-2mm]&&
\end{eqnarray}
We then obtain, after some simple manipulations,
\begin{eqnarray}
\hspace*{-9mm}
E_\gamma(\bbeta^0\!,\lambda_0)&=&  -\frac{\partial}{\partial\gamma}\lim_{n\to 0} \frac{1}{Nn}\log
\int\{\rmd\lambda_1\ldots\rmd\lambda_n\}\int\!\frac{\rmd\bC\rmd\hat{\bC}~\rme^{\rmi p\sum_{\alpha\rho=0}^n \hat{C}_{\alpha\rho}C_{\alpha\rho}}}{(2\pi/p)^{(n+1)^2}}
\nonumber
\\
\hspace*{-9mm}
&&\times
\Big\{\int\!\frac{\rmd\by~\rme^{-\frac{1}{2}\by\cdot\bC^{-1}\by}}{\sqrt{(2\pi)^{n+1}{\rm Det}\bC}}\int\!\rmd t~p(t|y_0,\lambda_0)
 \prod_{\alpha=1}^n\Big[
\frac{p(t|y_\alpha,\lambda_\alpha)}{p(t|y_0,\lambda_0)}
\Big]^\gamma
\Big\}^N
\nonumber
\\
\hspace*{-9mm}
&&\times 
\int\!\rmd\bbeta^1\ldots \rmd\bbeta^n~
\rme^{-\rmi\sum_{\alpha\rho=0}^n \hat{C}_{\alpha\rho}\bbeta^\alpha\cdot\bbeta^\rho}
\label{eq:Cox_finite_N}
\end{eqnarray}
For finite $N$, expressions such as (\ref{eq:Cox_finite_N}) are of course not easy to use, but as with 
all statistical theories  we will be able to progress upon assuming $N$ to be large\footnote{Note that the standard use of Cox regression away from the overfitting regime, including its formulae for confidence intervals and for p-values (which require Gaussian approximations that build on large $N$ expansions around the most probable parameter values, and assume that uncertainty in base hazard rates can be neglected), is similarly valid only when $N$ is sufficiently large.}. We therefore focus on the asymptotic behaviour of (\ref{eq:Cox_finite_N}) for $N\to\infty$, but with a fixed ratio $p/N$, and will confirm {\em a posteriori} the extent to which the resulting theory describes what is observed for large but finite sample sizes. 

\section{Asymptotic analysis of overfitting in the Cox model}

\subsection{Conversion to a saddle-point problem}

Following extensive experience with the replica method in other disciplines, with similar definitions, we 
assume that the two limits $N\to\infty$  and $n\to 0$ commute. 
The invariance of the right-hand side of (\ref{eq:Cox_finite_N}) under all permutations of the sample indices $i\in\{1,\ldots,N\}$ implies that $E(\bbeta^0,\lambda_0)$ can depend on the true association parameters 
$\bbeta^0$ only via the distribution $P(\beta_0)=p^{-1}\sum_{\mu=1}^p\delta[\beta_0-\beta_\mu^0]$. With a modest amount of foresight we define $S^2=p^{-1}\sum_{\mu=1}^p (\beta_\mu^0)^2$, and 
obtain
\begin{eqnarray}
\hspace*{-20mm}
E_\gamma(P,\lambda_0)&=&  -\frac{\partial}{\partial\gamma}\lim_{n\to 0} \frac{1}{Nn}\log
\int\{\rmd\lambda_1\ldots\rmd\lambda_n\}\int\!\frac{\rmd\bC\rmd\hat{\bC}~\rme^{\rmi p\big(\sum_{\alpha\rho=0}^n \hat{C}_{\alpha\rho}C_{\alpha\rho}-\hat{C}_{00}S^2\big)}}{(2\pi/p)^{(n+1)^2}}
\nonumber
\\
\hspace*{-20mm}
&&\times
\Big\{
\int\!\frac{\rmd\by~\rme^{-\frac{1}{2}\by\cdot\bC^{-1}\by}}{\sqrt{(2\pi)^{n+1}{\rm Det}\bC}}
\int\!\rmd t~p(t|y_0,\lambda_0)
 \prod_{\alpha=1}^n\Big[
\frac{p(t|y_\alpha,\lambda_\alpha)}{p(t|y_0,\lambda_0)}
\Big]^\gamma
\Big\}^N
\nonumber
\\
\hspace*{-20mm}
&&\times~
 \rme^{p\int\!\rmd\beta_0~P(\beta_0)\log 
\int\!\rmd\beta_1\ldots \rmd\beta_n~
\rme^{-2\rmi\beta_0\sum_{\rho=1}^n \hat{C}_{0\rho}\beta_\rho
-\rmi\sum_{\alpha\rho=1}^n \hat{C}_{\alpha\rho}\beta_\alpha\beta_\rho
}}
\end{eqnarray}
Writing the ratio of covariates over samples as $p/N=\zeta$, to be kept fixed in the limit $N\to\infty$, we may take the limit $N\to\infty$ and obtain an integral that can be evaluated using steepest descent:
\begin{eqnarray}
\hspace*{-20mm}
\lim_{N\to\infty}
E_\gamma(P,\lambda_0)&=&  -\frac{\partial}{\partial\gamma}\lim_{n\to 0}
\lim_{N\to\infty}
 \frac{1}{Nn}\log
\int\{\rmd\lambda_1\ldots\rmd\lambda_n\}
\nonumber
\\
\hspace*{-20mm}
&&\times
\rme^{-\frac{1}{2}N\log [(2\pi)^{n+1}{\rm Det}\bC]} \int\!\rmd\bC\rmd\hat{\bC}~\rme^{\rmi\zeta N \big(\sum_{\alpha\rho=0}^n \hat{C}_{\alpha\rho}C_{\alpha\rho}-\hat{C}_{00}S^2\big)}
\nonumber
\\
\hspace*{-20mm}
&&\times
\rme^{N\log \int\!\rmd\by~\rme^{-\frac{1}{2}\by\cdot\bC^{-1}\by}\int\!\rmd t~p(t|y_0,\lambda_0)
 \prod_{\alpha=1}^n\big[
\frac{p(t|y_\alpha,\lambda_\alpha)}{p(t|y_0,\lambda_0)}
\big]^\gamma}
\nonumber
\\
\hspace*{-20mm}
&&\times~
 \rme^{\zeta N\int\!\rmd\beta_0~P(\beta_0)\log 
\int\!\rmd\beta_1\ldots \rmd\beta_n~
\rme^{-2\rmi\beta_0\sum_{\rho=1}^n \hat{C}_{0\rho}\beta_\rho
-\rmi\sum_{\alpha\rho=1}^n \hat{C}_{\alpha\rho}\beta_\alpha\beta_\rho
}}
\nonumber
\\[1mm]
\hspace*{-20mm}
&=&  \frac{\partial}{\partial\gamma}\lim_{n\to 0}
 \frac{1}{n}
 {\rm extr}_{\bC,\hat{\bC},\lambda_1,\ldots,\lambda_n}\Psi[\bC,\hat{\bC};\lambda_1,\ldots,\lambda_n]
 \end{eqnarray}
 in which the function to be extremized is
 \begin{eqnarray}
 \hspace*{-15mm}
 \Psi[\ldots]&=& 
-\rmi\zeta\Big[\sum_{\alpha\rho=0}^n \hat{C}_{\alpha\rho}C_{\alpha\rho}-\hat{C}_{00}S^2\Big]
+\frac{1}{2}(n\!+\!1)\log (2\pi)
+\frac{1}{2}\log{\rm Det}\bC
\nonumber
\\
 \hspace*{-15mm}
&&-~\zeta \int\!\rmd\beta_0~P(\beta_0)\log 
\int\!\rmd\beta_1\ldots \rmd\beta_n~
\rme^{-2\rmi\beta_0\sum_{\rho=1}^n \hat{C}_{0\rho}\beta_\rho
-\rmi\sum_{\alpha\rho=1}^n \hat{C}_{\alpha\rho}\beta_\alpha\beta_\rho}
\nonumber
\\
 \hspace*{-15mm}
&&
-\log \int\!\rmd\by~\rme^{-\frac{1}{2}\by\cdot\bC^{-1}\by}\int\!\rmd t~p(t|y_0,\lambda_0)
 \prod_{\alpha=1}^n\Big[
\frac{p(t|y_\alpha,\lambda_\alpha)}{p(t|y_0,\lambda_0)}
\Big]^\gamma
\end{eqnarray}
Differentiation with respect to $\hat{C}_{00}$ immediately gives $C_{00}=S^2$. Moreover, 
for various integrals to be well-defined, the relevant saddle-point must (after contour deformation in the complex plane) be of a form where
\begin{eqnarray}
\alpha,\rho=1\ldots n:&~~~~& \hat{C}_{\alpha\rho}=-\frac{1}{2}\rmi D_{\alpha\rho},~~~\hat{C}_{0\rho}=-\frac{1}{2}\rmi d_{\rho}
\end{eqnarray}
with $D_{\alpha\rho},d_\rho\in\R$, and where the $n\times n$ matrix $\bD=\{D_{\alpha\rho}\}$ is positive definite. 
Thus at the relevant saddle-point we will have
 \begin{eqnarray}
 \hspace*{-10mm}
 \Psi[\ldots]&=& 
-\frac{1}{2}\zeta\sum_{\alpha\rho=1}^n D_{\alpha\rho}C_{\alpha\rho}
-\zeta\sum_{\rho=1}^n d_{\rho}C_{0\rho}
+\frac{1}{2}(n\!+\!1)\log (2\pi)
+\frac{1}{2}\log{\rm Det}\bC
\nonumber
\\
 \hspace*{-10mm}&&
-\log \int\!\rmd\by~\rme^{-\frac{1}{2}\by\cdot\bC^{-1}\by}\int\!\rmd t~p(t|y_0,\lambda_0)
 \prod_{\alpha=1}^n\Big[
\frac{p(t|y_\alpha,\lambda_\alpha)}{p(t|y_0,\lambda_0)}
\Big]^\gamma
\nonumber
\\
 \hspace*{-10mm}
&&-~\zeta \int\!\rmd\beta_0~P(\beta_0)\log 
\int\!\rmd\beta_1\ldots \rmd\beta_n~
\rme^{-\beta_0\sum_{\rho=1}^n d_\rho\beta_\rho
-\frac{1}{2}\sum_{\alpha\rho=1}^n D_{\alpha\rho}\beta_\alpha\beta_\rho}
\nonumber
\\
 \hspace*{-10mm}
&=&  
-\frac{1}{2}\zeta\sum_{\alpha\rho=1}^n D_{\alpha\rho}C_{\alpha\rho}
-\zeta\sum_{\rho=1}^n d_{\rho}C_{0\rho}
-\frac{1}{2}\zeta S^2\sum_{\alpha\rho=1}^n d_\alpha(\bD^{-1})_{\alpha\rho} d_\rho
\nonumber
\\
\hspace*{-10mm}
&&
+\frac{1}{2}(n\!+\!1)\log (2\pi)
+\frac{1}{2}\log{\rm Det}\bC
\nonumber
\\
 \hspace*{-10mm}
&&
-\log \int\!\rmd\by~\rme^{-\frac{1}{2}\by\cdot\bC^{-1}\by}\int\!\rmd t~p(t|y_0,\lambda_0)
 \prod_{\alpha=1}^n\Big[
\frac{p(t|y_\alpha,\lambda_\alpha)}{p(t|y_0,\lambda_0)}
\Big]^\gamma
\nonumber
\\
 \hspace*{-10mm}
&&-~\zeta \log 
\int\!\rmd\beta_1\ldots \rmd\beta_n~
\rme^{-\frac{1}{2}\sum_{\alpha\rho=1}^n D_{\alpha\rho}\beta_\alpha\beta_\rho}
\end{eqnarray}
Variation with respect to the $n$ components $\{d_\alpha\}$ gives $d_\alpha=-S^{-2}\sum_\rho D_{\alpha\rho}C_{0\rho}$, so
 \begin{eqnarray}
  \hspace*{-10mm}
 \Psi[\ldots]&=& 
 -\frac{1}{2}\zeta\sum_{\alpha\rho=1}^n D_{\alpha\rho}\Big[C_{\alpha\rho}
\!-\!\frac{C_{0\alpha}C_{0\rho }}{S^2}\Big]
+\frac{1}{2}(n\!+\!1)\log (2\pi)
+\frac{1}{2}\log{\rm Det}\bC
\nonumber
\\
 \hspace*{-10mm}&&
-\log \int\!\rmd\by~\rme^{-\frac{1}{2}\by\cdot\bC^{-1}\by}\int\!\rmd t~p(t|y_0,\lambda_0)
 \prod_{\alpha=1}^n\Big[
\frac{p(t|y_\alpha,\lambda_\alpha)}{p(t|y_0,\lambda_0)}
\Big]^\gamma
\nonumber
\\
 \hspace*{-10mm}
&&-~\zeta \log 
\int\!\rmd\beta_1\ldots \rmd\beta_n~
\rme^{-\frac{1}{2}\sum_{\alpha\rho=1}^n D_{\alpha\rho}\beta_\alpha\beta_\rho}
\end{eqnarray}
This intermediate result confirms that $\lim_{N\to\infty}E_\gamma(P,\lambda_0)$ indeed depends on the distribution $P(\beta_0)$ only via 
$S^2=\int\!\rmd\beta_0 ~P(\beta_0)\beta_0^2$, hence we may henceforth 
write the former quantity as $E_\gamma(S,\lambda_0)$. 
Variation with respect to $\bD$ finally gives $(\bD^{-1})_{\alpha\rho}=C_{\alpha\rho}\!-\!C_{0\alpha}C_{0\rho}/S^2$. Hence we arrive 
at  the following expression, in which the short-hand $\bC^\prime$ denotes the $n\times n$ matrix with 
entries $C^\prime_{\alpha\rho}=C_{\alpha\rho}\!-\!C_{0\alpha}C_{0\rho}/S^2$ (for $\alpha,\rho=1\ldots n$):
 \begin{eqnarray}
 \hspace*{-15mm}
 E_\gamma(S,\lambda_0)&=&  
 \frac{\partial}{\partial\gamma}\lim_{n\to 0}
 \frac{1}{n}
 {\rm extr}_{\bC;\lambda_1,\ldots,\lambda_n}\Psi[\bC;\lambda_1,\ldots,\lambda_n]
 \label{eq:Cox_RSB1}
 \\
  \hspace*{-15mm}
 \Psi[\bC;\lambda_1,\ldots,\lambda_n]&=& 
\frac{1}{2}\log{\rm Det}\bC-\frac{1}{2}\zeta \log{\rm Det}\bC^\prime
 \label{eq:Cox_RSB2}
\\
 \hspace*{-15mm}
&&
-\log \int\!\frac{\rmd\by}{\sqrt{2\pi}}~\rme^{-\frac{1}{2}\by\cdot\bC^{-1}\by}\!\int\!\rmd t~p(t|y_0,\lambda_0)
 \prod_{\alpha=1}^n\Big[
\frac{p(t|y_\alpha,\lambda_\alpha)}{p(t|y_0,\lambda_0)}
\Big]^\gamma
\nonumber
\end{eqnarray}
The extremisation over $\bC$ is to be done subject to $C_{00}=S^2$, and 
we have removed from $\Psi[\ldots]$ those terms that will vanish after taking $n\to 0$ and differentiating with respect to $\gamma$. 

\subsection{Replica symmetric extrema}

The replica symmetry ansatz (RS) can be translated into the statement that the solution space of the regression algorithm is ergodic \cite{ParisiMezardVirasoro,Nishimoribook,NNbook}, i.e. the typical set of equivalent minima in regression parameter space is connected. 
Replica symmetric saddle-points of  (\ref{eq:Cox_RSB2}) are of the following form:
\begin{eqnarray}
 \forall \alpha,\rho=1\ldots n:&~~~&  \lambda_\alpha(t)=\lambda(t),~~~C_{00}=S^2,~~~C_{0\alpha}=c_0,
 \\
 &&C_{\alpha\rho}=C\delta_{\alpha\rho}+c(1\!-\!\delta_{\alpha\rho})
\end{eqnarray}
{\red In \ref{app:RS} we derive the equations corresponding to the RS ansatz for the stochastic generalization of the Cox model. 
With  the short-hand $\rmD y=(2\pi)^{-1/2}\rme^{-\frac{1}{2}y^2}\rmd y$, 
 and upon removing terms that vanish upon differentiation by $\gamma$,  we can summarise these equations in the limit of large data sets, by the following compact expression:
 \begin{eqnarray}
  \hspace*{-15mm} 
 E_{\gamma}(S,\lambda_0)&=&  
 \int\! \rmD y_0
\int\!\rmd t~p(t|Sy_0,\lambda_0)\left\{\room
\log p(t|Sy_0,\lambda_0)
\label{eq:Cox_RS_final_E}
\right.
\\
 \hspace*{-15mm} 
&&\hspace*{5mm}
\left.
-\int\!\rmD z \left[\frac{\int\!\rmD y~p^\gamma(t|uy\!+\!wy_0\!+\!vz,\lambda)\log p(t|uy\!+\!wy_0\!+\!vz,\lambda)}
{\int\!\rmD y~p^\gamma(t|uy\!+\!wy_0\!+\!vz,\lambda)}
\right]
\right\}
\nonumber
\end{eqnarray}
in which the order parameters $\{u,v,w;\lambda\}$, which are related to the RS order parameters $\{C,c_0,c\}$ via 
\begin{eqnarray}
c_0=Sw,~~~~c=v^2+w^2,~~~~C=u^2+v^2+w^2,
\end{eqnarray}
 are to be evaluated at the saddle point of 
\begin{eqnarray}
\hspace*{-10mm}
\Psi_{\rm RS}(u,v,w;\lambda)&=& \zeta \Big(\frac{v^2}{2u^2}
+\log u\Big)
\label{eq:Cox_RS_final}
\\
\hspace*{-10mm}&&\hspace*{-5mm}
+\int\!\rmD z \rmD y_0
\int\!\rmd t~p(t|Sy_0,\lambda_0)
\log \int\!\rmD y~p^\gamma(t|uy\!+\!wy_0\!+\!vz,\lambda)
\nonumber
\end{eqnarray}}

\subsection{Physical interpretation of order parameters}

The physical meaning of the order parameters in the replica symmetric matrix $\bC$ is found in the usual manner for replica calculations \cite{ParisiMezardVirasoro}, by direct application of our manipulations to the calculation of observables. We will write averages over the stochastic maximization of the data log-likelihood at finite $\gamma$, for a fixed training set $\Data$, as $\bra \ldots \ket$, and averages over all data sets (as before) as $\bra \ldots\ket_{\Data}$. Since the relevant quantities in the theory are found asymptotically to depend on the true association vector $\bbeta^\star$ only via $S^2=p^{-1}\sum_{\mu=1}^p (\beta_\mu^\star)^2$, there is no need for explicit averages over $\bbeta^\star$. This results upon application to the Cox model in the following 
identifications, in the limit $n\to 0$:
\begin{eqnarray}
&&\hspace*{-10mm}
c_0=\lim_{p\to\infty} \frac{1}{p}\bbeta^{\star}\!\cdot\!\bra\bra \bbeta\ket \ket_{\Data},~~~~~ c=\lim_{p\to\infty} \frac{1}{p}\bra \bra\bbeta\ket^2\ket_{\Data} ,~~~~~
C=\lim_{p\to\infty}\frac{1}{p}\bra\bra \bbeta^2\ket\ket_{\Data} 
\label{eq:meaning}
\end{eqnarray}
In terms of the transformed order parameters $(u,v,w)$ this becomes
\begin{eqnarray}
u^2&=&\lim_{p\to\infty} \frac{1}{p}\bra
\bra \bbeta^2\ket- \bra\bbeta\ket^2\ket_{\Data}
\label{eq:interpretation_u}
\\
v^2&=&\lim_{p\to\infty}\frac{1}{p}\Big[\bra \bra\bbeta\ket^2\ket_{\Data} -\Big( \frac{\bbeta^{\star}\!\cdot\!\bra\bra \bbeta\ket \ket_{\Data}}{|\bbeta^\star|}\Big)^2\Big]
\label{eq:interpretation_v}
\\
w&=&\lim_{p\to\infty} \frac{1}{\sqrt{p}}\frac{\bbeta^{\star}\!\cdot\!\bra\bra \bbeta\ket \ket_{\Data}}{|\bbeta^\star|}
\label{eq:interpretation_w}
\end{eqnarray}
Here $\bbeta$ is the outcome of maximum likelihood regression for data set $\Data$ generated with true association parameters $\bbeta^{\star}$. Fully random parameter guessing would give  $c_0=c=0$ and $C>0$.  Perfect regression would imply $\bbeta=\bbeta^{\star}$ for all $\Data$ and all $\bbeta^{\star}$, and hence correspond to $c_{0}=c=C=S^2$, giving $u=v=0$ and $w=S$. It is reassuring to observe that  for $\zeta=0$, expression (\ref{eq:Cox_RS_final_E}) indeed reproduces $E_{\gamma}(S,\lambda_0)=0$ if in the right-hand side we substitute the values $u=v=0$ and $w=S$.

From  (\ref{eq:meaning})  follow useful inequalities that must hold at the relevant saddle-point in the limit $n\to 0$, which are consistent with our claim that $u,v,w\geq 0$:
\begin{eqnarray}
C\geq 0,~~~~c\geq 0,~~~~ c_{0}\geq 0,~~~~ C\geq c,~~~~ c \geq c_0^2/S^2
\label{eq:inequalities}
\end{eqnarray}
The first four inequalities are easy to derive.  The fifth follows from:
\begin{eqnarray}
c&=&\lim_{p\to\infty}\frac{1}{p}\bra \bra\bbeta\ket^2\ket_{\Data}~\geq ~ 
\lim_{p\to\infty}\frac{1}{p}\Big\bra \Big(\frac{\bbeta^{\star}}{|\bbeta^{\star}|}\cdot\bra\bbeta\ket\Big)^2\Big\ket_{\Data}
\nonumber
\\
&=& \frac{1}{p}\Big(\frac{p}{|\bbeta^{\star}|}c_0\Big)^2
~=~c_0^2/S^2
\end{eqnarray}

If, as suggested by the $\gamma\to\infty$ simulation results shown in Section 1,  $\bra\bbeta\ket\approx \kappa \bbeta^\star+\bxi$ for some $\kappa>0$, with a zero-average random vector $\bxi$ that reflects data set variability, such that $\bra \bxi\ket_{\Data}=\bnull$ and with amplitude $\lim_{p\to\infty}p^{-1}\sum_{\mu=1}^p\bra \xi_\mu^2\ket_{\Data}=\sigma^2$, then  we would find the RS saddle point  obeying 
$c_0= \kappa S^2$ and $c=\kappa^2 S^2+\sigma^2$. Hence we would find $v=\sigma$ and $\kappa=w/S$, and we would expect $\lim_{\gamma\to \infty}u=0$ for $\zeta<1$.  Note that the above relations are true given our definition of the event time distribution as $P(t|\bz,
\bbeta,\lambda)=-\frac{\rmd}{\rmd t}\exp[-\exp(\bbeta\cdot\bz/\sqrt{p})\Lambda(t)]$. If we were to define this distribution instead without the rescaling factor $\sqrt{p}$ as $P(t|\bz,
\bbeta,\lambda)=-\frac{\rmd}{\rmd t}\exp[-\exp(\bbeta\cdot\bz)\Lambda(t)]$ (which is the convention of \cite{Cox}), then the connection between regression of the form $\bra\bbeta\ket\approx \kappa \bbeta^\star+\bxi$ and our order parameters would be: 
\begin{eqnarray}
\kappa=w/S,~~~~~~\sigma=v/\sqrt{p}
\label{eq:link_with_experiment}
\end{eqnarray}
We conclude that from our RS equations we can extract  the dependence on the covariates/samples ratio $\zeta=p/N$ of the two main quantitative characteristics of the data clouds in Figure 
\ref{fig:overfitting2}: their angle $\kappa$ and their width $\sigma$. 

Finally, let us turn to the interpretation of equation (\ref{eq:Cox_RS_final_E}). We observe that this equation can be written as
\begin{eqnarray}
\hspace*{-10mm}
 E_{\gamma}(S,\lambda_0)&\!=\!& \int\!\rmd t \rmd x\rmd x^\prime ~\Prob_\gamma(x,x^\prime,t)
\log\Big[\frac{p(t|x,\lambda_0)}{ p(t|x^\prime,\lambda)}\Big]
\label{eq:compareE}
\\
\hspace*{-10mm}
\Prob_\gamma(x,x^\prime,t)&\!=\!& 
\int\!\rmD z \rmD y_0~\delta[x\!-\!Sy_0] p(t|Sy_0,\lambda_0)
\nonumber
\\
\hspace*{-10mm}
&&\times
\Big[
 \frac{\int\!\rmD y~p^\gamma(t|uy\!+\!wy_0\!+\!vz,\lambda)~\delta[x^\prime\!-\!uy\!-\!wy_0\!-\!vz]}
{\int\!\rmD y~p^\gamma(t|uy\!+\!wy_0\!+\!vz,\lambda)}
\Big]~~~
\label{eq:Wresult}
\end{eqnarray}
If we compare expression (\ref{eq:compareE})  with the definition of  $E_{\gamma}(S,\lambda_0)$, which for the Cox model is
\begin{eqnarray}
 E_{\gamma}(S,\lambda_0)&=&\lim_{N\to\infty}\Big\bra\Big\bra \frac{1}{N}\sum_{i=1}^N \log\Big[
 \frac{p(t_i|\bbeta^\star\cdot\bz_i/\sqrt{p},\lambda_0)}{p(t_i|\bbeta\cdot\bz_i/\sqrt{p},\lambda)}\Big]
 \Big\ket\Big\ket_{\Data}
 \end{eqnarray}
 we can infer that 
 \begin{eqnarray}
 \hspace*{-10mm}
 \Prob_\gamma(x,x^\prime,t)&=& \lim_{N\to\infty}\Big\bra\Big\bra \frac{1}{N}\sum_{i=1}^N \delta[t-t_i]~
 \delta\Big[x-\frac{\bbeta^\star\cdot\bz_i}{\sqrt{p}}\Big]
  \delta\Big[x^\prime\!-\frac{\bbeta\cdot\bz_i}{\sqrt{p}}\Big]
   \Big\ket\Big\ket_{\Data}
   \label{eq:Wmeaning}
   \end{eqnarray}
   As a consistency test one can confirm that, as an alternative to retracing the replica derivation,  the expressions (\ref{eq:meaning}) can also be derived explicitly
   from (\ref{eq:Wresult},\ref{eq:Wmeaning}).  
   
   \subsection{Derivation of RS saddle point equations}

The equations from which to solve the replica symmetric order parameters $(u,v,w,\lambda)$ are obtained by extremization of (\ref{eq:Cox_RS_final}). Using $\partial\log p(t|\xi)/\partial\xi=1-\rme^\xi\Lambda(t)$, the three scalar equations are found to be
\begin{eqnarray}
\hspace*{-25mm}
\frac{\zeta}{\gamma u} \Big(\frac{v^2}{u^2}-1\Big)
&=& \int\!\rmD z \rmD y_0
\int\!\rmd t~p(t|Sy_0,\lambda_0)
\nonumber
\\
\hspace*{-25mm}&&\hspace*{10mm}\times
\frac{\int\!\rmD y~y~p^\gamma(t|uy\!+\!wy_0\!+\!vz,\lambda)\Big[1-\rme^{uy\!+\!wy_0\!+\!vz}\Lambda(t)\Big]}
{\int\!\rmD y~p^\gamma(t|uy\!+\!wy_0\!+\!vz,\lambda)}
\label{eq:spe_u}
\\[1mm]
\hspace*{-25mm}
~~~~~~~~~\zeta \frac{v}{\gamma u^2}&=&
\int\!\rmD z \rmD y_0~z\!
\int\!\rmd t~p(t|Sy_0,\lambda_0)\Lambda(t)
\frac{\int\!\rmD y~p^\gamma(t|uy\!+\!wy_0\!+\!vz,\lambda)\rme^{uy\!+\!wy_0\!+\!vz}}
{\int\!\rmD y~p^\gamma(t|uy\!+\!wy_0\!+\!vz,\lambda)}
\nonumber
\\[-1mm]\hspace*{-25mm}&&
\label{eq:spe_v}
\\[1mm]
\hspace*{-25mm}
~~~~~~~~~~~~~0&=&\!\!
\int\!\rmD z \rmD y_0~y_0\!
\int\!\rmd t~p(t|Sy_0,\lambda_0)\Lambda(t)
\frac{\int\!\rmD y~p^\gamma(t|uy\!+\!wy_0\!+\!vz,\lambda)\rme^{uy\!+\!wy_0\!+\!vz}}
{\int\!\rmD y~p^\gamma(t|uy\!+\!wy_0\!+\!vz,\lambda)}
\nonumber
\\[-1mm]\hspace*{-25mm}&&
\label{eq:spe_w}
\end{eqnarray}
Upon integrating by parts over $y$, we can also write equation (\ref{eq:spe_u}) as
\begin{eqnarray}
\hspace*{-20mm}
\frac{\zeta}{\gamma u^2} \Big(\frac{v^2}{\gamma u^2}-\frac{1}{\gamma}\Big)
&=& \int\!\rmD z \rmD y_0
\int\!\rmd t~p(t|Sy_0,\lambda_0)
\label{eq:spe_u2}
\\
\hspace*{-20mm}
&&\hspace*{-20mm}
\times
\frac{\int\!\rmD y~p^\gamma(t|uy\!+\!wy_0\!+\!vz,\lambda)\Big[
[1-\rme^{uy+wy_0+vz}\Lambda(t)]^2-\gamma^{-1}\rme^{uy+wy_0+vz}\Lambda(t)
\Big]}
{\int\!\rmD y~p^\gamma(t|uy\!+\!wy_0\!+\!vz,\lambda)}
\nonumber
\end{eqnarray}
To work out the functional order parameter equation $\delta\Psi_{\rm RS}(u,v,w;\lambda)/\delta\lambda(s)=0$
we use  $\delta\log p(t|\xi)/\delta\lambda(s)=\delta(t\!-\!s)/\lambda(s)-\rme^{\xi}\theta(t\!-\!s)$, and the abbreviation $p(t)=\int\!\rmD y_0~p(t|Sy_0,\lambda_0)$. This gives
\begin{eqnarray}
\hspace*{-25mm}
0&=&\!
\int\!\rmD z \rmD y_0\!
\int\!\rmd t~p(t|Sy_0,\lambda_0)
\frac{\int\!\rmD y~p^\gamma(t|uy\!+\!wy_0\!+\!vz,\lambda)
\Big[\frac{\delta(t-s)}{\lambda(s)}-\rme^{uy+wy_0+vz}\theta(t\!-\!s)\Big]
}
{\int\!\rmD y~p^\gamma(t|uy\!+\!wy_0\!+\!vz,\lambda)}
\nonumber
\\
\hspace*{-25mm}
&=&
\frac{p(s)}{\lambda(s)}-
\int\!\rmD z \rmD y_0
\int_s^\infty\!\rmd t~p(t|Sy_0,\lambda_0)
\frac{\int\!\rmD y~p^\gamma(t|uy\!+\!wy_0\!+\!vz,\lambda)
\rme^{uy+wy_0+vz}
}
{\int\!\rmD y~p^\gamma(t|uy\!+\!wy_0\!+\!vz,\lambda)}
\label{eq:spe_lambda}
\end{eqnarray}
This latter equation can also be written in terms of the distribution 
(\ref{eq:Wresult}), giving a form that reduces to Breslow's \cite{Breslow} estimator when we subsequently use the interpretation identity (\ref{eq:Wmeaning}): 
\begin{eqnarray}
\lambda(t)&=& \frac{\int\!\rmd x\rmd x^\prime~\Prob_{\gamma}(x,x^\prime,t)}{\int_t^\infty\!\rmd t^\prime\int\!\rmd x\rmd x^\prime~\Prob_{\gamma}(x,x^\prime,t)\rme^{x^\prime}}
\end{eqnarray}
The remaining integrations over $y$ in our equations are for finite $\gamma$ quite nontrivial. They can be expressed in terms of the Laplace transform of the lognormal distribution \cite{Asmussen}, or mapped onto the core integral in the Random Energy Model \cite{Derrida}, both of 
which could in the past be evaluated analytically only in specific parameter limits.

\section{Analysis of the RS  equations for the Cox model}

\subsection{RS equations in the limit $\gamma\to\infty$}

The original Cox model \cite{Cox} corresponds to the limit $\gamma\to\infty$ of our equations. It turns out that the correct scaling with $\gamma$ of $u$ for $\gamma\to\infty$ is $u=\tilde{u}/\sqrt{\gamma}$; this is suggested by equation (\ref{eq:spe_u2}) and  confirms our expectation that follows from the physical meaning of $u$. 
Upon substituting $u=\tilde{u}/\sqrt{\gamma}$ as an ansatz into our equations, assuming the other order parameters to have finite $\gamma\to\infty$ limits, allows us to simplify the trio (\ref{eq:spe_v},\ref{eq:spe_w},\ref{eq:spe_u2}) and the functional equation (\ref{eq:spe_lambda})  to 
\begin{eqnarray}
 \frac{\zeta v}{\tilde{u}^2}&=&
\int\!\rmD z \rmD y_0~z
\int\!\rmd t~p(t|Sy_0,\lambda_0)\Lambda(t)A_1(wy_0+vz,t)
\label{eq:spe_v_Cox}
\\[1mm]
0&=&
\int\!\rmD z \rmD y_0~y_0
\int\!\rmd t~p(t|Sy_0,\lambda_0)\Lambda(t) A_1(wy_0+vz,t)
\label{eq:spe_w_Cox}
\\
\frac{\zeta v^2}{\tilde{u}^4} 
&=&1+ \int\!\rmD z \rmD y_0
\int\!\rmd t~p(t|Sy_0,\lambda_0)\Big[\Lambda^2(t) A_2(y_0,z,t)
\nonumber
\\
&&\hspace*{40mm}
-2\Lambda(t) A_1(wy_0+vz,t)\Big]
\label{eq:spe_uCox}
\\
\frac{p(t)}{\lambda(t)}&=& \int\!\rmD z \rmD y_0\int_t^\infty\!\rmd t^\prime~p(t^\prime|Sy_0,\lambda_0)A_1(wy_0+vz,t^\prime)
\end{eqnarray}
The remaining complexities of the limit are concentrated in 
\begin{eqnarray}
A_r(\eta,t)&=& \lim_{\gamma\to\infty}
\frac{\int\!\rmD y~p^\gamma(t|uy\!+\!\eta,\lambda)\rme^{r(uy+\eta)}}
{\int\!\rmD y~p^\gamma(t|uy\!+\!\eta,\lambda)}
\nonumber
\\
&=& \lim_{\gamma\to\infty}
\frac{\int\!\rmd y~\rme^{-\frac{1}{2}y^2 +\gamma\big[uy+\eta-\rme^{uy+\eta}\Lambda(t)\big]+r(uy+\eta)}}
{\int\!\rmd y~\rme^{-\frac{1}{2}y^2 +\gamma\big[uy+\eta-\rme^{uy+\eta}\Lambda(t)\big]}}
\nonumber
\\
&=& \lim_{\gamma\to\infty}
\frac{\int\!\rmd q~\rme^{\gamma\big[-\frac{1}{2}q^2 +\tilde{u}q+\eta-\rme^{\tilde{u}q+\eta}\Lambda(t)\big]+r(\tilde{u}q+wy_0+vz)}}
{\int\!\rmd q~\rme^{\gamma\big[-\frac{1}{2}q^2 +\tilde{u}q+\eta-\rme^{\tilde{u}q+\eta}\Lambda(t)\big]}}
\nonumber
\\
&=& \big[ \rme^{\varphi(wy_0+vz,t)\tilde{u}+wy_0+vz}\big]^r
\end{eqnarray}
with 
\begin{eqnarray}
\varphi(\eta,t)&=& {\rm argmax}_q\Big\{
-\frac{1}{2}q^2 +\tilde{u}q+\eta-\rme^{\tilde{u}q+\eta}\Lambda(t)
\Big\}
\end{eqnarray}
After differentiation and rewriting the resulting equation, we find that $\varphi(\eta,t)$ can be written in explicit form in terms of the Lambert W-function \cite{Lambert} as:
\begin{eqnarray}
\varphi(\eta,t)&=& \tilde{u}-\tilde{u}^{-1}W\Big(\tilde{u}^2\rme^{\tilde{u}^2+\eta}\Lambda(t)\Big)
\label{eq:phi_function}
\end{eqnarray}
Hence
\begin{eqnarray}
A_r(\eta,t)&=&\rme^{r\big[ \tilde{u}^2+\eta-W\big(\tilde{u}^2\exp(\tilde{u}^2+\eta)\Lambda(t)\big)\big]}
\end{eqnarray}
Using the identity $\rme^{-W(z)}=W(z)/z$, which follows directly from the definition of the Lambert $W$-function, we can simplify the above result to
\begin{eqnarray}
A_r(\eta,t)&=&  \Big(\frac{W\big(\tilde{u}^2\rme^{\tilde{u}^2+\eta}\Lambda(t)\big)}{\tilde{u}^2\Lambda(t)}\Big)^r
\end{eqnarray}
Substitution into our $\gamma\to\infty$ order parameter equations finally gives:
\begin{eqnarray}
\zeta v^2
&=& \int\!\rmD z \rmD y_0
\int\!\rmd t~p(t|Sy_0,\lambda_0)\Big[\tilde{u}^2-W\big(\tilde{u}^2\rme^{\tilde{u}^2+wy_0+vz}\Lambda(t)\big)
\Big]^2
\label{eq:spe_u_Cox_explicit_more}
\\
 \zeta v&=&
\int\!\rmD z \rmD y_0~z
\int\!\rmd t~p(t|Sy_0,\lambda_0)W\big(\tilde{u}^2\rme^{\tilde{u}^2+wy_0+vz}\Lambda(t)\big)
\label{eq:spe_v_Cox_explicit_more}
\\[1mm]
0&=&
\int\!\rmD z \rmD y_0~y_0
\int\!\rmd t~p(t|Sy_0,\lambda_0)W\big(\tilde{u}^2\rme^{\tilde{u}^2+wy_0+vz}\Lambda(t)\big)
\label{eq:spe_w_Cox_explicit_more}
\\
\frac{p(t)}{\lambda(t)}&=& \int\!\rmD z \rmD y_0\int_t^\infty\!\rmd t^\prime~p(t^\prime|Sy_0,\lambda_0)
\frac{W\big(\tilde{u}^2\rme^{\tilde{u}^2+wy_0+vz}\Lambda(t^\prime)\big)}{\tilde{u}^2\Lambda(t^\prime)}
\label{eq:spe_lambda_Cox_explicit_more}
\end{eqnarray}
We observe that the choice $v=0$ always solves (\ref{eq:spe_v_Cox_explicit_more}), but that for $\zeta>0$ it is ruled out by 
(\ref{eq:spe_u_Cox_explicit_more}). Upon doing integration by parts over $z$, using $\rmd W(z)/\rmd z= W(z)/z[1+W(z)]$
 and dismissing the solution $v=0$, we can simplify equation (\ref{eq:spe_v_Cox_explicit_more}) further to
\begin{eqnarray}
 \zeta &=&
\int\!\rmD z \rmD y_0
\int\!\rmd t~p(t|Sy_0,\lambda_0)\frac{W\big(\tilde{u}^2\rme^{\tilde{u}^2+wy_0+vz}\Lambda(t)\big)}{1+W\big(\tilde{u}^2\rme^{\tilde{u}^2+wy_0+vz}\Lambda(t)\big)}
\label{eq:spe_v_Cox_explicit_evenmore}
\end{eqnarray}
To compute the corresponding value of the overfitting measure $E(S,\lambda_0)=\lim_{\gamma\to \infty}E_\gamma(S,\lambda_0)$, we substitute $u=\tilde{u}/\sqrt{\gamma}$ into (\ref{eq:Cox_RS_final_E})  and take the limit $\gamma\to\infty$. This gives, using the short-hands (\ref{eq:phi_function}) and $p(t)=\int\!\rmD y_0~p(t|Sy_0,\lambda_0)$ and the identity $\exp[-W(z)]=W(z)/z$:
 \begin{eqnarray}
 \hspace*{-25mm}
 E(S,\lambda_0)&=&  
 \int\! \rmD y_0
\int\!\rmd t~p(t|Sy_0,\lambda_0)\left\{
\room
\log p(t|Sy_0,\lambda_0)-\log\lambda(t)
\right.
\nonumber
\\
 \hspace*{-25mm}
&&
\left.
\hspace*{-17mm}
-\lim_{\gamma\to\infty}\int\!\rmD z \frac{\int\!\rmd y~\rme^{\gamma[\tilde{u}y\!+\!wy_0\!+\!vz-\rme^{\tilde{u}y+wy_0+vz}\Lambda(t)-\frac{1}{2}y^2]}\Big[\tilde{u}y\!+\!wy_0\!+\!vz-\rme^{\tilde{u}y+wy_0+vz}\Lambda(t)\Big]}
{\int\!\rmd y~\rme^{\gamma[\tilde{u}y\!+\!wy_0\!+\!vz-\rme^{\tilde{u}y+wy_0+vz}\Lambda(t)-\frac{1}{2}y^2]}}
\right\}
\nonumber
\\
 \hspace*{-25mm}
&=&  
 \int\! \rmD y_0
\int\!\rmd t~p(t|Sy_0,\lambda_0)\left\{
\room
\log[\lambda_0(t)/\lambda(t)]-\rme^{Sy_0}\Lambda_0(t)
\right.
\nonumber
\\
 \hspace*{-25mm}
&&
\hspace*{30mm}
\left.
-\int\!\rmD z \Big[\tilde{u}\varphi(wy_0\!+\!vz,t)-\rme^{\tilde{u}\varphi(wy_0+vz,t)+wy_0+vz}\Lambda(t)\Big]
\right\}
\nonumber
\\
 \hspace*{-20mm}
&=& 
\int\!\rmd t~p(t)
\log\Big[\frac{\lambda_0(t)}{\lambda(t)}\Big]
-
 \int\! \rmD y_0
\int\!\rmd t~p(t|Sy_0,\lambda_0)\rme^{Sy_0}\Lambda_0(t)
-\tilde{u}^2
\nonumber
\\
 \hspace*{-20mm}
&&
\hspace*{10mm}
+~(1\!+\!\frac{1}{\tilde{u}^2})\int\!\rmD z  \rmD y_0
\int\!\rmd t~p(t|Sy_0,\lambda_0)
W\Big(\tilde{u}^2\rme^{\tilde{u}^2+wy_0+vz}\Lambda(t)\Big)
\end{eqnarray}
The second integral can be worked out explicitly:
\begin{eqnarray}
&&\hspace*{-20mm}
 \int\! \rmD y_0
\int_0^\infty\!\rmd t~p(t|Sy_0,\lambda_0)\rme^{Sy_0}\Lambda_0(t)\nonumber
\\
&=& 
- \int\! \rmD y_0
\int_0^\infty\!\rmd t~\rme^{Sy_0}\Lambda_0(t)\frac{\rmd}{\rmd t}\rme^{-\exp(Sy_0)\Lambda_0(t)}
\nonumber
\\
&=&
\int_0^\infty\!\rmd x~x\rme^{-x}~=~1
\end{eqnarray}
Therefore 
 \begin{eqnarray}
 \hspace*{-15mm}
 E(S,\lambda_0)&=&  
\int\!\rmd t~p(t)
\log\Big[\frac{\lambda_0(t)}{\lambda(t)}\Big]
\label{eq:Evalue_RS}
\\
\hspace*{-15mm}&&\hspace*{-5mm}
-(1\!+\!\tilde{u}^2)\Big[1-
\frac{1}{\tilde{u}^2}\int\!\rmD z  \rmD y_0
\int\!\rmd t~p(t|Sy_0,\lambda_0)
W\Big(\tilde{u}^2\rme^{\tilde{u}^2+wy_0+vz}\Lambda(t)\Big)
\Big]
\nonumber
\end{eqnarray}
{\red 
In \ref{app:limits} we study the behaviour of the above equations in the two limits $\zeta\to 0$ and $\zeta\to 1$. For $\zeta\to 0$ we recover the correct solution corresponding to perfect (overfitting-free) regression, as required. For $\zeta\to 1$ we find a phase transition, characterised by divergence of the order parameters $\{\tilde{u},v,w\}$. 
}

\subsection{Numerical and asymptotic solution of RS equations}

Solving the coupled order parameter equations (\ref{eq:spe_u_Cox_explicit_more},\ref{eq:spe_w_Cox_explicit_more},\ref{eq:spe_lambda_Cox_explicit_more},\ref{eq:spe_v_Cox_explicit_evenmore}) 
analytically seems for now too ambitious; solving them numerically is nontrivial, and requires some preparation.
To cast the equation for $w$ into a form similar to the others, we need to do partial integration over $y_0$:
\begin{eqnarray}
\hspace*{-10mm}
0&=&
w\int\!\rmD z \rmD y_0
\int\!\rmd t~p(t|Sy_0,\lambda_0)
\frac{W\big(\tilde{u}^2\rme^{\tilde{u}^2+wy_0+vz}\Lambda(t)\big)}{1\!+\!W\big(\tilde{u}^2\rme^{\tilde{u}^2+wy_0+vz}\Lambda(t)\big)}
\\
\hspace*{-10mm}&&
+~S
\int\!\rmD z \rmD y_0
\int\!\rmd t~p(t|Sy_0,\lambda_0)W\big(\tilde{u}^2\rme^{\tilde{u}^2+wy_0+vz}\Lambda(t)\big)\Big[
1\!-\!\rme^{Sy_0}\Lambda_0(t)\Big]
\nonumber
\end{eqnarray}
 We also rewrite the functional equation in a form that involves $\Lambda(t)$ only:
 \begin{eqnarray}
 \hspace*{-15mm}
\Lambda(t)&=&\int_0^t\!\rmd t^\prime~p(t^\prime)
\Big\{
 \int\!\rmD z \rmD y_0\int_{t^\prime}^\infty\!\rmd t^\pprime~p(t^\pprime|Sy_0,\lambda_0)
\frac{W\big(\tilde{u}^2\rme^{\tilde{u}^2+wy_0+vz}\Lambda(t^\pprime)\big)}{\tilde{u}^2\Lambda(t^\pprime)}
\Big\}^{-1}
\nonumber
\\[-1mm]
\hspace*{-15mm}&&
\end{eqnarray}
Numerical integration over $t>0$  can be transformed into integration over the survival function $s(t,y_0)=\exp[-\rme^{Sy_0}\Lambda_0(t)]\in[0,1]$, using $p(t|Sy_0,\lambda_0)\rmd t=-\rmd s$ and $t(s,y_0)=\Lambda_0^{\rm inv}(\rme^{-Sy_0}\log(1/s))$. We also define the short-hand $L(t)=\tilde{u}^2\rme^{\tilde{u}^2}\Lambda(t)$. 
These definitions transform our RS equations to:
\begin{eqnarray}
\hspace*{-10mm}
\zeta v^2&=& 
\int\! \rmD y_0\rmD z
\int_0^1\!\rmd s~\Big[\tilde{u}^2-W\Big(\rme^{wy_0+vz}L(t(s,y_0))\Big)
\Big]^2
\label{eq:RS_u_final}
\\
\hspace*{-10mm}
 \zeta &=&
\int\! \rmD y_0\rmD z
\int_0^1\!\rmd s ~
\left\{
\frac{W\Big(\rme^{wy_0+vz}L(t(s,y_0))\Big)}{1\!+\!W\Big(\rme^{wy_0+vz}L(t(s,y_0))\Big)}
\right\}
\label{eq:RS_v_final}
\\
\hspace*{-10mm}
\frac{\zeta w}{S}&=& -
\int\!\rmD y_0\rmD z
\int_0^1\!\rmd s~\big[
1\!+\!\log(s)\big]  W\Big(\rme^{wy_0+vz}L(t(s,y_0))\Big)
\label{eq:RS_w_final}
\\
\hspace*{-10mm}
L(t)
&=& \tilde{u}^2\int_0^t \!\rmd t^\prime~p(t^\prime)\nonumber
\\
\hspace*{-10mm}
&&\times \left\{
 \int\!\rmD y_0\rmD z\int_0^1\!\rmd s^\prime~\frac{\theta[t(s^\prime,y_0)\!-\!t^\prime]}{L(t(s^\prime,y_0))}
~W\Big(\rme^{wy_0+vz}L(t(s^\prime,y_0))\Big)
\right\}^{\!-1}~~
\label{eq:RS_L_final}
\end{eqnarray}

\begin{figure}[t]
\unitlength=0.38mm
\hspace*{-8mm}
\begin{picture}(400,140)
\put(20,0){\includegraphics[width=203\unitlength]{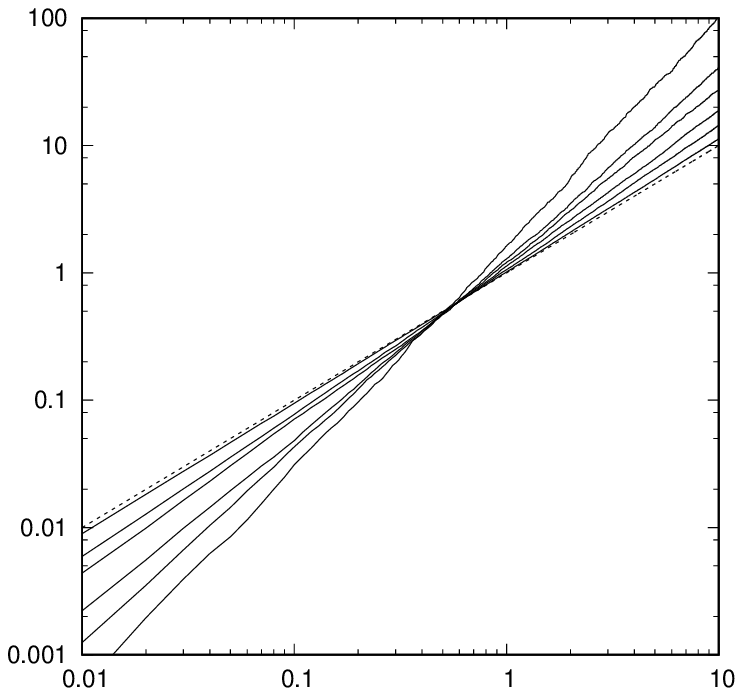}}
\put(180,0){\includegraphics[width=203\unitlength]{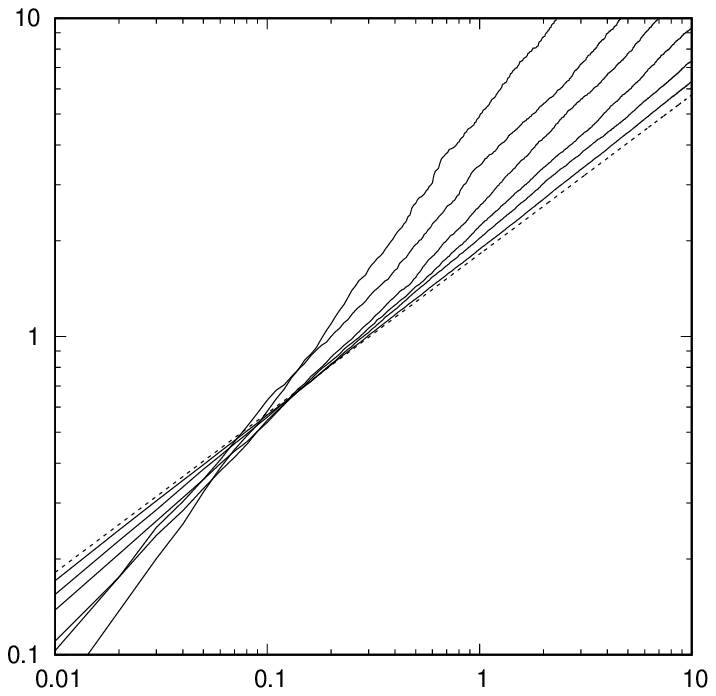}}

\put(115,3){\small\em $\Lambda_0(t)$}
\put(275,3){\small\em $\Lambda_0(t)$}
\put(34,85){\small $\hat{\Lambda}(t)$}
\put(194,85){\small $\hat{\Lambda}(t)$}
\put(68,122){\small $\lambda_0(t)\!=\!1$}
\put(228,122){\small $\lambda_0(t)\!=\!a/\sqrt{t}$}

\end{picture}
\vspace*{-5mm}

\caption{\small 
Here we show the simulation data of Figure \ref{fig:overfitting3} alternatively by drawing the inferred integrated base hazard rates $\hat{\Lambda}(t)$ versus the true values $\Lambda_0(t)$ in log-log plots. We observe that the curves for different values of $\zeta=p/N$ thereby become linear, with high accuracy, for both time-independent (left panel) and time-dependent base hazard rates (right panel). This suggests that $\hat{\Lambda}(t)\approx k\Lambda_0^\rho(t)$, with time-independent parameters $k$ and $\rho$ that depend on $\zeta$. {\red The power $\rho$ and the prefactor $k$  both increase with $\zeta$.} }
\label{fig:overfitting4}
\end{figure}

We next study the functional equation (\ref{eq:RS_L_final}) in more detail. We first rewrite it  by differentiation with respect to time, and some simple rearrangements, into the more suitable form
\begin{eqnarray}
\hspace*{-15mm}
\tilde{u}^2 \frac{p(t)}{
\frac{\rmd}{\rmd t}L(t)}
&=&
 \int\!\rmD y_0\rmD z\int_0^1\!\rmd s~\frac{\theta[t(s,y_0)\!-\!t]}{L(t(s,y_0))}
~W\Big(\rme^{wy_0+vz}L(t(s,y_0))\Big)
\end{eqnarray}
or, upon further differentiation:
\begin{eqnarray}
\hspace*{-15mm}
-\tilde{u}^2 L(t)\frac{\rmd}{\rmd t}\Big(\frac{p(t)}{
\frac{\rmd}{\rmd t}L(t)}\Big)
&=&\!
 \int\!\rmD y_0\rmD z~W\Big(\rme^{wy_0+vz}L(t)\Big)\int_0^1\!\rmd s~\delta[t(s,y_0)\!-\!t]~~
\end{eqnarray}
Using $\int_0^1\!\rmd s~\delta[t(s,y_0)\!-\!t]=p(t|Sy_0)$, and upon multiplying both sides by $\frac{\rmd}{\rmd t}L(t)/p(t)$, this becomes
\begin{eqnarray}
\hspace*{-15mm}
\tilde{u}^2\frac{\rmd}{\rmd t}\log\Big(\frac{
\rmd L(t)/\rmd t}{p(t)}\Big)
&=&\frac{\rmd \log L(t)}{\rmd t}
 \int\!\rmD y_0~\frac{p(t|Sy_0)}{p(t)}\int\!\rmD z~W\Big(\rme^{wy_0+vz}L(t)\Big)
 \label{eq:L_before_variational}
 \end{eqnarray}
 We write $L(t)$ in the form $L(t)=\Phi(\Lambda_0(t))$, which is always possible since both $L(t)$ and $\Lambda_0(t)$ are monotonic functions of time, and we write $p(t)=\lambda_0(t)g(\Lambda_0(t))$ with
 \begin{eqnarray}
g(x)&=& \int\!\rmD y_0~\rme^{Sy_0-x\exp(Sy_0)}
 \end{eqnarray}
 Substitution of these conventions, and working out the various time derivatives, then leads to the following equation from which to solve $\Phi(x)$:
  \begin{eqnarray}
  \hspace*{-10mm}
\frac{\tilde{u}^2 g(x)}{\rmd\log\Phi(x)/\rmd x}~\frac{\rmd}{\rmd x}
\log\Big(\frac{\rmd \Phi(x)/\rmd x}{g(x)}\Big)
&=& 
\int\!\rmD y_0~\rme^{Sy_0-x\exp(Sy_0)}
\nonumber
\\
  \hspace*{-10mm}
&&\times
\int\!\rmD z~W\Big(\rme^{wy_0+vz}\Phi(x)\Big)
 \end{eqnarray}
 We now proceed to calculate the solution $\Phi(x)$ of the above equation, which gives us the form of the inferred integrated base hazard rates $\Lambda(t)$ as shown in Figure \ref{fig:overfitting3}, for  large times, i.e. in the regime where $x\to \infty$ and $\Phi(x)\to \infty$. 
 Here we can use use the asymptotic form of the Lambert $W$-function \cite{Lambert}:  $W(z)=\log z - \log(\log z) +\order(\log(\log z)/\log z)$ (for $z\to\infty$), to obtain
   \begin{eqnarray}
   \hspace*{-25mm}
\frac{\tilde{u}^2 g(x)}{\rmd\log\Phi(x)/\rmd x}~\frac{\rmd}{\rmd x}
\log\Big(\frac{\rmd \Phi(x)/\rmd x}{g(x)}\Big)
&=& g(x)
\log\Big(\frac{\Phi(x)}{\log\Phi(x)}\Big)
+w
\int\!\rmD y_0~y_0\rme^{Sy_0-x\exp(Sy_0)}
\nonumber\\
  \hspace*{-25mm}
&&\hspace*{-15mm}
+
\int\!\rmD y_0~\rme^{Sy_0-x\exp(Sy_0)}
\order\Big(\frac{y_0}{\log \Phi(x)},\frac{\log\log\Phi(x)}{\log\Phi(x)}
\Big)
 \end{eqnarray}
 We can do the remaining integral over $y_0$ via integration by parts, giving
 \begin{eqnarray}
 \int\!\rmD y_0~y_0\rme^{Sy_0-x\exp(Sy_0)}&=&   S[g(x)+x\frac{\rmd}{\rmd x}g(x)]
 \end{eqnarray}
 Hence
   \begin{eqnarray}
   \hspace*{-15mm}
\frac{\tilde{u}^2\Phi}{\rmd\Phi/\rmd x}~\frac{\rmd}{\rmd x}
\Big[\log\Big(\frac{\rmd \Phi}{\rmd x}\Big)\!-\!\log g\Big]
&=&
\log\Big(\frac{\Phi}{\log\Phi}\Big)
+w
S\Big(1\!+\!x\frac{\rmd}{\rmd x}\log g\Big)
\nonumber
\\
\hspace*{-15mm}
&&
+
\order\Big(\frac{x~\rmd\log g/\rmd x}{\log \Phi},\frac{\log\log\Phi}{\log\Phi}\Big)
 \end{eqnarray}
 To proceed we need the leading {\red orders of $g(x)$. These are} derived in \ref{app:Asymptotic_g}: 
 {\red
 \begin{eqnarray}
  \hspace*{-15mm}
 \log g(x)&=& -\frac{1}{2S^2}(\log x)^2 +\frac{1}{S^2}\log x.\log(\log x)+\order(\log x)~~~~~~(x\to\infty)
 \label{eq:p_large_times}
 \end{eqnarray}
 }
 Our asymptotic equation for $\Phi(x)$ thereby becomes
 {\red 
    \begin{eqnarray}
    \hspace*{-15mm}
\frac{\tilde{u}^2\Phi}{\rmd\Phi/\rmd x}\Big[\frac{\rmd}{\rmd x}
\log\Big(\frac{\rmd \Phi}{\rmd x}\Big)
\!+\!
\frac{\log x}{xS^2}\!-\!\frac{\log\log x}{xS^2}\!+\!\order(\frac{1}{x})
\Big]
&=&
\log\Big(\frac{\Phi}{\log\Phi}\Big)
\nonumber
\\
&&
\hspace*{-55mm}
+\frac{w}{S}\Big(\log\log x\!-\!\log x\Big)
+~
\order\Big(1,\frac{\log x}{\log \Phi},\frac{\log\log\Phi}{\log\Phi},\frac{\Phi}{x\rmd\Phi/\rmd x}\Big)
\label{eq:IntermediatePhi_largex}
 \end{eqnarray}
Inspection of this equation shows that 
 the leading {\red orders of the solution are }
  \begin{eqnarray}
  \Phi(x)&=&\rho\log x+(1\!-\!\rho) \log\log x+{\it o}(\log\log x)\\
  \rho&=& \frac{w}{2S}\Big(1\!+\!\sqrt{1\!+\!4\tilde{u}^2/w^2}\Big)
  \label{eq:large_t_rho}
  \end{eqnarray}
 or
 \begin{eqnarray}
t\gg 1: &~&  \log\Lambda(t)=\rho\log \Lambda_0(t)+(1\!-\!\rho)\log(\log \Lambda_0(t)) +\ldots
\label{eq:large_t_Lambda}
 \end{eqnarray}}
This remarkably simple expression, linking the true and the inferred integrated base hazard rates $\Lambda(t)$ and $\Lambda_0(t)$, predicts that the relation between the two should {\red approach} a straight line when shown in a log-log plot. It is not only confirmed by simulations for large times (for which it was derived from our theory) but is in fact found to be quite accurate for all times. 
This is shown in Figure \ref{fig:overfitting4}, and forms the basis of our variational approximations below. 

\subsection{Variational approximation}

The main complexity of the RS theory is in solving the functional order parameter equation (\ref{eq:L_before_variational}). This is the motivation for investigating variational approximations for $\Lambda(t)$. Since our equations were obtained by solving an extremization problem, variational approaches are in the present context both natural and conceptually straightforward. The simulation data in Figure \ref{fig:overfitting4} suggest writing the functional order parameter in the form $\Lambda(t)=k\Lambda^\rho_0(t)$. To compute  the new scalar order parameters $k$ and  $\rho$  we substitute this expression for $\Lambda(t)$ into the quantity  (\ref{eq:Cox_RS_final}) to be extremized. 
As before we then put $u=\tilde{u}/\sqrt{\gamma}$ and take the limit $\gamma\to\infty$, and find that we need to extremize the following quantity over $(\tilde{u},v,w,k,\rho)$:
\begin{eqnarray}
\hspace*{-20mm}
\Psi(\tilde{u},v,w,k,\rho)&=& \frac{\zeta v^2}{2\tilde{u}^2}
+\log k+\log\rho+\int\!\rmd t~p(t)\log \Big[\lambda_0(t)\Lambda^{\rho-1}_0(t)\Big]
\nonumber
\\
\hspace*{-20mm}
&&
+\int\!\rmD z \rmD y_0
\int\!\rmd t~p(t|Sy_0,\lambda_0)
\nonumber
\\
\hspace*{-20mm}&&\hspace*{15mm}
\times
\lim_{\gamma\to\infty}\frac{1}{\gamma}
\log \int\!\rmd y~\rme^{\gamma[\tilde{u}y+wy_0+vz-k\rme^{\tilde{u}y+wy_0+vz}\Lambda^\rho_0(t)
-\frac{1}{2}y^2]}
\nonumber
\\
\hspace*{-20mm}
&=& 
 \frac{\zeta v^2}{2\tilde{u}^2}
+\log k+\log\rho+\int\!\rmd t~p(t)\log \Big[\lambda_0(t)\Lambda^{\rho-1}_0(t)\Big]
\nonumber
\\
\hspace*{-20mm}
&&
+\int\!\rmD z \rmD y_0\!
\int\!\rmd t~p(t|Sy_0,\lambda_0)
\nonumber
\\
\hspace*{-20mm}&&
\hspace*{15mm}
\times{\rm max}_y\Big[\tilde{u}y\!+\!wy_0\!+\!vz\!-\!
k
\rme^{\tilde{u}y+wy_0+vz}\Lambda^\rho_0(t)
\!-\!\frac{1}{2}y^2\Big]
\nonumber
\\
\hspace*{-20mm}
&=& 
 \frac{\zeta v^2}{2\tilde{u}^2}
+\log k+\log\rho+\int\!\rmd t~p(t)\log \Big[\lambda_0(t)\Lambda^{\rho-1}_0(t)\Big]
\nonumber
\\
\hspace*{-20mm}&&
+\int\!\rmD z \rmD y_0\!
\int\!\rmd t~p(t|Sy_0,\lambda_0)
\Big[\tilde{u}\varphi(wy_0\!+\!vz,t)
\!+\!wy_0\!+\!vz
\nonumber
\\
\hspace*{-20mm}
&&
\hspace*{10mm}
-k
\rme^{\tilde{u}\varphi(wy_0\!+\!vz,t)+wy_0+vz}\Lambda^\rho_0(t)
\!-\!\frac{1}{2}\varphi^2(wy_0\!+\!vz,t)\Big]~~~~
\label{eq:Psi_variational}
\end{eqnarray}
in which 
\begin{eqnarray}
\varphi(\eta,t)&=&
\tilde{u}-\frac{1}{\tilde{u}}W\Big(k\tilde{u}^2\rme^{\tilde{u}^2+\eta}\Lambda^\rho_0(t)\Big)
\end{eqnarray}
It is now easy to derive our order parameter equations, since all contributions to partial derivatives that involve  
$\varphi(wy_0\!+\!vz,t)$ vanish, by virtue of $\varphi(wy_0\!+\!vz,t)$ maximising the factor between the square brackets. Extremizing (\ref{eq:Psi_variational}) over $(\tilde{u},v,w)$ recovers our earlier equations (\ref{eq:RS_u_final},\ref{eq:RS_v_final},\ref{eq:RS_w_final}), with $L(t)=k\tilde{u}^2\rme^{\tilde{u}^2}\Lambda^\rho_0(t)$, as expected. Extremizing (\ref{eq:Psi_variational}) over the new order parameters $k$ and $\rho$ gives:
\begin{eqnarray}
\hspace*{-17mm}
\frac{1}{k}&=& 
\int\!\rmD y_0\rmD z \!
\int\!\rmd t~p(t|Sy_0,\lambda_0)
\Lambda^\rho_0(t)
\rme^{
\tilde{u}^2+wy_0+vz-W\big(k\tilde{u}^2\rme^{\tilde{u}^2+wy_0+vz}\Lambda^\rho_0(t)\big)
}
\\
\hspace*{-17mm}
\frac{1}{\rho}
&=& k\!\int\!\rmD z \rmD y_0\!
\int\!\rmd t~p(t|Sy_0,\lambda_0)\Lambda^\rho_0(t)
\rme^{
\tilde{u}^2+wy_0+vz-W\big(k\tilde{u}^2\rme^{\tilde{u}^2+wy_0+vz}\Lambda^\rho_0(t)\big)
}\log\Lambda_0(t)
\nonumber
\\[-2mm]
\hspace*{-17mm}
&&\hspace*{60mm} -\int\!\rmd t~p(t)\log\Lambda_0(t)
\end{eqnarray}
Using $W(z)\exp[W(z)]=z$ and our definition of $L(t)$, these two equations can be rewritten as
\begin{eqnarray}
\tilde{u}^2 &=& 
\int\!\rmD y_0\rmD z \!
\int_0^1\!\rmd s~
W\Big(
\rme^{wy_0+vz}
L(t(s,y_0))\Big)
\\
\frac{\tilde{u}^2}{\rho}
&=& \int\!\rmD z \rmD y_0\!
\int_0^1\!\rmd s~W\big(\rme^{wy_0+vz}L(t(s,y_0))\big)
\Big[\log\log(1/s)\!-\!Sy_0\Big]
\nonumber
\\
&&\hspace*{10mm}
 -\tilde{u}^2\int\!\rmd t~p(t)\log\Lambda_0(t)
\end{eqnarray}
In the second equation we rewrite the term with the explicit factor $y_0$, using
\begin{eqnarray}
\hspace*{-10mm}
&&\hspace*{-15mm}
 \int\!\rmD z \rmD y_0~y_0
\int_0^1\!\rmd s~W\big(\rme^{wy_0+vz}L(t(s,y_0))\big)
\nonumber
\\
\hspace*{-10mm}
&=&  \int\!\rmD z \rmD y_0\!\int_0^1\!\rmd s~\frac{\partial}{\partial y_0}W\big(\rme^{wy_0+vz}L(t(s,y_0))\big)
\nonumber
\\
\hspace*{-10mm}
&=&
\int\!\rmD z \rmD y_0\!\int_0^1\!\rmd s~
\frac{W\big(\rme^{wy_0+vz}L(t(s,y_0))\big)}{1\!+\!W\big(\rme^{wy_0+vz}L(t(s,y_0))\big)}
\frac{\partial}{\partial y_0}\log\big(\rme^{wy_0+vz}L(t(s,y_0))\big)
\nonumber
\\[1mm]
\hspace*{-10mm}
&=&
(w\!-\!\rho S)
\int\!\rmD z \rmD y_0\!\int_0^1\!\rmd s~
\frac{W\big(\rme^{wy_0+vz}L(t(s,y_0))\big)}{1\!+\!W\big(\rme^{wy_0+vz}L(t(s,y_0))\big)}
\end{eqnarray}
We thus arrive at five relatively simple closed  equations from which to solve $(\tilde{u},v,w,k,\rho)$ in our variational approximation.  Upon substituting the definition $t(s,y_0)
=\Lambda_0^{\rm inv}(\rme^{-Sy_0}\log(1/s))$ we can simplify the argument of Lambert's $W$-function, which appears in all equations, further to 
\begin{eqnarray}
W\Big(\rme^{wy_0+vz}L(t(s,y_0))\Big)&=& 
 W\Big(k\tilde{u}^2\rme^{\tilde{u}^2+(w-\rho S)y_0+vz}\log^\rho(1/s)\Big)
\end{eqnarray}
This enables us to combine the two Gaussian integrals appearing in each order parameter equation by a single zero-average Gaussian integral, 
with width
\begin{eqnarray}
\sigma(v,w)=\sqrt{(w\!-\!\rho S)^2+v^2}
\end{eqnarray}
We finally transform the variational order parameter $k$ to $q=k\tilde{u}^2 \rme^{\tilde{u}^2}$, and evaluate $\int\!\rmd t~p(t)\log\Lambda_0(t) = \int_0^\infty\!\rmd x~\rme^{-x}\log x=-C_{\rm E}$ \cite{GR}, which involves Euler's constant $C_{\rm E}=
0.5772156649015\ldots $. We then obtain
\begin{eqnarray}
\zeta v^2&=& 
\int\! \rmD x\!
\int_0^1\!\!\rmd s~\Big[\tilde{u}^2-W\Big(q\rme^{x\sigma(v,w,\rho)}\log^\rho(1/s)\Big)
\Big]^2
\label{eq:var1}
\\
 \zeta &=&
\int\! \rmD x\!
\int_0^1\!\!\rmd s ~
\frac{W\Big(q\rme^{x\sigma(v,w,\rho)}\log^\rho(1/s)\Big)}{1\!+\!W\Big(q\rme^{x\sigma(v,w,\rho)}\log^\rho(1/s)\Big)}
\label{eq:var2}
\\
\frac{\zeta w}{S}&=& -
\int\!\rmD x\!
\int_0^1\!\!\rmd s~\big[
1\!+\!\log(s)\big]~  W\Big(q\rme^{x\sigma(v,w,\rho)}\log^\rho(1/s)\Big)
\label{eq:var3}
\\[1mm]
\tilde{u}^2 &=& 
\int\!\rmD x\!
\int_0^1\!\!\rmd s~
W\Big(q\rme^{x\sigma(v,w,\rho)}\log^\rho(1/s)\Big)
\label{eq:var4}
\\[1mm]
\frac{\tilde{u}^2}{\rho}
&=& \int\!\rmD x\!
\int_0^1\!\!\rmd s~W\Big(q\rme^{x\sigma(v,w,\rho)}\log^\rho(1/s)\Big)
\log\log(1/s)
\nonumber
\\
&&\hspace*{30mm}
-S(w\!-\!\rho S)\zeta +\tilde{u}^2 C_{\rm E}
\label{eq:var5}
\end{eqnarray}
In the same way we can work out the value of $E(S,\lambda_0)$ for the variational solution, and find:
 \begin{eqnarray}
 E(S,\lambda_0)&=&  
\int\!\rmd t~p(t)
\log\Big[\frac{\lambda_0(t)}{\lambda(t)}\Big]
~=~-\int\!\rmd t~p(t)
\log\Big[k\rho \Lambda^{\rho-1}_0(t)\Big]
\nonumber
\\
&=& -\log k-\log \rho
 -(\rho\!-\!1)\int\!\rmd t~p(t)
\log \Lambda_0(t)
\nonumber
\\
&=& -\log k-\log \rho
 -(\rho\!-\!1)\int_0^\infty\!\rmd x~\rme^{-x}
\log x
\nonumber
\\[1mm]
&=& -\log k-\log \rho
 +(\rho\!-\!1)C_{\rm E}
 \end{eqnarray}
For $q\to 0$ we may replace $W(q\rme^{\sigma x}\log^\rho(1/s))\approx q\rme^{\sigma x}\log^\rho(1/s)$
and use the integral $\int_0^1\!\rmd s~\log(1/s)\log\log(1/s)=1-C_{\rm E}$, 
 to recover after some simple expansions the correct $\zeta\to 0$ solution:
 $\lim_{\zeta\to 0}v=\lim_{\zeta\to 0}\tilde{u}=0$, 
$\lim_{\zeta\to 0}w= S$, $\lim_{\zeta\to 0}\rho=\lim_{\zeta\to 0}k= 1$, and $\lim_{\zeta\to 0}E(S,\lambda_0)=0$. 

\begin{figure}[t]
\unitlength=0.45mm
\hspace*{-25mm}{\red 
\begin{picture}(400,125)
\put(120,0){\includegraphics[width=180\unitlength]{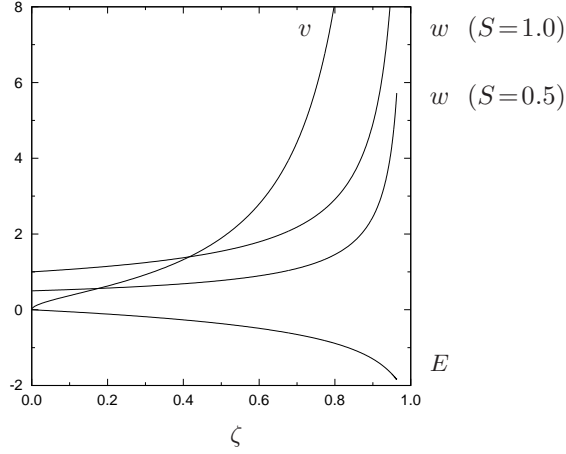}}
\put(210,-8){\small\em $\zeta$}
\put(230,112){$v$}
\put(269,13){\small $E$}
\put(269,92){$w~~{\small (S\!=\!0.5)}$}
\put(269,112){$w~~{\small (S\!=\!1.0)}$}

\end{picture}}
\vspace*{0mm}

\caption{\small 
Result of solving numerically the variational equations  (\ref{eq:compact1},\ref{eq:compact2},\ref{eq:compact3}). The values of $v$, $k,$ $\rho=w/S$ and $E$ are independent of the strength $S$ of the true associations and independent of the true base hazard rate $\lambda_0(t)$. For $\zeta=0$ we recover the overfitting-free state $w=S$ and $v=E=0$. At $\zeta=1$ a phase transition occurs, marked by divergence of $v$ and $w$. }
\label{fig:variational}
\end{figure}

We observe that our above closed variational equations (\ref{eq:var1}--\ref{eq:var5}) are completely independent of the true base hazard rate $\lambda_0(t)$. Hence they predict that the key quantities required for overfitting correction in the Cox model (the slope of the data cloud, and the deformation parameters of the base hazard rate) are independent of the true shape of the base hazard rate.

The easiest protocol for solving our equations numerically  is to regard $q$ as an independent parameter, and compute $(\zeta,v,w,\tilde{u},\rho)$ for each $q$ by iterative mapping. Upon doing so (see Figure \ref{fig:variational}),  one finds that the solution always exhibits $\rho=w/S$, within numerical accuracy limitations. 
We have not yet been able to confirm this analytically, as that would require proving that the solution of our equation obeys
\begin{eqnarray}
\int\!\rmD x\!\int_0^1\!\rmd s~W\Big(q\rme^{xv}\log^\rho(1/s)\Big)\Big[\log\log(1/s)\!+\!C_{\rm E}\!-\!\frac{1}{\rho}\Big]=0
\end{eqnarray}
 but it is for small $\zeta$ in agreement with (\ref{eq:large_t_rho}) (as it should be). If $\rho=w/S$ is indeed generally  true for the solution of our variational equations, it implies  that $\rho$ is identical to the slope of the data clouds in Figure \ref{fig:overfitting2}, and that the values of $(v,\rho,q)$ (hence also of the slope and the width of the data clouds in Figure \ref{fig:overfitting2}) are not only independent of $\lambda_0(t)$ but also independent of $S$.
It would also allow us to obtain a more compact closed theory in terms of just three scalar order parameters, as we will show now. Upon making directly the variational ansatz $\Lambda(t)=k\Lambda^\rho_0(t)$ with $w=\rho S$, 
we need to extremize 
\begin{eqnarray}
\hspace*{-20mm}
\Psi(\tilde{u},v,k,\rho)&=& 
 \frac{\zeta v^2}{2\tilde{u}^2}
+\log k+\log\rho+\int\!\rmd t~p(t)\log \Big[\lambda_0(t)\Lambda^{\rho-1}_0(t)\Big]
\nonumber
\\
\hspace*{-20mm}
&&
+\int\!\rmD z \rmD y_0\!
\int\!\rmd t~p(t|Sy_0,\lambda_0)\Big[\tilde{u}\varphi(\rho Sy_0\!+\!vz,t)
\!+\!\rho Sy_0\!+\!vz
\nonumber
\\
\hspace*{-20mm}
&&
\hspace*{10mm}
-
k
\rme^{\tilde{u}\varphi(\rho Sy_0\!+\!vz,t)+\rho Sy_0+vz}\Lambda^\rho_0(t)
\!-\!\frac{1}{2}\varphi^2(\rho Sy_0\!+\!vz,t)\Big]~~~~
\label{eq:Psi_variational2}
\end{eqnarray}
in which again 
$\varphi(\eta,t)=
\tilde{u}-\tilde{u}^{-1}W(k\tilde{u}^2\rme^{\tilde{u}^2+\eta}\Lambda^\rho_0(t))$.
Following similar manipulations as used for the first variational analysis, and with the previous short-hand $q=k\tilde{u}^2\rme^{\tilde{u}^2}$, we find upon extremization of $\Psi(\tilde{u},v,k,\rho)$ and after elimination of $\tilde{u}$ the following three closed equations for $(v,k,\rho)$:
\begin{eqnarray}
\hspace*{-10mm}
\zeta v^2&=& 
\int\!\rmD x\!
\int_0^1\!\rmd s~\Big[\tilde{u}^2-W\Big(q\rme^{vx}\log^\rho(1/s)\Big)\Big]^2
\label{eq:compact1}
\\
\hspace*{-10mm}
\zeta &=&
\int\!\rmD x\!
\int_0^1\!\rmd s~\frac{W\Big(q\rme^{vx}\log^\rho(1/s)\Big)}{1\!+\!W\Big(q\rme^{vx}\log^\rho(1/s)\Big)}
\label{eq:compact2}
\\
\hspace*{-10mm}
\zeta\rho &=& 
-\frac{1}{S^2}\int\!\rmD x\!
\int_0^1\!\rmd s~
W\Big(q\rme^{vx}\log^\rho(1/s)\Big)\log\log(1/s)
\nonumber
\\
\hspace*{-10mm}
&&
-\int\!\rmD x \!
\int_0^1\!\rmd s~
\Big[1\!+\!\log(s)\!+\!(C_{\rm E}\!-\!\frac{1}{\rho})/S^2\Big]W\Big(q\rme^{vx}\log^\rho(1/s)\Big)
\label{eq:compact3}
\end{eqnarray}
Upon solving the trio (\ref{eq:compact1},\ref{eq:compact2},\ref{eq:compact3}),  
the values of $\tilde{u}$, $w$ and $k$ then follow via
\begin{eqnarray}
\hspace*{-10mm}
\tilde{u}^2= 
\int\!\rmD x\!
\int_0^1\!\rmd s~
W\Big(q\rme^{vx}\log^\rho(1/s)\Big),~~~~~~k=\frac{q}{\tilde{u}^2}\rme^{-
\tilde{u}^2},~~~~~~w=\rho S
\end{eqnarray}
Finally we note that all our equations in this section can also be written in a form that involves only integrations over the interval $[0,1]$, using the general identity
\begin{eqnarray}
\int\!\rmD x~f(x)&=& \int_0^1\!\rmd s~\frac{f\big(\sqrt{2\log(1/s)}\big)+f\big(-\!\sqrt{2\log(1/s)}\big)}{2\sqrt{\pi\log(1/s)}}
\end{eqnarray}

\begin{figure}[t]
\unitlength=0.375mm
\hspace*{-9mm}{\red
\begin{picture}(400,119)
\put(0,0){\includegraphics[width=170\unitlength]{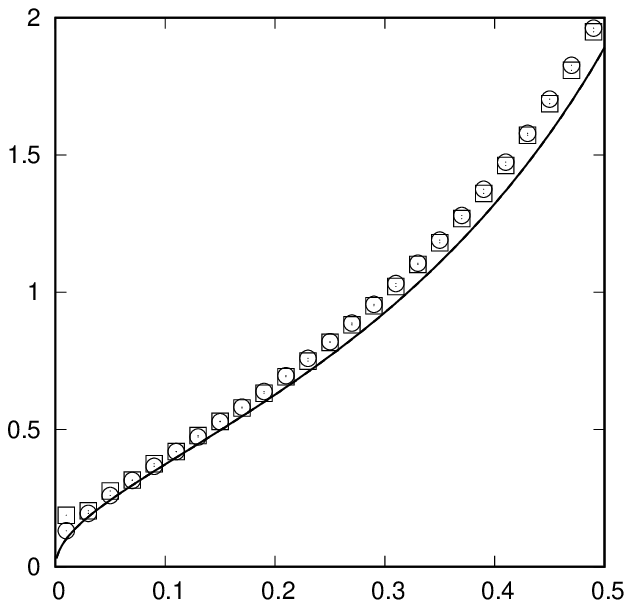}}
\put(120,0){\includegraphics[width=170\unitlength]{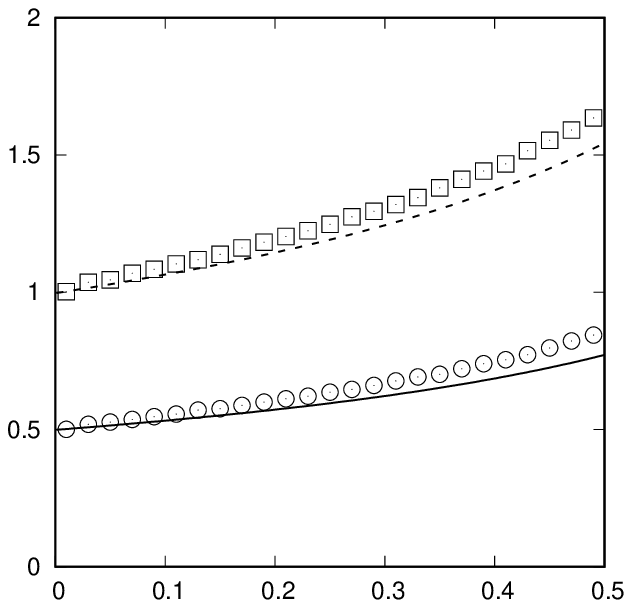}}
\put(240,0){\includegraphics[width=170\unitlength]{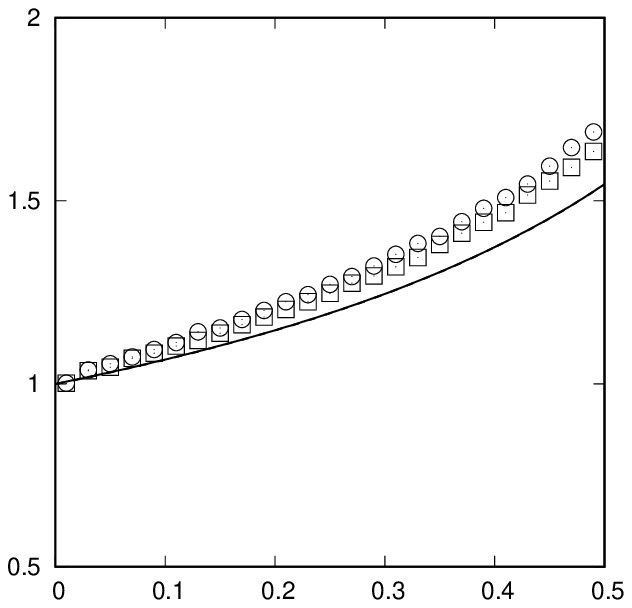}}
\put(20,70){\small $v$}
\put(80,5){\small $\zeta$}
\put(140,70){\small $w$}
\put(200,5){\small $\zeta$}
\put(256,70){\small $w/S$}
\put(320,5){\small $\zeta$}

\end{picture}}
\vspace*{-6mm}

\caption{\small 
Test of the predictions of the variational equations (\ref{eq:compact1},\ref{eq:compact2},\ref{eq:compact3}) against numerical simulations of Cox regression, with $N=200$, $\lambda_0(t)=1$, and either $S=0.5$ (circles) or $S=1.0$ (squares). Left: order parameter $v$ (solid line) versus $v(r,N)$, see equation  (\ref{eq:v_measured}). Middle: order parameter $w$ (solid line: $S=0.5$; dashed: $S=1.0$) versus $w(r,N)$, see equation  (\ref{eq:w_measured}). Right: the corresponding values of {\red $w/S$. In all cases $r=10^4$. } The simulations confirm the predictions of the theory that both $v$ and $w/S$ are independent of $S$.
}
\label{fig:variational_test}
\end{figure}

It is instructive at this stage to test the predictions of the above simple variational equations (\ref{eq:compact1},\ref{eq:compact2},\ref{eq:compact3}) against numerical simulations of Cox regression on synthetic data.  According to (\ref{eq:interpretation_u},\ref{eq:interpretation_v},\ref{eq:interpretation_w}),
we must expect to find in our simulations that $v=\lim_{r,N\to\infty}v(r,N)$ and 
$w=\lim_{r,N\to\infty}w(r,N)$, where
\begin{eqnarray}
v(r,N)&=&\frac{1}{\zeta N}\Big[\sum_{\mu=1}^{\zeta N}\bra \hat{\beta}_\mu^2\ket_{\Data} -
\frac{1}{|\bbeta^\star|^2}\Big(\sum_{\mu=1}^{\zeta N}\beta_\mu^{\star}\bra \hat{\beta}_\mu \ket_{\Data}\Big)^2
\Big]
\label{eq:v_measured}
\\
w(r,N)&=&\frac{1}{\zeta N}\sum_{\mu=1}^{\zeta N} \frac{\beta_\mu^{\star}\!\cdot\!\bra \hat{\beta}_\mu \ket_{\Data}}{|\bbeta^\star|} 
\label{eq:w_measured}
\end{eqnarray}
Here $\{\hat{\beta}_\mu\}$ denotes the inferred values of the (rescaled) regression parameters, and the averages $\bra \ldots\ket_{\Data}$ are over $r$ randomly generated data  sets.  The results of measuring $v(r,N)$ and $w(r,N)$ in numerical simulations are shown in Figure \ref{fig:variational_test} together with the variational predictions. In spite of the modest values in our simulations of $N=200$ and the finite number of training sets over which inferred parameters are averaged in evaluating (\ref{eq:v_measured},\ref{eq:w_measured})  (which one expects to generate excess variability), the agreement  between the variational predictions and the simulations is seen to be surprisingly good.

\section{Tests and applications}

\begin{figure}[t]
\unitlength=0.37mm
\hspace*{5mm}
\begin{picture}(400,278)
\put(20,0){\includegraphics[width=203\unitlength]{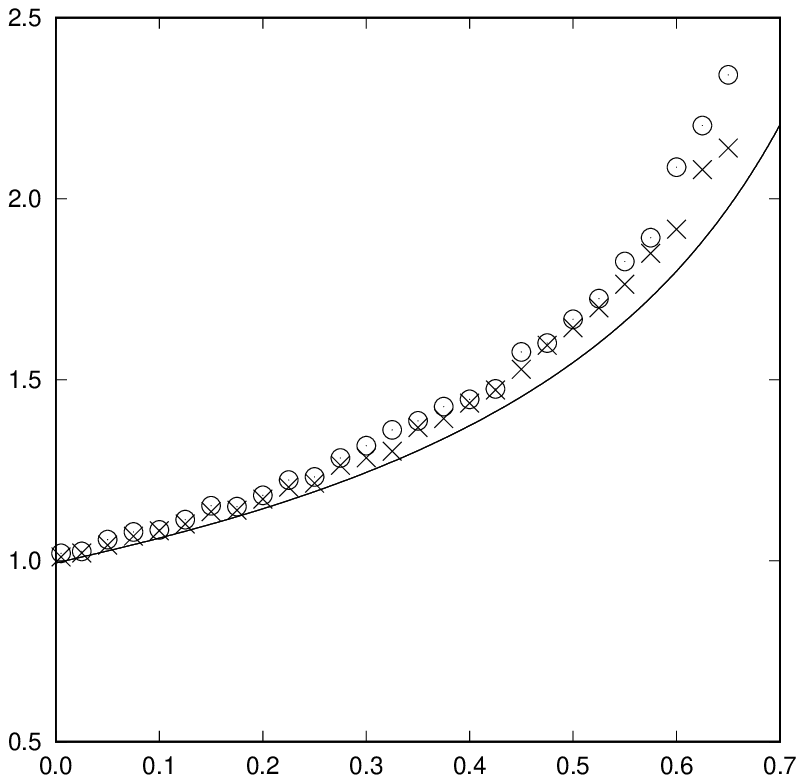}}
\put(155,0){\includegraphics[width=203\unitlength]{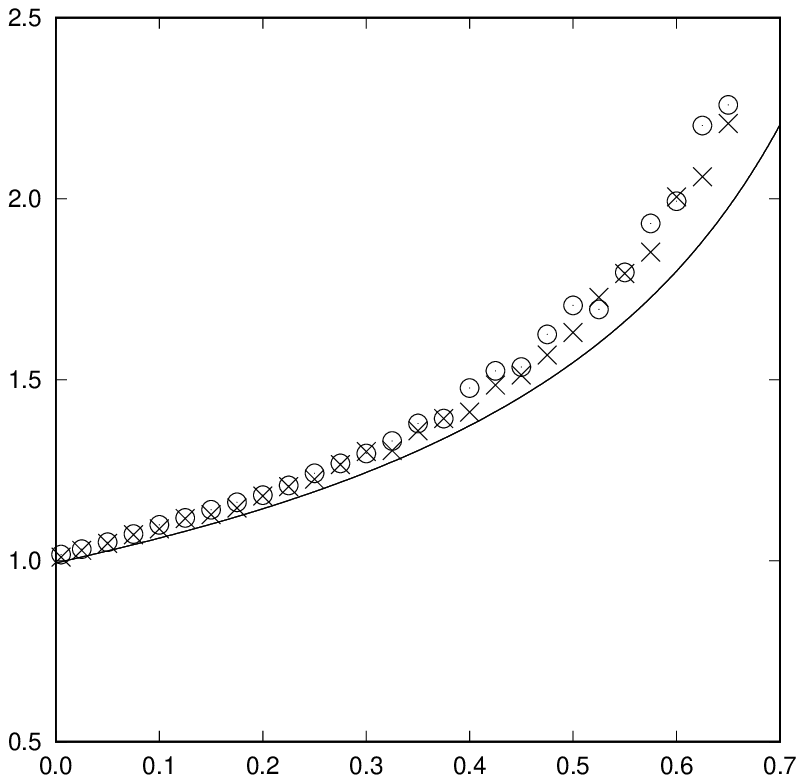}}
\put(20,135){\includegraphics[width=203\unitlength]{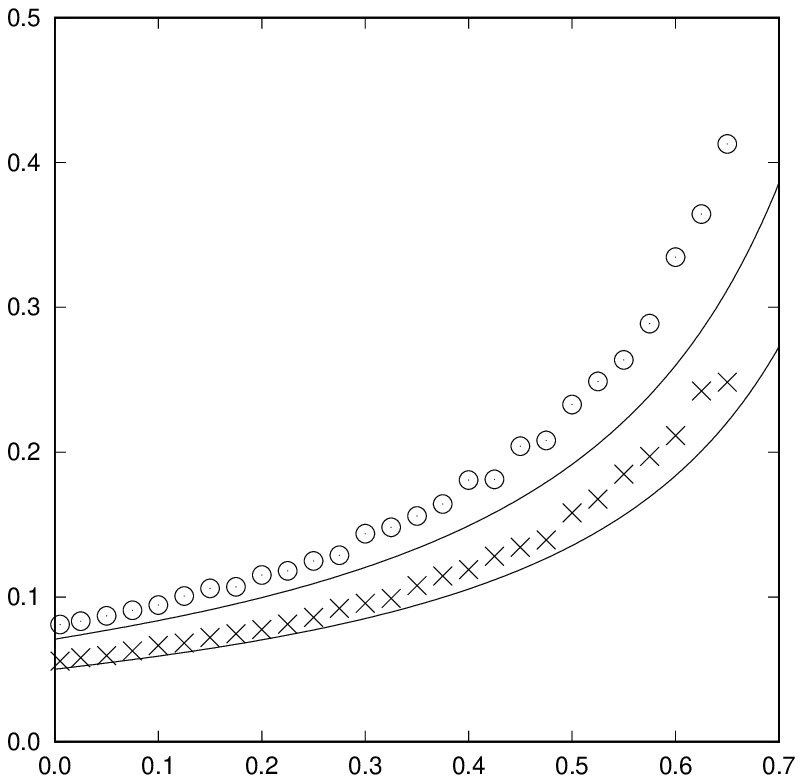}}
\put(155,135){\includegraphics[width=203\unitlength]{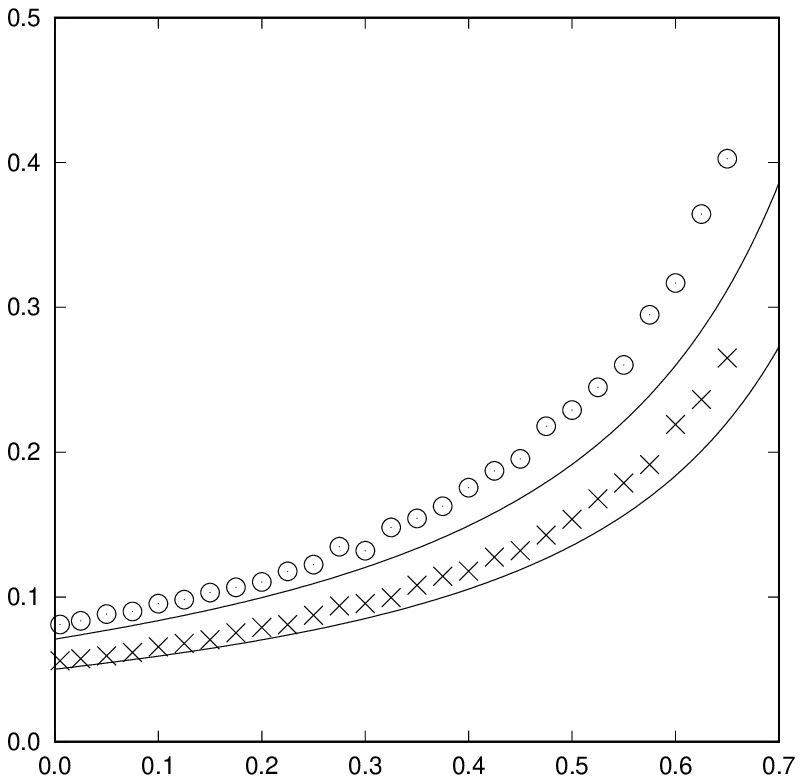}}

\put(118,3){\small $\zeta$}
\put(253,3){\small $\zeta$}
\put(42,220){$\sigma$}
\put(42,85){\small $\kappa$}
\put(67,122){\small $\lambda_0(t)\!=\!1$}\put(67,257){\small $\lambda_0(t)\!=\!1$}
\put(202,122){\small $\lambda_0(t)\!=\! a/\sqrt{t}$}\put(202,257){\small $\lambda_0(t)\!=\!a/\sqrt{t}$}

\end{picture}
\vspace*{-6mm}

\caption{\small 
We show the slopes $\kappa$ and the widths $\sigma$ of the association parameter data clouds of Figure 
 \ref{fig:overfitting2}, computed from regression simulations carried out on synthetic survival data via least squares fitting, for $N=200$ (circles) and $N=400$ (crosses). In all cases $S=0.5$. Solid lines: predictions of the variational theory, viz. $\sigma=v/\sqrt{p}$ and $\kappa=\rho$ (both of which are independent of $\lambda_0(t)$ and of $S$). Top row: widths $\sigma$, for constant (left) and time-dependent (right) base hazard rates, with $a=\exp(S^2)/\sqrt{2}$ defined such that $\int\!\rmd t~p(t)t=1$. Bottom row: slopes $\kappa$, for constant (left) and time-dependent (right) base hazard rates. Each marker is an average over $r$ independent simulation experiments, such that the product $pr$ is the same for all markers.
 }
\label{fig:test_slopes_and_errors}
\end{figure}

We will now test the variational RS theory (\ref{eq:compact1},\ref{eq:compact2},\ref{eq:compact3}) further  
 against numerical simulations, focusing on the the dependence on the ratio $\zeta$ of the main characteristics of the regression parameter data clouds of Figure 
 \ref{fig:overfitting2} (i.e. their slope $\kappa$ and their width $\sigma$), and of the integrated base hazard rates as shown e.g. in Figure 
 \ref{fig:overfitting3}. We know  (\ref{eq:link_with_experiment}) that the theory predicts 
$\kappa=\rho$ and $\sigma=v/\sqrt{p}$ (for the standard scaling convention of the Cox model \cite{Cox}, i.e. for $p(t|\bz)=-\frac{\rmd}{\rmd t}\exp[-\exp(\bbeta\cdot\bz)\Lambda(y)]$), and these predictions are plotted in Figure \ref{fig:test_slopes_and_errors} as solid lines, together with the values obtained in regression simulations 
of the Cox model on synthetic data (markers), for $N=200$ and $N=400$, and for two distinct choices for the true base hazard rate $\lambda_0(t)$. Modulo finite size effects, which increase as we approach the phase transition point $\zeta =1$, there is again good agreement between theory and simulations. The data confirm also the prediction of the variational theory that both $\kappa$ and $\sigma$ are independent of the true base hazard rate $\lambda_0(t)$. 

\begin{figure}[t]
\unitlength=0.38mm
\hspace*{-3mm}
\begin{picture}(400,136)
\put(20,-7){\includegraphics[width=203\unitlength]{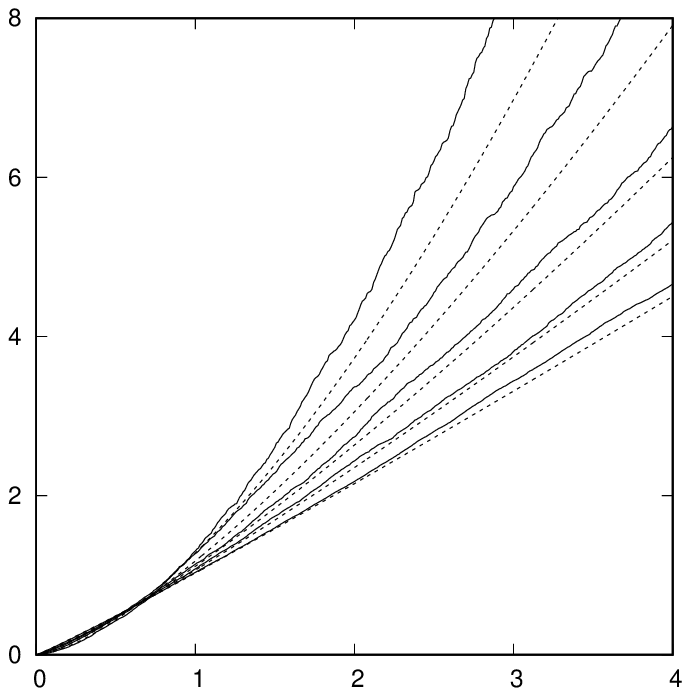}}
\put(180,-7){\includegraphics[width=203\unitlength]{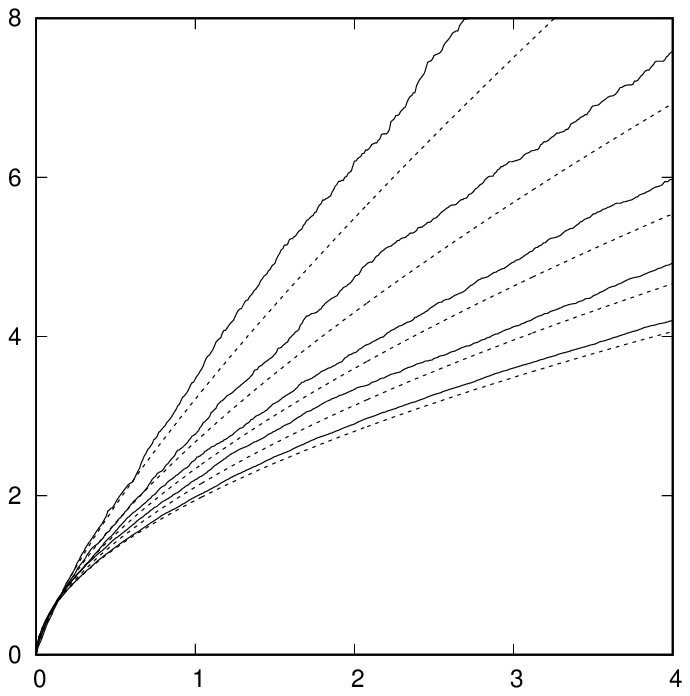}}

\put(119,-3){\small $t$}
\put(279,-3){\small $t$}
\put(34,75){\small $\hat{\Lambda}(t)$}
\put(194,75){\small $\hat{\Lambda}(t)$}
\put(69,115){\small $\lambda_0(t)\!=\!1$}
\put(229,115){\small $\lambda_0(t)\!=\!a/\sqrt{t}$}

\end{picture}
\vspace*{-4mm}

\caption{\small 
Inferred integrated base hazard rates $\hat{\Lambda}(t)$ (solid curves, averaged over multiple experiments) for synthetic survival data, shown together with the predictions of the variational RS theory (dashed curves) for $\zeta\in\{0.1, 0.2, 0.3, 0.4, 0.5\}$ (lower to upper curves). 
In all simulations $N=400$, $S=0.5$, and  $a$ is defined such that $\int\!\rmd t~p(t)t=1$.
}
\label{fig:compare_baserates}
\end{figure}

In Figure \ref{fig:compare_baserates} we compare the inferred integrated base hazard rates $\hat{\Lambda}(t)$, obtained for synthetic data with $N=400$, with the predictions of the variational RS theory (\ref{eq:compact1},\ref{eq:compact2},\ref{eq:compact3}), for two choices of the base hazard rate. The agreement is satisfactory for times of the order of the typical event times in the data. For larger times (where the theory has to extrapolate to times where available data are at best sparse) one observes increasing 
deviations, with the variational theory underestimating the impact of overfitting; this is indeed consistent with (\ref{eq:large_t_Lambda}), since the variational approximation captures only the first (leading) term of the exact expansion (\ref{eq:large_t_Lambda}). We can {\red in principle} obtain more accurate integrated base hazard rate predictions within the current framework, but this requires that we either solve (numerically) the full RS equations (\ref{eq:RS_u_final},\ref{eq:RS_v_final},\ref{eq:RS_w_final},\ref{eq:RS_L_final}), or develop a more refined variational ansatz for the function $L(t)$.

We found in our simulations that as the ratio $\zeta =p/N$ increases, higher numerical precision is required in solving Cox's equations. For values $N\sim 10^2-10^3$ and $\zeta>0.4$, using conventional C-code compiled with gcc at double floating point precision (data type `double') will occasional lead to   degeneracies in the equations that cause the association parameters $\hat{\bbeta}$ to be ill-defined. Upon switching to quadruple floating point precision (data type `long double') these degeneracies disappeared. 

\begin{figure}[t]
\unitlength=0.34mm
\hspace*{-10mm}
\begin{picture}(400,199)

\put(0,100){\includegraphics[width=143\unitlength]{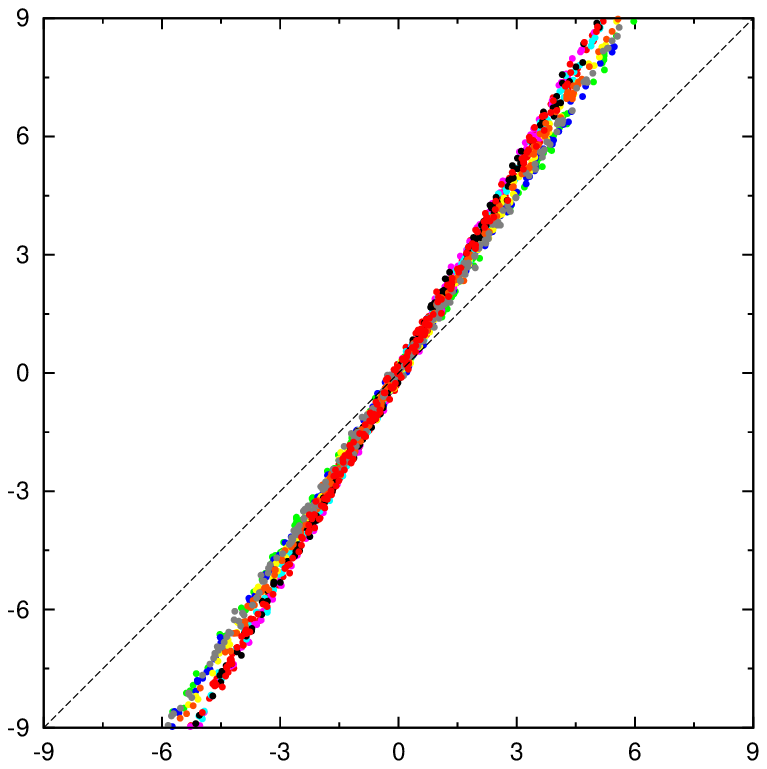}}
\put(100,100){\includegraphics[width=143\unitlength]{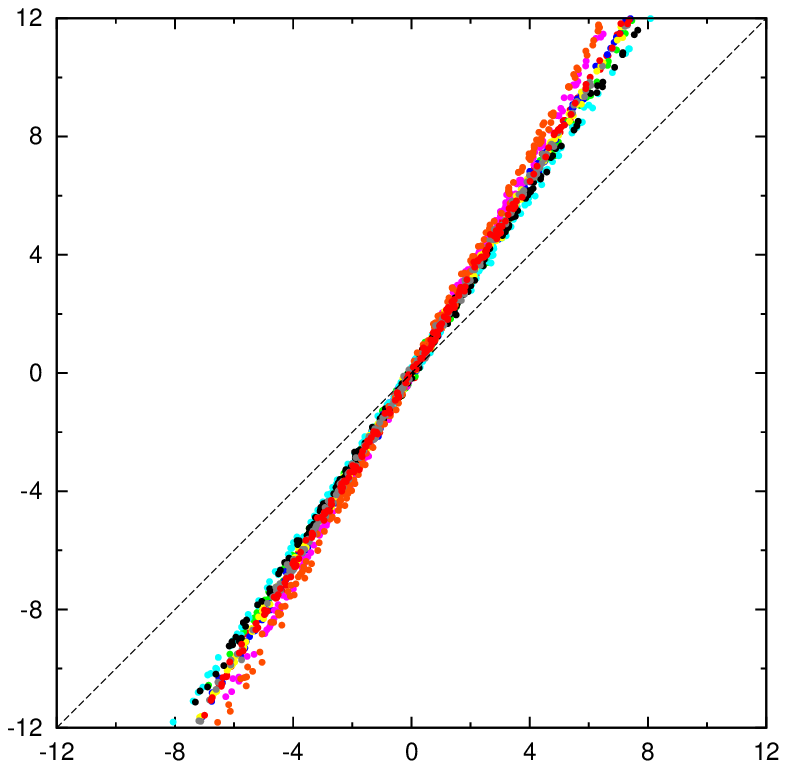}}
\put(200,100){\includegraphics[width=143\unitlength]{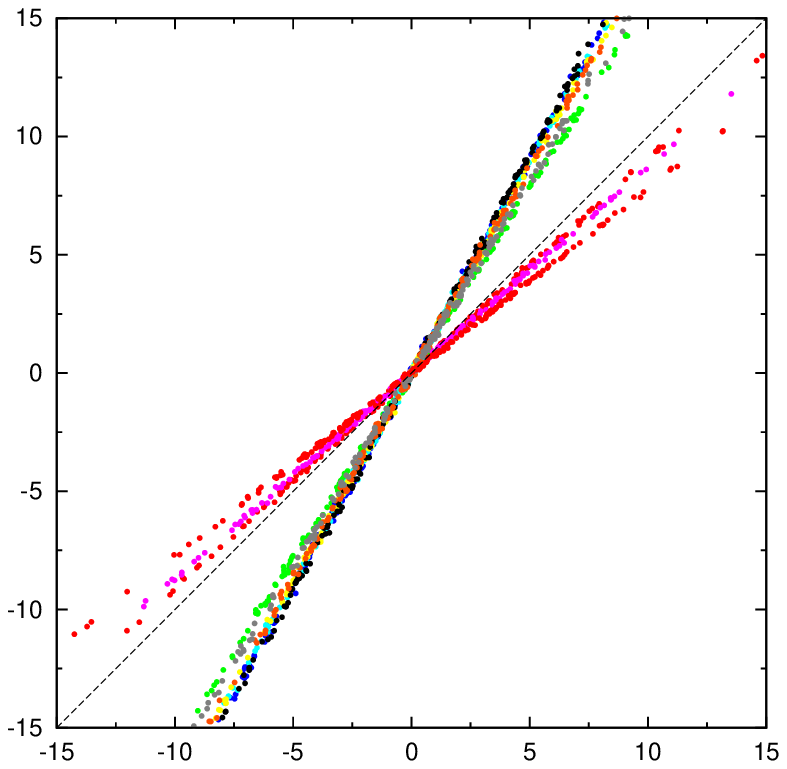}}
\put(300,100){\includegraphics[width=143\unitlength]{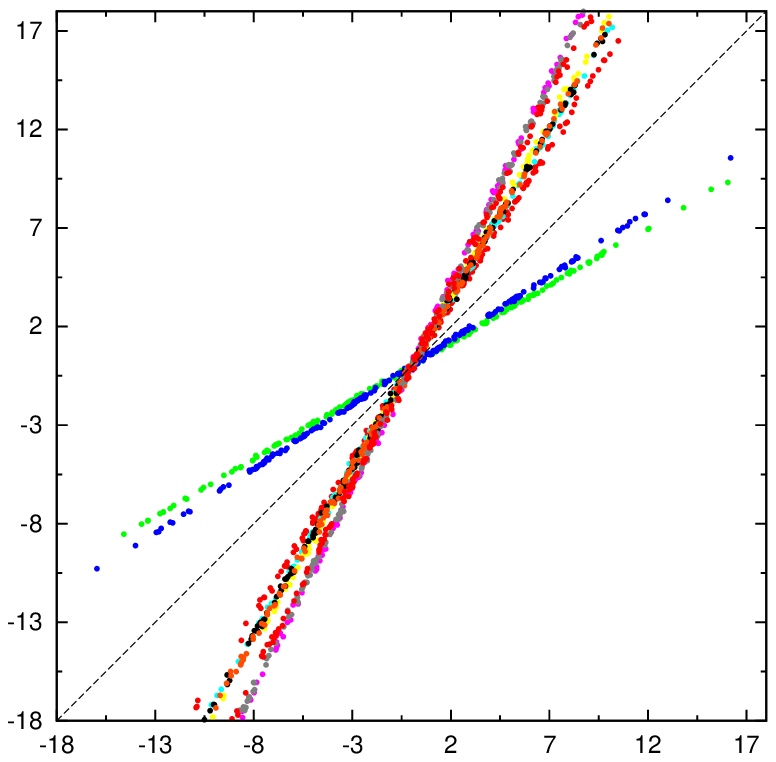}}

\put(0,0){\includegraphics[width=143\unitlength]{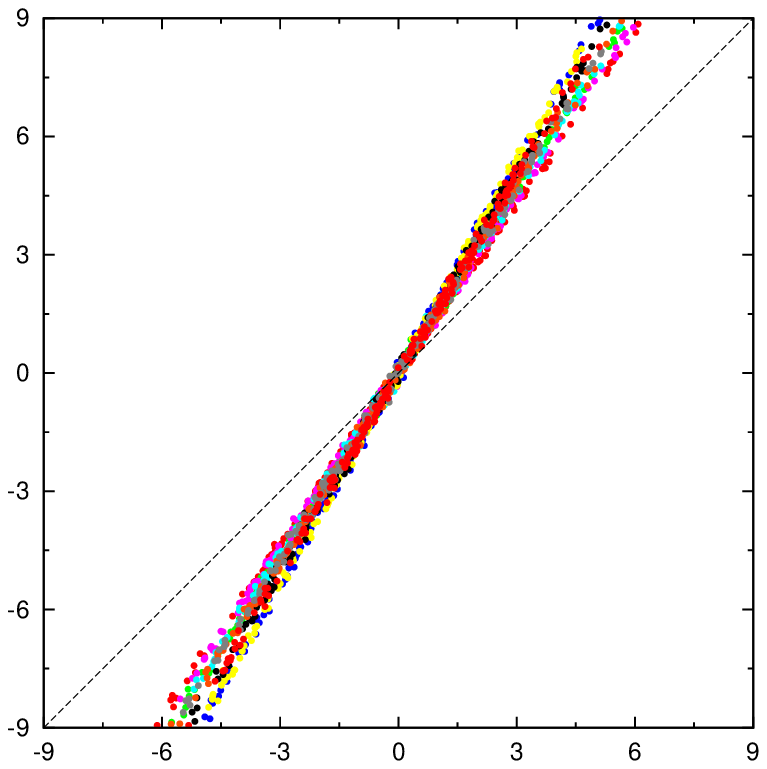}}
\put(100,0){\includegraphics[width=143\unitlength]{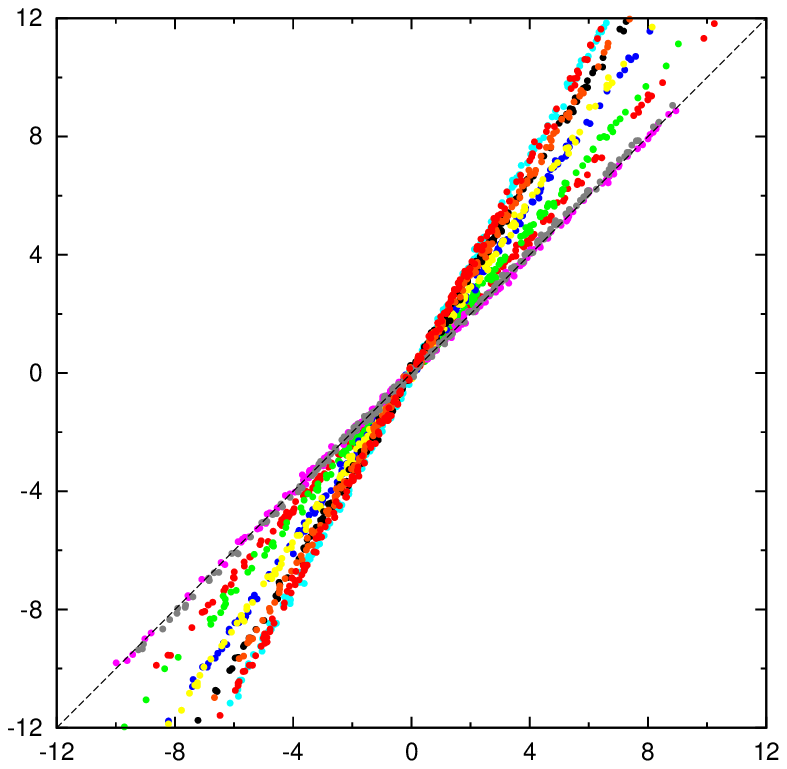}}
\put(200,0){\includegraphics[width=143\unitlength]{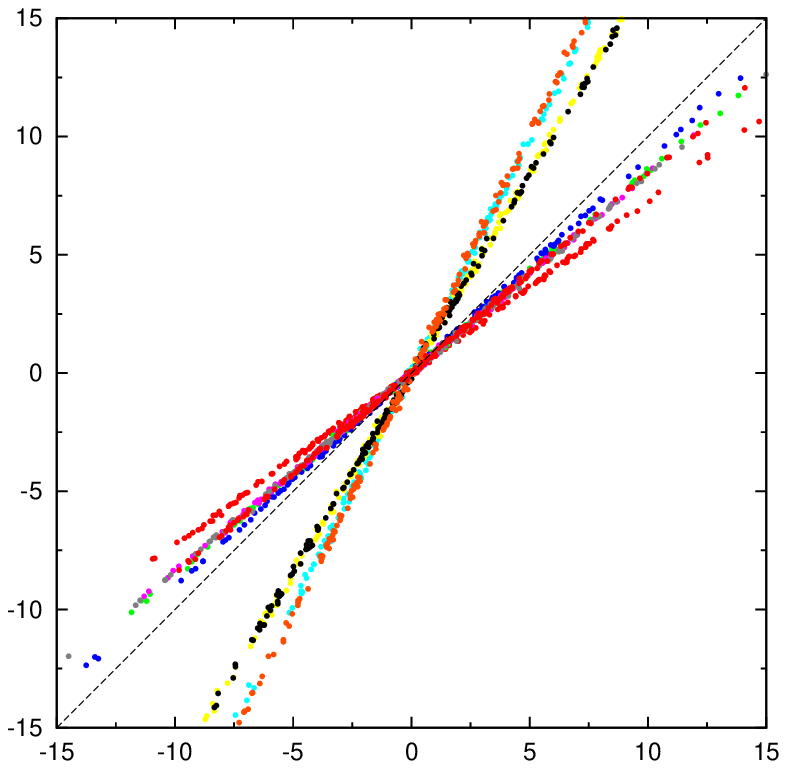}}
\put(300,0){\includegraphics[width=143\unitlength]{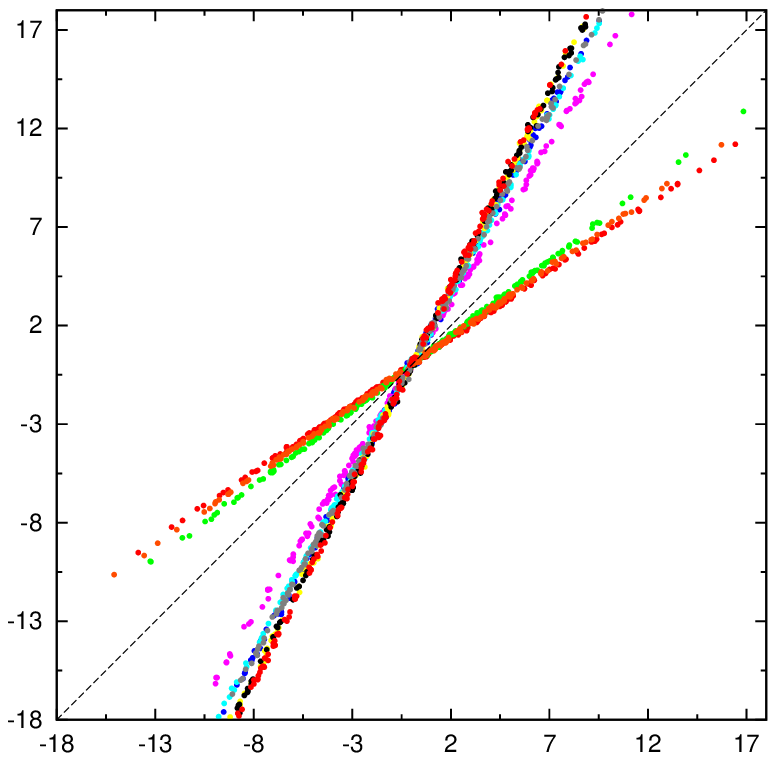}}

\put(60,-8){\small\em $\beta_\mu$}
\put(160,-8){\small\em $\beta_\mu$}
\put(260,-8){\small\em $\beta_\mu$}
\put(360,-8){\small\em $\beta_\mu$}
\put(30,86){\fs $S\!=\!3\sqrt{p}$}\put(30,75){\fs $\lambda(t)\!= \! a/\sqrt{t}$}
\put(130,86){\fs $S\!=\!4\sqrt{p}$}\put(130,75){\fs $\lambda(t)\!= \! a/\sqrt{t}$}
\put(230,86){\fs $S\!=\!5\sqrt{p}$}\put(230,75){\fs $\lambda(t)\!= \! a/\sqrt{t}$}
\put(330,86){\fs $S\!=\!6\sqrt{p}$}\put(330,75){\fs $\lambda(t)\!= \! a/\sqrt{t}$}
\put(30,186){\fs $S\!=\!3\sqrt{p}$}\put(30,175){\fs $\lambda(t)\!=\!1$}
\put(130,186){\fs $S\!=\!4\sqrt{p}$}\put(130,175){\fs $\lambda(t)\!=\!1$}
\put(230,186){\fs $S\!=\!5\sqrt{p}$}\put(230,175){\fs $\lambda(t)\!=\!1$}
\put(330,186){\fs $S\!=\!6\sqrt{p}$}\put(330,175){\fs $\lambda(t)\!=\!1$}
\put(5,55){\small\em $\hat{\beta}_\mu$}
\put(5,155){\small\em $\hat{\beta}_\mu$}

\end{picture}
\vsp

\caption{\small 
Examples of non-ergodicity   in Cox regression, for large values of $\zeta$ and $S$, signalled by the breaking up of the single linear data cloud found for small $S$ into  multiple linear clouds, each with distinct slopes (that depend on the realisation of the data set). As in Figure \ref{fig:overfitting2}, we show true versus inferred association coefficients. In all cases  $N=500$, $\zeta=0.4$ and $S/\sqrt{p}\in\{3,4,5,6\}$, and all plots show  data from 10 independent simulations (where each simulation  is given a different colour). Top row: $\lambda_0(t)=1$; bottom row: $\lambda(t)=a/\sqrt{t}$, with $a$ such that $\int\!\rmd t~p(t)t=1$ 
}
\label{fig:RSB}
\end{figure}

The present RS theory has so far been tested only for `normal' regimes  for the parameter $S$, which represents the typical width of the sum $\sum_{\mu}\beta_\mu^\star z_\mu/\sqrt{p}$, and hence the typical scale of the covariate-conditioned hazard rates. It turns out that upon carrying out Cox regression for synthetic survival data with large values of $\zeta$ and very large values of $S$, we  observe ergodicity breaking: upon plotting true versus inferred association parameters, as in Figure \ref{fig:overfitting3}, for different simulation experiments with the same parameters $N$ and $p$, we now find multiple data clouds with distinct slopes, as opposed to a single data cloud with unique reproducible characteristics. This  suggest that the relevant saddle points in the replica calculation will no longer be replica-symmetric. This phenomenology, of which examples are shown in Figure \ref{fig:RSB},  can be studied in a natural way within the replica formalism, but it requires so-called RSB (replica symmetry breaking) ans\"{a}tze for the overlap matrix $\bC$. One anticipates that for sufficiently large values of $\zeta$ there may be a critical value of $S/\sqrt{p}$ that marks an RSB transition, i.e. the onset of non-ergodicity; the preliminary data in Figure \ref{fig:RSB} suggest that this critical value may also depend on the shape of the true base hazard rate.
 Computing these critical values of $S$ from the replica formalism, in terms of the parameters $\zeta$, $S$ and $\lambda(t)$, will be the subject of a future study.  

\section{Discussion}

The Cox model has been by far  the most popular and effective statistical tool for the analysis of time-to-event data
in medicine, since its publication nearly half a century ago. However, the demands on statistical methods in 21st century medicine are changing. We can now take measurements on individual patients of unprecedented dimensionality $p$, such as gene expressions and high-resolution imaging data, but the typical number of samples $N$ in our medical data bases has not  grown in proportion. As a result, the condition for maximum likelihood (ML) multivariate regression methods (including the model of Cox) to be applicable, being $p/N\ll 1$ in order to avoid overfitting, is nowadays very often not met. 
Apart from a few early (and modest) simulation experiments, there appear not to have been any published studies aimed at modelling mathematically the mechanism of overfitting in Cox regression,  which is a prerequisite for the development of methods to deal with the overfitting problem. 
When the dimensionality of the data, relative to the number of available samples, is too high to justify using the multivariate Cox model, medical statisticians and epidemiologists are presently left having to resort to poor  alternatives for proper regression: they can either limit a priori the number of covariates used in regression (and thereby limit outcome prediction potential), or switch to univariate analysis (which is undesirable since we know that univariate estimates of association parameters correlate poorly with their multivariate counterparts), or work with so-called `risk signatures' (which tend to involve ad-hoc definitions, 
and ad-hoc recipes for interpretation). Thus, expensive and potentially informative high-dimensional clinical data remain under-utilised.

Our regression simulations with synthetic survival data show clearly  that the mechanism of overfitting in Cox regression is surprisingly reproducible and consistent: it always leads to a  clear bias, which reports association parameter values that are more extreme than their true values, underestimates base hazard rates for short times, and over-estimates base hazard rates for large times.  This consistency suggests  that it must in principle be possible to model overfitting mathematically, and that (if such modelling is successful) one should be able to correct the outcomes of Cox regression systematically for the impact of overfitting.  This, in turn, would allow us to do multivariate regression reliably for significantly larger ratios of the number of covariates over the number of samples, and obtain more accurate and reproducible predictions of clinical outcomes. 

In this paper we have presented such a theory, which is built on the mathematical methods of statistical mechanics and inspired by Gardner's famous analysis of binary classifiers \cite{Gardner}. It assumes that $N$ is large, but with $p/N$ finite, and it combines three ideas: (i) the formulation of an information-theoretic measure of overfitting in time-to-event regression, (ii) translating the calculation of this quantity into computing the ground state of a statistical mechanical system, and (iii) dealing with the heterogeneity in the problem (here: the realisation of the data set) with the replica method. 
Our modeling approach is generic. It is developed initially for arbitrary parametrised time-to-event regression models, but we devote most of our paper to the Cox model, in recognition of its importance and dominance in the medical statistics field. 
We show that by combining the above three ideas, it is possible to derive explicit macroscopic equations, exact in the asymptotic limit, with which to characterise the regression process for finite values of the ratio $p/N$. In this paper we assume that the regression process is ergodic, and make the so-called replica symmetric (RS) ansatz for the solution of our equations; this assumption is supported by numerical simulations, provided the true association parameters are not too large. 

{\red For the Cox model,} the order parameters of the RS theory contain all the relevant information required to quantify the impact of overfitting, but since one of them is a function (the inferred integrated base hazard rate), we introduced a suitable variational approximation, which resulted in a much simpler three-parameter theory. The simplified theory makes various qualitative predictions that are confirmed by regression simulations with synthetic data: that the `inflation' of inferred association parameters is independent of the amplitude of the true association parameters and of the true base hazard rate, that there is a phase transition when $p/N\to 1$, that the base hazard rate  is underestimated for short times and over-estimated for large times, and that the relation between inferred and true integrated base hazard rate is for large times of the form $\log \hat{\Lambda}(t)\sim \rho\log \Lambda_0(t)$, with a parameter $\rho$ that increases with the ratio $\zeta=p/N$.  
The quantitative agreement between our variational theory and regression simulations with synthetic data is generally very good, modulo finite size fluctuations, including the predicted overfitting-induced bias in association parameters. The only exception is the integrated base rate at large times, where available data are sparse, and where the variational ansatz (which incorporates only the leading order time dependence) under-estimates the impact of overfitting. 
Upon increasing the values of $\zeta$ and $S$, we observe new phenomenology, such as ergodicity breaking in the regression process (which requires order parameters with broken replica symmetry, or RSB). The calculation of the RSB transition line will be the subject of a subsequent paper.  
 
 The present study represents only a first step. It demonstrates that it is possible to model overfitting in Cox regression mathematically,  using the replica formalism. We envisage many direct extensions, such as {\red increasing the precision of our predictions by constructing full non-variational solutions to our RS order parameter equations (analytically or numerically)}, the incorporation of censoring,  and the addition of MAP-type regulariser terms. More technical potential follow-up studies could investigate RSB phenomena, including the calculation of the ergodicity breaking transition line, or the impact of having covariate distributions for which the sums $\sum_\mu\beta_\mu z_\mu$ no longer have Gaussian statistics. 
 Casting the net somewhat wider, and given our more general initial formulation of the theory, we expect that there will be other survival analysis models for which a similar overfitting analysis can be done. 
 
Last but certainly not least, we would now like to explore the potential of our methodology 
to provide practical tools with which to correct {\red multivariate} Cox regression analyses of real time-to-event data in medicine for the impact of overfitting. Such tools could be used retrospectively, to determine objectively which past results in the medical literature that were obtained with the Cox method can be trusted, and which perhaps cannot.  They should hopefully also lead to more accurate clinical outcome predictions in the future, by allowing medical statisticians to include  more covariates in multivariate regression by default, without overfitting danger, and enable the construction of sample size tables for multivariate regression that allow overfitting effects to be taken into account in the design of clinical trials.  The results {\red presented} in this paper suggest that in the near future, with proper overfitting corrections,  reliable multivariate  regression for time-to-event data at ratios of up to $p/N\approx 0.5$ or higher will be quite feasible. 
\vspace*{\fill}

\noindent
{\bf Acknowledgements}
\\[0.5mm]
We would like to thank Bryan Lutchmanen for contributing to the 
regression simulation studies, and Anita Grigoriadis for the data used to produce Figure \ref{fig:overfitting}. We are also grateful for  support from Saddle Point Science, the Engineering and Physical Sciences Research Council (EPSRC), and the Medical Research Council (MRC) of the United Kingdom.

\appendix

{\red 
\section{Covariate correlations in Cox regression} 
\label{app:correlations}

In the absence of censoring, the equations from which to compute the inferred base hazard rate $\hat{\lambda}(t)$ and the inferred association parameters $\hat{\bbeta}\in\R^p$ in Cox regression are the following \cite{Cox}:
\begin{eqnarray}
\hat{\lambda}(t)&=& \frac{\sum_{i=1}^N\delta(t-t_i)}{\sum_{i=1}^N \theta(t_i-t)\rme^{\hat{\bbeta}\cdot\bz_i}}
\\
\hat{\bbeta}& =& {\rm argmax}_{\bbeta} ~\sum_{i=1}^N\Big\{  \bbeta\!\cdot\! \bz_i
- \log\Big[\sum_{j=1}^N \theta(t_j-t_i)\rme^{\bbeta\cdot\bz_j}\Big]
\Big\}
\end{eqnarray}
Let us define the average values and correlations of the covariates as $\bra \bz\ket=\bar{\bz}$  and $\bra (z_\mu\!-\!\bar{z}_\mu) (z_\nu\!-\!\bar{z}_\nu)\ket=A_{\mu\nu}$, with $\bra f(\bz)\ket=N^{-1}\sum_{i=1}^N f(\bz_i)$. 
We can then simply write the original $\{\bz_i\}$ in terms of zero-average and uncorrelated covariate vectors $\{\tilde{\bz}_i\}$, by writing $\bz_i=\bar{\bz}+\bA^{\frac{1}{2}}\tilde{\bz}_i$. The equation for the regression parameters thereby becomes
\begin{eqnarray}
\hspace*{-10mm}
\hat{\bbeta}&=& {\rm argmax}_{\bbeta} ~\sum_{i=1}^N\Big\{  \bbeta\!\cdot\! \bar{\bz}+  \bbeta\!\cdot\! \bA^{\frac{1}{2}}\tilde{\bz}_i
- \log\Big[\sum_{j=1}^N \theta(t_j-t_i)\rme^{\bbeta\cdot\bar{\bz}+\bbeta\cdot\bA^{\frac{1}{2}}\tilde{\bz}_j}\Big]
\Big\}
\nonumber
\\
\hspace*{-10mm}
&=& {\rm argmax}_{\bbeta} ~\sum_{i=1}^N \Big\{  (\bA^{\frac{1}{2}}\bbeta)\!\cdot\! \tilde{\bz}_i
- \log\Big[\sum_{j=1}^N \theta(t_j-t_i)\rme^{(\bA^{\frac{1}{2}}\bbeta)\cdot\tilde{\bz}_j}\Big]
\Big\}
\end{eqnarray}
Hence $\hat{\bbeta}= \bA^{-\frac{1}{2}}\tilde{\bbeta}$, in which $\tilde{\bbeta}$ is the regression outcome of the Cox method applied to the {\em zero-average, uncorrelated and normalized} covariates $\{\tilde{\bz}_i\}$, i.e. 
\begin{eqnarray}
\tilde{\bbeta}&=& {\rm argmax}_{\bbeta} ~\sum_{i=1}^N \Big\{ \bbeta\!\cdot\! \tilde{\bz}_i
- \log\Big[\sum_{j=1}^N \theta(t_j-t_i)\rme^{\bbeta\cdot\tilde{\bz}_j}\Big]
\Big\} 
\end{eqnarray}
Similarly, for the base hazard rate we find:
\begin{eqnarray}
\hspace*{-10mm}
\hat{\lambda}(t)&=& \frac{\sum_{i=1}^N\delta(t-t_i)}{\sum_{i=1}^N \theta(t_i-t)\rme^{\hat{\bbeta}\cdot\bar{\bz}+\hat{\bbeta}\cdot \bA^{\frac{1}{2}}\tilde{\bz}_i}}
~=~\rme^{-\hat{\bbeta}\cdot\bar{\bz}}\frac{\sum_{i=1}^N\delta(t-t_i)}{\sum_{i=1}^N \theta(t_i-t)\rme^{\tilde{\bbeta}\cdot\bz_i}}
\end{eqnarray}
Hence $\hat{\lambda}(t)=\tilde{\lambda}(t)\exp(-\tilde{\bbeta}\!\cdot\!\bA^{-\frac{1}{2}}\bar{\bz})$, in which $\tilde{\lambda}(t)$ is given by Breslow's formula (the regression outcome for the base hazard rate of the Cox method) applied once more to the zero-average uncorrelated and normalised covariates $\{\tilde{\bz}_i\}$, i.e. 
\begin{eqnarray}
\tilde{\lambda}(t)&=&
\frac{\sum_{i=1}^N\delta(t-t_i)}{\sum_{i=1}^N \theta(t_i-t)\rme^{\tilde{\bbeta}\cdot\tilde{\bz}_i}}
\end{eqnarray}
We conclude that for the Cox model one can always express the regression outcomes for any choice of covariate vectors in terms of the regression outcomes for zero-average, normalized and uncorrelated covariates, where $\bra z_\mu\ket=0$ and $\bra z_\mu z_\nu\ket=\delta_{\mu\nu}$.

\section{Deriviation of the replica symmetric equations}
\label{app:RS}

Assuming replica symmetry to hold 
 converts our problem into calculating
 \begin{eqnarray}
 \hspace*{-15mm}
 E_{\gamma}(S,\lambda_0)&=&  
  \frac{\partial}{\partial\gamma}
 {\rm extr}_{C,c,c_0;\lambda}\Psi_{\rm RS}[C,c,c_0;\lambda]
 \label{eq:Cox_towards_RS1}
 \\
  \hspace*{-15mm}
 \Psi_{\rm RS}[C,c,c_0;\lambda]&=& \lim_{n\to 0}\frac{1}{n}\Big\{
\frac{1}{2}\log{\rm Det}\bC
-\frac{1}{2}\zeta \log{\rm Det}\bC^\prime
 \label{eq:Cox_towards_RS2}
\\&&
 \hspace*{-0mm}
-\log \int\!\frac{\rmd\by}{\sqrt{2\pi}}~\rme^{-\frac{1}{2}\by\cdot\bC^{-1}\by}\!\int\!\rmd t~p(t|y_0,\lambda_0)
 \prod_{\alpha=1}^n\Big[
\frac{p(t|y_\alpha,\lambda)}{p(t|y_0,\lambda_0)}
\Big]^\gamma
\Big\}
\nonumber
\end{eqnarray}
To proceed we need the determinant and inverse of the $(n\!+\!1)\times(n\!+\!1)$ covariance matrix $\bC$, and the determinant of the 
$n\times n$ matrix $\bC^\prime$.
Both $\bC$ and $\bC^{-1}$  will inherit the assumed replica-symmetric (RS) structure of the saddle-point. Hence they must have the respective forms
\begin{eqnarray}
\hspace*{-15mm}
\bC=\left(\!\!\begin{array}{ccccc}
S^2   & c_{0} & \cdots & \cdots & c_{0}\\
c_{0}       & C      &  c         & \cdots & c  \\
 \vdots   & c       & C         &   \cdots & c  \\
\vdots    & \vdots & \vdots &  \ddots &\vdots \\
c_0        & c & \cdots & c          & C
\end{array}   
\!\!\right)
~~~~~~
\bC^{-1}=\left(\!\!\begin{array}{ccccc}
d_{00}     & d_{0} & \cdots & \cdots & d_{0}\\
d_{0}       & D      &  d         & \cdots & d \\
 \vdots   & d       & D        &   \cdots & d  \\
\vdots    & \vdots & \vdots &  \ddots &\vdots \\
d_0        & d & \cdots & d         & D
\end{array}   
\!\!\right)
\label{eq:RSmatrices}
\end{eqnarray}
The RS eigenvectors $\bx$ and eigenvalues $\mu$ of $\bC$ are calculated easily:
\begin{eqnarray}
\hspace*{-15mm}
\bx=(u,v,\ldots,v):&& \mu_\pm=\frac{1}{2}\Big\{
C+(n\!-\!1)c\!+\!S^2\pm\!\sqrt{[C\!+\!(n\!-\!1)c-S^2]^2\!+\!4nc_0^2}
\Big\}
\nonumber
\\[-1mm]
\hspace*{-15mm}&&\\
\hspace*{-15mm}
\bx=(0,v_1,\ldots,v_n):&~&~\sum_{\alpha=1}^n v_\alpha=0,~~~\mu=C\!-\!c~~~({\rm multiplicity}~n\!-\!1)
\end{eqnarray}
It follows that 
\begin{eqnarray}
\log{\rm Det}\bC&=&\log[(C\!-\!c)^{n-1}\mu_+\mu_-]
\nonumber
\\[1mm]&=&
\log\Big[S^2
(C\!-\!c)^{n-1}
\Big(
C\!-\!c+n(
c\!-\!c_0^2/S^2)
\Big)
\Big]
\\
&=&\log S^2+n\log(C\!-\!c)
+n
\frac{c\!-\!c_0^2/S^2}{C\!-\!c}+\order(n^2)
\label{eq:logdetC}
\end{eqnarray}
We obtain the parameters $(D,d,d_{00},d_0)$  by multiplying the two matrices in (\ref{eq:RSmatrices}) and demanding that this gives the identity matrix. After some simple algebra this results in:
\begin{eqnarray}
&&\hspace*{-5mm} 
d_{00}= \frac{C+(n\!-\!1)c}{S^2(C+(n\!-\!1)c)-nc_0^2},~~~~~~~~
d_0= - \frac{c_0}{S^2(C+(n\!-\!1)c)-nc_0^2}
\label{eq:inverseentries1}
\\
&&\hspace*{-5mm} 
d= \frac{1}{C\!-\!c}~\frac{c^2_0-cS^2}{S^2(C+(n\!-\!1)c)-nc_0^2},
~~~~
D= d+\frac{1}{C\!-\!c}
\label{eq:inverseentries2}
\end{eqnarray}
It is now a trivial matter to calculate also the quantity $\log{\rm Det} \bC^\prime $, since the RS form of $\bC$ implies that for $\alpha,\rho=1\ldots n$ we have $C^\prime_{\alpha\rho}=\delta_{\alpha\rho}(C\!-\!c)+c-(c_0/S)^2$. It has one eigenvector $(1,\ldots,1)$ with eigenvalue $C\!-\!c\!-\!nc_0^2/S^2+nc$, and an $(n\!-\!1)$-fold degenerate eigenspace with eigenvalue $C\!-\!c$. 
Hence
\begin{eqnarray}
\log{\rm Det}\bC^\prime &=& (n\!-\!1)\log (C\!-\!c)+\log\Big(C\!-\!c+n[c\!-\!c_0^2/S^2]\Big)
\nonumber
\\
&=& n\log(C\!-\!c)+\log \Big(1+n\frac{c\!-\!c_0^2/S^2}{C\!-\!c}\Big)
\nonumber
\\
&=& n\Big[\log(C\!-\!c)+\frac{c\!-\!c_0^2/S^2}{C\!-\!c}\Big]+\order(n^2)
\end{eqnarray}
Inserting these  results into (\ref{eq:Cox_towards_RS2}) gives, with the short-hand $\rmD y=(2\pi)^{-1/2}\rme^{-\frac{1}{2}y^2}\rmd y$, 
and upon carrying out successive Taylor expansions for small $n$:
\begin{eqnarray}
\hspace*{-20mm}
\Psi_{\rm RS}[C,c,c_0;\lambda]&=& \lim_{n\to 0}\Big\{
\frac{1}{2}(1\!-\!\zeta)\Big[\log(C\!-\!c)
+\frac{c\!-\!c_0^2/S^2}{C\!-\!c}\Big]
+
\frac{1}{n}
\log S
\nonumber
\\[2mm]
\hspace*{-20mm}
&&\hspace*{-10mm}
-\frac{1}{n}\log \int\!\frac{\rmd\by}{\sqrt{2\pi}}\rme^{-\frac{1}{2}d_{00}y_0^2-\frac{1}{2}(D-d)\sum_{\alpha=1}^ny_\alpha^2-\frac{1}{2}d(\sum_{\alpha=1}^n y_\alpha)^2-d_0y_0\sum_{\alpha=1}^n y_\alpha}
\nonumber
\\[-1mm]
\hspace*{-20mm}
&&\hspace*{30mm} \times
\int\!\rmd t~p(t|y_0,\lambda_0)
 \prod_{\alpha=1}^n\Big[
\frac{p(t|y_\alpha,\lambda)}{p(t|y_0,\lambda_0)}
\Big]^\gamma
\Big\}
\nonumber
\\[1mm]
\hspace*{-20mm}
&=& \lim_{n\to 0}\Big\{
\frac{1}{2}(1\!-\!\zeta)\Big[\log(C\!-\!c)
+\frac{c\!-\!c_0^2/S^2}{C\!-\!c}\Big]
+
\frac{1}{2n}
\log (S^2 d_{00})
\nonumber
\\
\hspace*{-20mm}
&&
-\frac{1}{n}\log\int\!\rmD z \rmD y_0
\int\!\rmd t~p(t|\frac{y_0}{\sqrt{d_{00}}},\lambda_0)
\nonumber
\\
\hspace*{-20mm}
&&\hspace*{5mm}
\times
 \Big[ \int\!\rmd y~\rme^{-\frac{1}{2}(D-d)y^2-y(d_0y_0/\sqrt{d_{00}}+\rmi z\sqrt{d})}
\Big(\frac{p(t|y,\lambda)}{p(t|\frac{y_0}{\sqrt{d_{00}}},\lambda_0)}
\Big)^{\!\gamma}
\Big]^n
\Big\}
\nonumber
\\
\hspace*{-20mm}
&=&
\frac{1}{2}(1\!-\!\zeta)\Big[\log(C\!-\!c)
+\frac{c\!-\!c_0^2/S^2}{C\!-\!c}\Big]
\nonumber
\\\hspace*{-20mm}&&
+ \lim_{n\to 0}
\frac{1}{2n}
\log \Big[ \frac{1+nc/(C\!-\!c)}{1+n[c-c_0^2/S^2]/(C\!-\!c)}\Big]
\nonumber
\\
\hspace*{-20mm}
&&
\hspace*{0mm}
- \lim_{n\to 0}\int\!\rmD z \rmD y_0
\int\!\rmd t~p(t|\frac{y_0}{\sqrt{d_{00}}},\lambda_0)
\nonumber
\\
\hspace*{-20mm}
&&\hspace*{5mm}\times
\log \int\!\rmd y~\rme^{-\frac{1}{2}y^2/(C-c)-y(d_0y_0/\sqrt{d_{00}}+\rmi z\sqrt{d})}
\Big(\frac{p(t|y,\lambda)}{p(t|\frac{y_0}{\sqrt{d_{00}}},\lambda_0)}
\Big)^{\!\gamma}
\nonumber
\\
\hspace*{-20mm}
&=&
\frac{1}{2}(1\!-\!\zeta)\Big[\log(C\!-\!c)
+\frac{c\!-\!c_0^2/S^2}{C\!-\!c}\Big]
+\frac{1}{2}\frac{c_0^2/S^2}{C\!-\!c}-\frac{1}{2}\log (C\!-\!c)
\nonumber
\\
\hspace*{-20mm}
&&
\hspace*{0mm}
-\frac{1}{2}\log(2\pi)
-\int\!\rmD z \rmD y_0
\int\!\rmd t~p(t|Sy_0,\lambda_0)
\nonumber
\\
\hspace*{-20mm}
&&\hspace*{0mm}\times
\log \int\!\rmD y~\rme^{y[y_0c_0/S\sqrt{C-c}+ z\sqrt{(c-c_0^2/S^2)/(C-c)}]}
\Big(\frac{p(t|y\sqrt{C\!-\!c},\lambda)}{p(t|Sy_0,\lambda_0)}
\Big)^{\!\gamma}
\nonumber
\\[0mm]
\hspace*{-20mm}&&
\end{eqnarray}
This expression takes a simpler form if we introduce the following transformation of the trio $\{C,c,c_0\}$ to new non-negative variables $\{u,v,w\}$:
\begin{eqnarray}
u=\sqrt{C-c},~~~~v=\sqrt{c-c_0^2/S^2},~~~~w=c_0/S
\end{eqnarray}
with inverse transformation
\begin{eqnarray}
c_0=Sw,~~~~c=v^2+w^2,~~~~C=u^2+v^2+w^2
\end{eqnarray}
With these definitions, and upon removing terms that vanish upon differentiation by $\gamma$,  we can summarise the current state of our RS calculations for the stochastic generalization of the Cox model, in the limit of large data sets, by the following compact expression:
 \begin{eqnarray}
  \hspace*{-15mm} 
 E_{\gamma}(S,\lambda_0)&=&  
 \frac{\partial}{\partial\gamma}
 {\rm extr}_{u,v,w;\lambda}\Big\{
\frac{1}{2}(1\!-\!\zeta)v^2/u^2
+\frac{1}{2}w^2/u^2-\zeta\log u
\label{eq:Cox_RS}
\\
 \hspace*{-15mm} &&\hspace*{-3mm}
-\int\!\rmD z \rmD y_0
\int\!\rmd t~p(t|Sy_0,\lambda_0)
\log \int\!\rmD y~\rme^{y(wy_0+ vz)/u}
\Big(\frac{p(t|uy,\lambda)}{p(t|Sy_0,\lambda_0)}
\Big)^{\!\gamma}
\Big\}
\nonumber
\end{eqnarray}
If we transform $y\to y+(wy_0+vz)/u$, we can write this result equivalently as
 \begin{eqnarray}
 \hspace*{-15mm} 
 E_{\gamma}(S,\lambda_0)&=&  
 \int\! \rmD y_0
\int\!\rmd t~p(t|Sy_0,\lambda_0)
\log p(t|Sy_0,\lambda_0)
\nonumber
\\
 \hspace*{-15mm} 
&&
-
\frac{\partial}{\partial\gamma}
 {\rm extr}_{u,v,w;\lambda}\Big\{
\zeta \Big(\frac{v^2}{2u^2}
+\log u\Big)
\label{eq:Cox_nearly_final}
\\
 \hspace*{-15mm} 
 &&
 \hspace*{8mm}
+\int\!\rmD z \rmD y_0
\int\!\rmd t~p(t|Sy_0,\lambda_0)
\log \int\!\rmD y~p^\gamma(t|uy\!+\!wy_0\!+\!vz,\lambda)
\Big\}
\nonumber
\end{eqnarray}
 At the relevant saddle point, the order parameter derivative of the function that is being extremized will by definition be zero, 
 so
 \begin{eqnarray}
  \hspace*{-15mm} 
 E_{\gamma}(S,\lambda_0)&=&  
 \int\! \rmD y_0
\int\!\rmd t~p(t|Sy_0,\lambda_0)\left\{\room
\log p(t|Sy_0,\lambda_0)
\right.
\\
 \hspace*{-15mm} 
&&\hspace*{5mm}
\left.
-\int\!\rmD z \left[\frac{\int\!\rmD y~p^\gamma(t|uy\!+\!wy_0\!+\!vz,\lambda)\log p(t|uy\!+\!wy_0\!+\!vz,\lambda)}
{\int\!\rmD y~p^\gamma(t|uy\!+\!wy_0\!+\!vz,\lambda)}
\right]
\right\}
\nonumber
\end{eqnarray}
in which the order parameters $\{u,v,w;\lambda\}$ are to be evaluated at the saddle point of 
\begin{eqnarray}
\hspace*{-10mm}
\Psi_{\rm RS}(u,v,w;\lambda)&=& \zeta \Big(\frac{v^2}{2u^2}
+\log u\Big)
\\
\hspace*{-10mm}&&\hspace*{-5mm}
+\int\!\rmD z \rmD y_0
\int\!\rmd t~p(t|Sy_0,\lambda_0)
\log \int\!\rmD y~p^\gamma(t|uy\!+\!wy_0\!+\!vz,\lambda)
\nonumber
\end{eqnarray}

\section{The limits $\zeta\to 0$ and $\zeta\to 1$}
\label{app:limits}

For $\zeta\to 0$, the limit of no overfitting, we immediately find from (\ref{eq:spe_u_Cox_explicit_more},\ref{eq:spe_v_Cox_explicit_evenmore}) that $\tilde{u},v\to 0$. To find also $w$ and $\lambda(t)$ we need to go to the next order in $\zeta$, using $W(z)=z+\order(z^2)$. This results in 
\begin{eqnarray}
\frac{\zeta v^2}{\tilde{u}^4}
&=& \int\!\rmD z \rmD y_0
\int\!\rmd t~p(t|Sy_0,\lambda_0)\Big[1-\rme^{wy_0+vz}\Lambda(t)\Big]^2+\order(\tilde{u}^2)
\\
 \frac{\zeta}{\tilde{u}^2} &=&
\int\!\rmD z \rmD y_0
\int\!\rmd t~p(t|Sy_0,\lambda_0)\rme^{wy_0+vz}\Lambda(t)+\order(\tilde{u}^2)
\\
0&=&
\int\!\rmD z \rmD y_0~y_0
\int\!\rmd t~p(t|Sy_0,\lambda_0)\rme^{wy_0+vz}\Lambda(t)+\order(\tilde{u}^2)
\\
\frac{p(t)}{\lambda(t)}&=& \int\!\rmD z \rmD y_0\int_t^\infty\!\rmd t^\prime~p(t^\prime|Sy_0,\lambda_0)
\rme^{wy_0+vz}+\order(\tilde{u}^2)
\end{eqnarray}
It follows that $v=\order(\tilde{u})$ and $\tilde{u}=\order(\sqrt{\zeta})$ for $\zeta\to 0$, and that $\lim_{\zeta\to 0}w$ and $\lim_{\zeta\to 0}\lambda(t)$ are to be solved from the following two coupled equations: 
\begin{eqnarray}
0&=&
\int\!\rmD y_0~y_0
\int\!\rmd t~p(t|Sy_0,\lambda_0)\rme^{wy_0}\Lambda(t)
\\
\frac{p(t)}{\lambda(t)}&=& \int\! \rmD y_0\int_t^\infty\!\rmd t^\prime~p(t^\prime|Sy_0,\lambda_0)
\rme^{wy_0}
\end{eqnarray}
After some simple rewriting and integration by parts over time, they take the alternative forms
\begin{eqnarray}
0&=&
\int\!\rmD y_0~y_0\rme^{(w-S)y_0}
\int\!\rmd t~p(t|Sy_0,\lambda_0)\frac{\lambda(t)}{\lambda_0(t)}
\\
p(t)&=& \int\! \rmD y_0~\rme^{(w-S)y_0}p(t|Sy_0,\lambda_0)\frac{\lambda(t)}{\lambda_0(t)}
\rme^{wy_0}
\end{eqnarray}
From this we immediately confirm the correct solution $\lim_{\zeta\to 0}w=S$ and $\lim_{\zeta\to 0}\lambda(t)=\lambda_0(t)$, which describes perfect inference, as expected for $\zeta \to 0$. From the pair (\ref{eq:compareE},\ref{eq:Wresult}) we also find the correct  corresponding value for $\lim_{\zeta\to 0}\lim_{\gamma\to\infty}  E_{\gamma}(S,\lambda_0)$:
\begin{eqnarray}
\lim_{\zeta\to 0}\lim_{\gamma\to\infty}\Prob_\gamma(x,x^\prime,t)&=& 
\int\! \rmD y_0~p(t|Sy_0,\lambda_0)\delta[x\!-\!Sy_0] \delta[x^\prime\!-\!Sy_0] 
\\
\lim_{\zeta\to 0}\lim_{\gamma\to\infty} E_{\gamma}(S,\lambda_0) & =& 0
\end{eqnarray}

Next we turn to the limit $\zeta\to 1$. Here it follows from (\ref{eq:spe_v_Cox_explicit_evenmore}) that $\tilde{u}\to\infty$, and we need  the expansion of $W(z)$ for large arguments, i.e. $W(z)=\log z-\log(\log z)+\ldots$. With a modest amount of foresight we make the ansatz $\tilde{u}=\kappa/\sqrt{1-\zeta}+\order(\log (1/(1-\zeta))$ and $v,w=\order(\log (1/(1-\zeta))$ for $\zeta\to 1$. Using 
\begin{eqnarray}
W\big(\tilde{u}^2\rme^{\tilde{u}^2+wy_0+vz}\Lambda(t)\big)&=& \frac{\kappa^2}{1-\zeta}+\order(\log(\frac{1}{1\!-\!\zeta}))
\end{eqnarray}
our $\gamma\to\infty$ order parameter equations then give
\begin{eqnarray}
\zeta v^2
&=& \int\!\rmD z \rmD y_0
\int\!\rmd t~p(t|Sy_0,\lambda_0)\Big[\order(\log(\frac{1}{1\!-\!\zeta}))
\Big]^2
\\
 \zeta &=&
\int\!\rmD z \rmD y_0
\int\!\rmd t~p(t|Sy_0,\lambda_0)[1-\order(1\!-\!\zeta)]
\\
0&=& \int\!\rmD z \rmD y_0~y_0
\int\!\rmd t~p(t|Sy_0,\lambda_0) ~\order\Big((1\!-\!\zeta)\log(\frac{1}{1\!-\!\zeta})\Big)
\\
\frac{p(t)}{\lambda(t)}&=& \int\!\rmD z \rmD y_0\int_t^\infty\!\rmd t^\prime~p(t^\prime|Sy_0,\lambda_0)
\frac{1}{\Lambda(t^\prime)}\nonumber
\\
&&\hspace*{30mm}\times \Big[1+\order\Big((1\!-\!\zeta)\log(\frac{1}{1\!-\!\zeta})\Big)\Big]
\end{eqnarray}
Our scaling ansatz is seen to be consistent with the three scalar order parameter equations. Hence $\tilde{u}$, $v$ and $w$ all diverge at a phase transition point $\zeta=1$, whereas for the functional order parameter equation we find in the limit $\zeta\to 1$:
\begin{eqnarray}
\frac{p(t)}{\lambda(t)}&=& \int_t^\infty\!\rmd t^\prime~\frac{p(t^\prime)}{
\Lambda(t^\prime)}
\end{eqnarray}
From this it follows after differentiation that $\frac{\rmd}{\rmd t}[p(t)\Lambda(t)/\lambda(t)]=0$, and after some further manipulations one arrives at the following degenerate solution for $\Lambda(t)$:
\begin{eqnarray}
\lim_{\zeta\uparrow 1}\lim_{\gamma\to \infty}\Lambda(t)=\left\{\begin{array}{lll}
0 & {\rm for} & t<\tau \\
1 & {\rm for} & t=\tau \\
\infty & {\rm for} & t>\tau 
\end{array}
\right.
\end{eqnarray}
Apparently, as one varies the ratio $\zeta$ of the number of covariates over the number of samples  in the deterministic Cox model,  the integrated inferred  base hazard rate changes from the correct shape 
$\Lambda_0(t)$ at $\zeta=0$ to a step function at the phase transition point $\zeta=1$, with the discontinuity at some time point $\tau$ that should follow from inspecting sub-leading orders in $1-\zeta$. Moreover, at this transition (if not even earlier) one expects to find breaking of the assumed replica symmetry. 

}

\section{Asymptotic form of the event time distribution}
\label{app:Asymptotic_g}

Here we calculate the asymptotic form of the function $g(x)=\int\!\rmD y~\rme^{Sy-x\exp(Sy)}$ for $x\to\infty$, and derive expression (\ref{eq:p_large_times}). Working out the definition gives
\begin{eqnarray}
\log g(x)&=&
\frac{1}{2}S^2+\log \int\!\frac{\rmd y}{\sqrt{2\pi}}~\rme^{-\frac{1}{2}y^2-x\exp(S^2+Sy)}
\nonumber
\\
&=&\frac{1}{2}S^2+\log \int\!\frac{\rmd y}{\sqrt{2\pi}}~\rme^{-\varphi(y,\rme^{S^2}x)}
\end{eqnarray}
with
\begin{eqnarray}
\varphi(y,\eta)&=&
\frac{1}{2}y^2+\eta \rme^{Sy}
\end{eqnarray}
Differentiation shows that the function  $\varphi(y,\eta)$ is mimimal at $y=-W(\eta S^2)$, where $W(z)$ is Lambert's $W$-function \cite{Lambert}. 
Expansion of $\varphi(y,\eta)$ around its minimum gives:
\begin{eqnarray}
\hspace*{-10mm}
\varphi(y,\eta)&=&
\frac{1}{2S^2}\Big(W(\eta S^2)\!+\!1\Big)^2\!-\frac{1}{2S^2}+\frac{1}{2}\Big[W(\eta S^2)\!+\!1\Big]\Big(y
\!+\!\frac{1}{S}W(\eta S^2)\Big)^2
\nonumber
\\
\hspace*{-10mm}
&&\hspace*{30mm}
+~\order(\big[W(\eta S^2)\!+\!1\big]\Big(y
\!+\!\frac{1}{S}W(\eta S^2)\Big)^3
\end{eqnarray}
This leads to the following Gaussian approximation of the integral over $y$:
\begin{eqnarray}
\hspace*{-0mm}
\log \int\!\frac{\rmd y}{\sqrt{2\pi}}~\rme^{-\varphi(y,\eta)}
&=& 
\frac{1}{2S^2}-\frac{1}{2S^2}\Big(W(\eta S^2)\!+\!1\Big)^2
\nonumber
\\&&
\hspace*{10mm}
+\order\Big(\log \Big[W(\eta S^2)\!+\!1\Big]\Big)
\end{eqnarray}
Application to $\eta=x\rme^{S^2}$ then gives:
{\red 
\begin{eqnarray}
\hspace*{-0mm}
\log g(x)&=&
-\frac{1}{2S^2}\big[W(xS^2\rme^{S^2})\!+\!1\big]^2\!
-\frac{1}{2}\log W(xS^2\rme^{S^2})
+\order(1)
\end{eqnarray}
Finally, for $x\to\infty$ we can use $W(z)=\log z-\log\log z+\order(\log\log z/\log z)$ to obtain
\begin{eqnarray}
\log g(x)&=&
-\frac{1}{2S^2}(\log x)^2+\frac{1}{S^2}\log x.\log\log x+\order(\log x)
\end{eqnarray}
}

\end{document}

%% file: commands.tex
\newcommand{\room}{\rule[-0.3cm]{0cm}{0.8cm}}

\newcommand{\here}{\makebox(0,0)}
\newcommand{\vsp}{\vspace*{3mm}}

\newcommand{\be}{\begin{equation}}
\newcommand{\ee}{\end{equation}}
\newcommand{\bd}{\begin{displaymath}}
\newcommand{\ed}{\end{displaymath}}
\newcommand{\bea}{\begin{eqnarray}}
\newcommand{\eea}{\end{eqnarray}}

\newcommand{\pprime}{{\prime\prime}}

\newcommand{\bra}{\langle}
\newcommand{\ket}{\rangle}

\newcommand{\order}{{\mathcal O}}

\newcommand{\bnull}{\mbox{\boldmath $0$}}

\newcommand{\R}{{\rm I\!R}}

\newcommand{\bx}{\mbox{\protect\boldmath $x$}}
\newcommand{\by}{\mbox{\protect\boldmath $y$}}
\newcommand{\bz}{\mbox{\protect\boldmath $z$}}
\newcommand{\bA}{\mbox{\protect\boldmath $A$}}

\newcommand{\bC}{\mbox{\protect\boldmath $C$}}
\newcommand{\bD}{\mbox{\protect\boldmath $D$}}

\newcommand{\bbeta}{\mbox{\protect\boldmath $\beta$}}

\newcommand{\btheta}{\mbox{\protect\boldmath $\theta$}}

\newcommand{\bxi}{\mbox{\protect\boldmath $\xi$}}

\newcommand{\ds}{\rmd s}